\documentclass[aps,prd,onecolumn,preprint,nofootinbib]{revtex4-2}

\pdfoutput=1

\usepackage{amsmath,amssymb}

\usepackage{enumerate}
\usepackage{comment}
\usepackage{cancel}
\usepackage{etex}
\usepackage{amsthm}
\usepackage{dcolumn}
\usepackage{epsf}
\usepackage{mathrsfs}
\usepackage{multirow}
\usepackage{booktabs}
\usepackage{array}
\usepackage{slashed}
\usepackage{float}
\usepackage{leftidx}
\usepackage{setspace}
\usepackage{verbatim}
\usepackage{adjustbox}
\usepackage{tikz}
\usepackage[caption=false]{subfig}
\usepackage{arydshln}
\usepackage{shuffle}
\usepackage{cases}
\usepackage{lipsum}
\usepackage{tabularx}
\usepackage{graphicx}
\usepackage{cancel}
\usepackage{ulem}

\numberwithin{equation}{section}
\usetikzlibrary{arrows,decorations.markings,shapes.arrows,patterns,positioning}

\def \be {\begin{equation}}
\def \ee {\end{equation}}
\def \nn {\nonumber\\}

\usepackage[dvipsnames,usenames]{xcolor}

\newcommand{\bea}{\begin{eqnarray}}
\newcommand{\eea}{\end{eqnarray}}
\newcommand{\bean}{\begin{eqnarray*}}
\newcommand{\eean}{\end{eqnarray*}}

\newcommand{\Sl}{\sum\limits}

\def\W #1{\widetilde{#1}}

\renewcommand{\eqref}[1]{eq.~(\ref{#1})}

\newcommand{\appref}[1]{appendix~\ref{#1}}

\begin{document}

\title{Algebraic Consistency and Explicit Construction of One-Loop BCJ Numerators of Yang–Mills and Related Theories}

\author{Yi-Jian Du}
\email{yijian.du@whu.edu.cn}
\affiliation{School of Physics and Technology, Wuhan University,
No.299 Bayi Road, Wuhan 430072, P.R. China}

\author{Chih-Hao Fu}
\email{chihhaofu@hnas.ac.cn}
\affiliation{Institute of Mathematics, Henan Academy of Sciences,
NO.228, Chongshi Village, Zhengzhou, Henan 450046, P.R. China}

\author{Yihong Wang}
\email{yihong.wang@hrbeu.edu.cn}
\affiliation{College of Mathematical Sciences, Harbin Engineering University,
145 Nantong Street, Nangang District, Harbin 150001, P. R. China}

\author{Chongsi Xie}
\email{chongsi.xie@whu.edu.cn}
\affiliation{College of Science, Hainan Tropical Ocean University,\\
Sanya, Hainan 572022,P.R. China
\\
School of Physics and Technology, Wuhan University,\\
No.299 Bayi Road, Wuhan 430072, P.R. China}

\begin{abstract}
We study the algebraic structure of one-loop BCJ numerators in Yang-Mills and related
theories. Starting from the propagator matrix that connects colour-ordered integrands to
numerators, we identify the consistency conditions that ensure the existence of
Jacobi-satisfying numerator solutions and determine the unique construction. The relation between one-loop numerators and
forward-limit tree numerators is clarified, together with the additional physical
conditions required for a consistent double-copy interpretation. 

We propose a two-step expansion strategy for obtaining explicit one-loop numerators.
The Yang-Mills integrand is first decomposed into scalar-loop Yang-Mills-scalar
building blocks, which are then expanded into bi-adjoint scalar integrands. We derive
explicit results for up to three external gluons, showing how the kinematic consistency
conditions uniquely determine the coefficients in each case. Similar results for
Einstein-Yang-Mills and gravity amplitudes are also presented.
\end{abstract}

\maketitle

\maketitle

\section{Introduction}

At tree level, the Bern-Carrasco-Johansson (BCJ) duality \cite{Bern:2008qj,Bern:2010ue} states that
colour-ordered amplitudes $A(1,2,\dots,n)$ can be expressed as a
sum over cubic tree diagrams whose leaves are labelled by the external
legs $1,2,\dots,n-1$. Each diagram $\Gamma$ contributes a term of
the form 
\begin{equation}
A(1,2,\dots,n)=\sum_{\Gamma}\frac{N_{\Gamma}}{D_{\Gamma}},\label{eq:bcj-amplitude}
\end{equation}
where $D_{\Gamma}$ denotes the product of all propagators associated
with the internal branches of the cubic tree, and $N_{\Gamma}$ is
a kinematic function depending on momenta and polarisation vectors.
The essential content of the BCJ duality is that the kinematic numerators
$N_{\Gamma}$ can be arranged to satisfy the same Jacobi identities
as the colour factors of Yang-Mills theory. In this way, the algebraic
structure underlying gauge amplitudes mirrors that of the colour algebra,
allowing gravity amplitudes to be obtained through a ``double copy''
construction by replacing colour factors with kinematic numerators.

At one-loop level, this organisation naturally generalises \cite{Bern:2008qj,Bern:2010ue}: The integrand
can be written as a sum over cubic one-loop diagrams, each diagram
being a polygon whose vertices are decorated by cubic tree subgraphs
with the same cyclic ordering of external legs. Schematically,
\begin{equation}
\mathcal{I}_{n}^{\text{1-loop}}=\sum_{\Gamma}\frac{N_{\Gamma}(\ell)}{D_{\Gamma}(\ell)},\label{eq:bcj-integrand}
\end{equation}
where the denominator $D_{\Gamma}(\ell)$ is the product of the Feynman
propagators corresponding to all branches of the cubic one-loop diagram,
and $N_{\Gamma}(\ell)$ is a loop-momentum-dependent kinematic numerator
satisfying the Jacobi identities among neighbouring graphs. The central
question is then how to determine such numerators consistently, and
in particular, how their existence constrains the analytic structure
of the integrand.

In this work we focus on the algebraic consistency conditions that ensure
the existence and uniqueness of such one-loop numerators, and on how
these consistency principles can be used to construct explicit one-loop
numerators for Yang--Mills and related theories.  We also analyse the
compatibility of these constructions with the
Kawai-Lewellen-Tye
(KLT) relations \cite{Kawai:1985xq,Mafra:2011nw,Bjerrum-Bohr:2010pnr} at one loop.
In this paper, we address the following three issues related
to the problem of solving kinematic numerators and consequences on
the integrands due to the numerator structure at one loop.
\begin{enumerate}[(i)]
\item  The first issue is how to solve numerators from integrands. Recall
at tree level, the solution is based on the observation that Kleiss-Kuijf
(KK) relations \cite{Kleiss:1988ne} and Jacobi identities happen to result
in the same number of independent amplitudes and numerators, respectively \cite{Bjerrum-Bohr:2012kaa}.
So that given a KK satisfying theory, assuming the colour-ordered
amplitudes have been already computed from Feynman rules, the numerators
can be determined from linear equations with equal number of known
and unknown variables. Take the $4$-point case as an example, a KK
satisfying theory has $2$ independent colour-ordered amplitudes,
as well as $2$ independent numerators (the half-ladders \cite{Naculich:2014naa}). The two
sets are related by (\ref{eq:4pt-matrix-eqn}) with the propagator
matrix being the coefficients \cite{ref-Kiermaier}.
\begin{equation}
\left[\begin{array}{c}
A(1234)\\
A(1324)
\end{array}\right]=\left[\begin{array}{cc}
\frac{1}{s}+\frac{1}{t} & \frac{-1}{t}\\
\frac{-1}{t} & \frac{1}{u}+\frac{1}{t}
\end{array}\right]\left[\begin{array}{c}
N^{\text{tree}}(1234)\\
N^{\text{tree}}(1324)
\end{array}\right].\label{eq:4pt-matrix-eqn}
\end{equation}
The Mandelstam variables $s$, $t$ and $u$ are defined by $s=2k_1\cdot k_2$, $t=2k_1\cdot k_3$ and  $u=2k_1\cdot k_4$, where $k_i$ is the momentum of the particle $i$.
For massless theories, the propagator matrix is known to be singular
and cannot be simply inverted. The kernel of this matrix corresponds
to the BCJ relations of colour-ordered amplitudes \cite{Boels:2012sy,HenryTye:2010tcy}.
One then concludes that the necessary and sufficient condition for
an arbitrary massless theory to be allowed an consistent numerator
formulation is that the amplitudes should satisfy BCJ relations, in
which case the kinematic numerators, regarded as unknowns, can be
determined by inverting the propagator matrix (the momentum kernel
\cite{Bjerrum-Bohr:2010pnr,Mafra:2011nw,Bern:1998sv}).

From this perspective it would be interesting\footnote{This is especially so when the rank of propagator matrix has been revealed
to be further reduced from those at tree level \cite{Ochirov:2017jby}.} to see if a similar generalised argument can be used to determine
the numerators from the integrand, where the cubic structure is assumed
to reside at one loop level\footnote{Note however, for the purpose of calculating gravity amplitudes using
double-copy, which is usually the motive for studying colour-kinematics
duality, one does not actually need to fully determine the numerators.
The knowledge of the existence of kinematic numerators suffices \cite{Bern:2017yxu}.}. 

This question motivates the first part of the present paper, in which we 
analyse the algebraic consistency conditions of the one-loop propagator 
matrix and demonstrate how these conditions ensure the existence and 
uniqueness of consistent numerators.

To further illustrate the problem, let us consider a $2$-point
integrand, which is the first non-trivial example. BCJ's formulation
(\ref{eq:bcj-integrand}) demands that if the solution of numerators
exists, the $2$-point colour-ordered integrand $\mathcal{I}_{2}$
needs to be expressible as the following sum over cubic diagrams.
\begin{equation}
\mathcal{I}_{2}(\ell,k_{1},k_{2})={1\over \ell^2}{1\over (\ell+k_1)^2}\times\begin{minipage}{2cm}\includegraphics[width=2cm]{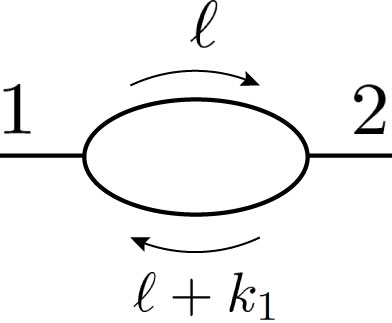}\end{minipage}+{1\over \ell^2}{1\over i\epsilon}\times\begin{minipage}{1cm}\includegraphics[width=1cm]{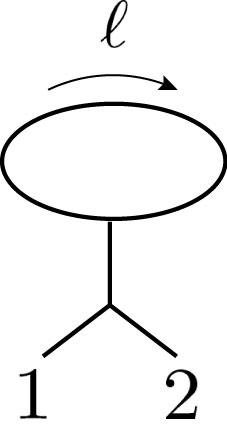}\end{minipage}+{1\over (\ell+k_1)^2}{1\over i\epsilon}\times\begin{minipage}{1cm}\includegraphics[width=1cm]{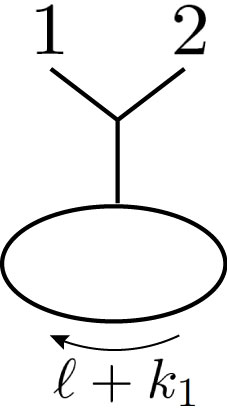}\end{minipage},
 \label{eq:defn-2pt}
\end{equation}
where $\ell$ is the loop momentum. The tree propagators in the second and the third terms have the form ${1\over (k_1+k_2)^2+i\epsilon}$, which can be written as ${1\over i\epsilon}$ due to momentum conservation\footnote{In most situations of this paper, the $i\epsilon$ is omitted for convenience.}. In view of the argument at tree level, let us re-write the right hand
side of the equation in terms of basis numerators. At one loop a commonly
used basis at $n$-point is the polygons attached with $n$ external
legs. In the $2$-point case these are the bubble graphs with various
loop momentum values $\ell$ flowing from leg $1$ towards $2$.\footnote{Note the convention for loop momentum used in this paper is slightly
different from the one naturally inherited from string theory clarified
in \cite{Tourkine:2019ukp}, and used in related problems, for
example, in \cite{Geyer:2015bja,Boels:2011tp,Du:2012mt}. }
\begin{equation}
\begin{minipage}{2.2cm}\includegraphics[width=2.2cm]{2ptEq1a}\end{minipage}
  \label{eq:bubble-numerator}
\end{equation}
In this paper we symbolically represent this numerator by $N(\ell,1,2,-\ell)$,
as a single cut from the above internal line yields a familiar half-ladder.
We would like to emphasise this imaginary ``cut'' is merely a convenient
way to understand the notation. The symbol $N(\ell,1,2,-\ell)$ represents
the whole bubble numerator (\ref{eq:bubble-numerator}). Alternatively,
one may introduce an imaginary cut from below. Therefore in this slightly
redundant notation, the following two expressions both represent the
same bubble graph.
\begin{equation}
N(\ell,1,2,-\ell)=N(\ell+k_{1},2,1,-(\ell+k_{1})).\label{eq:consistency-1}
\end{equation}
Using this notation, the Jacobi identity satisfied by the tadpole
graph can be written as the following.
\begin{equation}
 \begin{minipage}{1cm}\includegraphics[width=1cm]{2ptEq1b}\end{minipage}
 = N(\ell,1,2,-\ell)- N(\ell,2,1,-\ell). \label{eq:jacobi-1}
\end{equation}
And similarly for the other tadpole numerator. The $2$-pt integrand
$\mathcal{I}_{2}$ is then expressible symbolically as
\begin{align}
\mathcal{I}_{2}(\ell,k_{1},k_{2})= & \frac{1}{\ell^{2}(\ell+k_{1})^{2}}N(\ell,1,2,-\ell)\label{eq:2pt-integrand}\\
 & +\frac{1}{\ell^{2}}\,\frac{1}{i\epsilon}\Biggl(N(\ell,1,2,-\ell)-N(\ell-k_{1},1,2,-(\ell-k_{1}))\Biggr)\nonumber \\
 & +\frac{1}{(\ell+k_{1})^{2}}\,\frac{1}{i\epsilon}\Biggl(N(\ell,1,2,-\ell)-N(\ell+k_{1},1,2,-(\ell+k_{1}))\Biggr).\nonumber 
\end{align}

From this equation we see that the problem of solving numerators at
one loop is much more involved than its tree level counterpart (\ref{eq:4pt-matrix-eqn}).
The above linear equation can be more or less rigorously solved by
manually imposing periodic boundary conditions $\ell+Mk_{1}\sim\ell$,
for some large integer $M$ \cite{ref-jj,Carrasco:2015iwa}. At $2$-point momentum
conservation requires there is only one independent external momentum,
therefore all shifted numerators can be labelled by the amount of
shifting in the $k_{1}$ direction. Suppose if we further introduce
shorthand notations for the integrand and the numerators, $\mathcal{I}_{2}(m):=\mathcal{I}_{2}(\ell+mk_{1},k_{1},k_{2})$
and $N(m):=N(\ell+mk_{1},1,2,-(\ell+mk_{1}))$, and note that the
integrand-numerator relation (\ref{eq:2pt-integrand}) takes the following
form,
\begin{equation}
\mathcal{I}_{2}(m)=A(m)\,N(m)+B(m)\,N(m+1)+C(m)\,N(m-1).
\end{equation}
 Assuming again all integrands are known from Feynman rules, in principle
the numerators can be determined from the $M\rightarrow\infty$ solution
of the matrix equation

\begin{equation}
\resizebox{0.85\textwidth}{!}{$
\left[
\begin{array}{c}
\mathcal{I}_{2}(0)\\
\mathcal{I}_{2}(1)\\
\mathcal{I}_{2}(2)\\
\vdots\\
\mathcal{I}_{2}(M-1)
\end{array}
\right]
=
\left[
\begin{array}{cccccc}
A(0) & B(0) &        &        &        & C(0) \\
C(1) & A(1) &  B(1)  &        &        &      \\
     & C(2) &  A(2)  &  B(2)  &        &      \\
     &      & \ddots & \ddots & \ddots &      \\
     &      &        & \ddots & \ddots & \ddots \\
B(M\!-\!1) &        &        &        & C(M\!-\!1) & A(M\!-\!1)
\end{array}
\right]
\left[
\begin{array}{c}
N(0)\\
N(1)\\
N(2)\\
\vdots\\
N(M-1)
\end{array}
\right]
$}
\label{eq:infinite-matrix-eqn}
\end{equation}
The solution of this particular matrix problem is known in the mathematics
literature \cite{Gray:2006,Sarayi:2020}. However at $3$-point we have
two independent momenta, and the set of unknown variables $N(\ell+m_{1}k_{1}+m_{2}k_{2},\sigma,-(\ell+m_{1}k_{1}+m_{2}k_{2}))$
form a two dimensional lattice \cite{ref-jj,Carrasco:2015iwa}. One then needs to solve a corresponding
$(M_{1}\times M_{2})\times(M_{1}\times M_{2})$ matrix problem and
so on. It is preferably not to resort to such involved technique,
even for the purpose of understanding the problem. Furthermore, it
is difficult to check if a coefficient matrix of such a large dimension
is singular, and whether the kernel of the matrix results in relations
between colour-ordered integrands like tree level\footnote{See \cite{Boels:2011tp,Du:2012mt}, however, for
discussions long this line of thoughts.}. In section \ref{sec:shifting-used-explained} we show this difficulty
can actually be circumvented by introducing physical conditions. This
leads to a one loop generalisation of the linear relation (\ref{eq:3pt-matrix-eqn})
between integrands and numerators. In section \ref{sec:forward-limit-propagators}
we explain how the propagator matrix in this generalised linear relation
is related to the original tree propagator matrix. Additionally, using a technique based on Sherman-Morrison formula \cite{ref-Sherman-Morrison}, we
explain in \ref{sec:momentum-kernel} how the structure of the inverse propagator
matrix (the momentum kernel) is affected by this generalisation.
\item  The second issue concerns the relation between the numerator
solutions solved from loop integrand and those solved from tree amplitudes,
particularly the forward limit amplitudes.
BCJ's integrand formulation
(\ref{eq:bcj-integrand}) at one loop can be alternatively understood
as a sum over various polygons with sub-trees attached. It was observed
by \cite{Geyer:2015bja,He:2016mzd} that if we substitute the
propagators of all $n$-gons appear in the integrand using the following
partial fraction identity,
\begin{equation}
\frac{1}{\prod_{i=1}^{n}D_{i}}=\sum_{i=1}^{n}\frac{1}{D_{i}\prod_{j\neq i}(D_{j}-D_{i})}\label{eq:partial-fraction-npt},
\end{equation}
the resulting integrand can be re-grouped as a cyclic permutation sum of
forward limit tree amplitudes \cite{He:2015yua,Cachazo:2015aol,He:2016mzd,He:2017spx,Geyer:2015jch,Geyer:2017ela,Edison:2020uzf}. Explicitly at $2$-point this is done
by replacing the propagator of the bubble graph in (\ref{eq:2pt-integrand})
by
\begin{align}
\frac{1}{\ell^{2}(\ell+k_{1})^{2}} & =\frac{1}{\ell^{2}\left[(\ell+k_{1})^{2}-\ell^{2}\right]}-\frac{1}{(\ell+k_{1})^{2}\left[(\ell+k_{1})^{2}-\ell^{2}\right]} \nonumber \\
 & =\frac{1}{\ell^{2}}\,\frac{1}{2\ell\cdot k_{1}}+\frac{1}{(\ell+k_{1})^{2}}\,\frac{1}{(\ell+k_{1})\cdot k_{2}}, \nonumber
\end{align}
and then re-group the first and the second term above with the first
and the second tadpoles in (\ref{eq:2pt-integrand}) respectively,
yielding
\begin{equation}
\mathcal{I}_{2}(\ell,k_{1},k_{2})=\frac{1}{\ell^{2}}\,A_{\text{BCJ}}^{\text{tree}}(\ell,1,2,-\ell)+\frac{1}{(\ell+k_{1})^{2}}\,A_{\text{BCJ}}^{\text{tree}}(\ell+k_{1},2,1,-(\ell+k_{1})).\label{eq:tree-decomposition-2pt}
\end{equation}
Here by forward limit tree amplitude  we mean that if we construct
colour-order tree amplitudes using BCJ's formula (\ref{eq:bcj-amplitude})
for tree amplitude, formally writing down tree numerators, identifying
external legs according to the forward limit, and additionally, replacing
the tree numerator $N^{\text{tree}}(\ell,\sigma(1),\sigma(2),\dots,\sigma(n),-\ell)$
formally just written down by their corresponding symbols introduced
to denote polygon basis numerators\footnote{As discussed for instance in
\cite{He:2016mzd,He:2017spx,Geyer:2017ela,Porkert:2022efy}, where the tree-level and forward-limit structures are treated in the same sense.}, $N(\ell,\sigma(1),\sigma(2),\dots,\sigma(n),-\ell)$,
then the forward limit tree amplitudes such as $A_{\text{BCJ}}^{\text{tree}}(\ell,1,2,-\ell)$
obtained after re-grouping will be identical to the ones we just wrote
down through this procedure. Graphically speaking, these can be understood
as welding two parallel legs of the forward limit tree amplitude.
All propagators appear in these tree amplitudes are just linear in
loop momentum $\ell$, which is formally identical to tree propagator
in a massless theory. Here we introduce subscripts $A_{\text{BCJ}}^{\text{tree}}$
to indicate these are built from the cubic formulation of BCJ and
contain numerators, regarded as unknowns, which is not to be confused
with amplitudes built from Feynman rules that generically carry more
than cubic structure.

The fact that these forward limit tree amplitudes appear in the integrand
decomposition (\ref{eq:tree-decomposition-2pt}) bear such remarkable
resemblance to tree amplitudes however, seem to suggest that they
could be identified their values with the actual forward limit tree
amplitudes obtained from Feynman rules, which we shall denote as $A_{\text{Feyn}}^{\text{tree}}$
in the following. In view of the linear relation (\ref{eq:4pt-matrix-eqn})
between tree amplitudes and numerators, this is equivalent to suggesting
that we identify polygon basis numerators, $N(\ell,\sigma(1),\sigma(2),\dots,\sigma(n),-\ell)$
such as (\ref{eq:bubble-numerator}) and tree numerators in the same
theory with both ends welded together. Therefore the existence of
numerator solution at one loop requires the existence of numerator
solution at tree level, which seems to be a reasonable requirement.

A few questions then arise. One naturally wonders the relation between
the numerator solutions obtained from the knowledge of forward limit
amplitudes just described, and the numerator solutions obtained directly
from integrands (mentioned in (i) and will be explained in section
\ref{sec:shifting-used-explained}). Additionally one may ask when
such a solution exists. We will show in section \ref{sec:forward-limit-numerators}
that more conditions need to be met to guarantee a solution built
from the welded forward limit tree numerators, in addition to requiring
tree amplitudes of the same theory to satisfy BCJ relations. We shall
see that, for non-linear sigma model (NLSM), these conditions happen
to be met, and therefore yields a consistent set of kinematic numerators
(part of the verification detail will be given in appendix \ref{sec:nlsm-numerators}).
The fact that NLSM satisfies the extra conditions is not surprising,
as the theory has a natural cubic action that satisfies the Jacobi
identity \cite{Cheung:2016prv}. For the same reason this is also true for
self-dual Yang-Mills theory in general dimensions \cite{Monteiro:2011pc}.
From this  perspective, a physically  interesting BCJ structure seems to
require stronger conditions than the bypassing method of (i). In section
\ref{sec:forward-limit-numerators} we will also explain the inconsistency
caused by identifying polygon basis numerators with the welded forward
limit tree numerators in the case when these extra conditions are
violated.

Another question is how or whether the one loop numerators related
to the welded forward limit tree numerators, in the case when extra
conditions are not met, so that we cannot simply identify the two.
In section \ref{sec:loop-and-tree-relation} we will write down the
relation between these two numerators, therefore for a generic theory,
one can calculate one loop numerators if tree level numerators are
already known, and vice versa.
\item  The third issue concerns compatibility between the Kawai-Lewellen-Tye
(KLT) \cite{Kawai:1985xq,Mafra:2011nw,Bjerrum-Bohr:2010pnr} relation,
the double-copy formulation of BCJ, and related expressions of gravity
integrands. Specifically, we find that at one loop level, starting
from a double-copy formulation of the gravity integrand, suppose if
we order integrands by the method explained in section \ref{sec:forward-limit-propagators},
the resulting KLT formula either contain graphs with miss-aligned
loop momenta $\ell$, or that one copy of the integrand carries an
extra difference term $\Delta_{\ell}$,
\begin{align}
\mathcal{\tilde{I}}_{GR}(\ell,12) & =\left(\mathcal{I}_{YM}(\ell,12)+\Delta_{\ell}\right)\,S[1|1]_{\ell}\,\bar{\mathcal{I}}_{YM}(\ell,12),
\end{align}
where $\Delta_{\ell}=f(\ell)-f(\ell+c)$ . The reverse is also true.
Starting from a KLT formula with the loop momenta in both of the gauge theory
integrands defined in the conventional way, one obtains a double-copy
formula with the loop momenta of the two copies related by a shift.
Therefore cautions must be made when reproducing gravity integrands
using this approach. We discuss this issue in section \ref{sec:double-copy}\footnote{While preparing this manuscript, we became aware of a related
work \cite{Cao:2025ygu} that happens to overlap with this part
of our work.}.
\end{enumerate}
Additionally, we present in appendix \ref{sec:nlsm-numerators} a
symmetry property of the nonlinear sigma model numerator, which supplement
the discussion in section \ref{sec:forward-limit-numerators}. A $3$-point
example solving the numerators using the method explained in section
\ref{sec:nlsm-numerators} is presented in appendix \ref{sec:3pt-example}. 

\medskip

Beside analysing general BCJ structure at one loop level, in this
paper we present explicit numerator formulas at one loop level for
Yang-Mills-scalar (YMS) and related theories (pure Yang-Mills (YM), gravity and Einstein-Yang-Mills (EYM)). The practical problems we tackle in
this paper are the following.
\begin{enumerate}[(a)]
\item  We shall see in section \ref{sec:shifting-used-explained} that,
at least for the low-point cases we have checked explicitly, the propagator
matrices we derive using the bypassing method mentioned above (issue
(i)) is full-rank\footnote{ This phenomenon---that algebraic relations implied by the Jacobi identity
impose stricter constraints at one loop than at tree level---is not isolated.
In fact, as the loop order increases, it becomes increasingly difficult to
construct numerators that satisfy exact Jacobi identities. Starting from two
loops, various contact-term contributions that integrate to zero must be
introduced in order to build numerators obeying generalized forms of Jacobi
relations~\cite{Bern:2017yxu,Ochirov:2017jby}}. So that given a colour-ordered integrand, in principle
one should be able to solve the numerators by inverting the propagator
matrix, and such a solution is unique\footnote{By uniqueness here we mean with respect to this particular integrand,
which is in general defined up terms that vanish after the loop integration.}. Inverting the propagator matrix, however, could be a rather formidable
task in practical calculations. In this paper we present an alternative,
constructive approach to this problem, specifically tailored for the
YMS theory. This part of the discussions, presented in sections \ref{sec:yms-numerators}
to \ref{sec:three-gluons}, runs in parallel with the general discussion
given in sections \ref{sec:shifting-used-explained} to \ref{sec:momentum-kernel}
(issue (i)). \\
Recall that at tree level, in the special case when only one gluon
participates the scattering with an arbitrary number of bi-adjoint
scalars, the colour-ordered amplitudes are built from cubic Feynman
graphs only \cite{Bern:1999bx}, which naturally define a Jacobi
satisfying set of numerators \cite{Fu:2017uzt,Chiodaroli:2017ngp}. Take the $1$-gluon
and $3$-scalar amplitudes for example, equation (\ref{eq:4pt-matrix-eqn})
in this case can be written as the following.
\begin{equation}
A^{\text{YMS}}(1^{s},2^{g},3^{s},4^{s})=\sum_{\sigma\in S_{2}}m(1^{s},2^{g},3^{s},4^{s}|1^{s}\sigma4^{s})\,N^{\text{tree}}(1^{s}\sigma4^{s}),\label{eq:yms-4pt-1}
\end{equation}
where we used superscripts to indicate particle species. The bi-adjoint
scalars carry two copies of the colour factors, so that when stripped
off one copy, they contribute one colour factor to the BCJ numerator.
On the other hand a gluon carries only one copy of the colour factor,
therefore after stripping off contributes just a kinematic factor.
Explicitly, in our example these are
\begin{align}
N^{\text{tree}}(1^{s},2^{g},3^{s},4^{s}) & =N^{(K)}(1234)\,N^{(C)}(1234)=\epsilon_{2}\cdot k_{1}\,f^{134},\label{eq:yms-4pt-solution}\\
N^{\text{tree}}(1^{s},3^{s},2^{g},4^{s}) & =N^{(K)}(1324)\,N^{(C)}(1324)=\epsilon_{2}\cdot(k_{1}+k_{3})\,f^{134}.
\end{align}
At higher points the kinematic dependent part of the BCJ numerator
generalises to a product of the polarisation with a sum of momenta
that appear on one side of the gluon, defined by the colour order,
$\epsilon\cdot X=\epsilon\cdot(\sum_{\text{all momenta on the left}}k_{i})$ \cite{Fu:2017uzt,Chiodaroli:2017ngp}.
In this paper we shall 
denote the  kinematic dependent part of the 
numerator $N^{(K)}(\sigma)$ as 
 $C(\sigma)$, to be consistent with the notation used in \cite{Hou:2018bwm,Du:2019vzf,Wu:2021exa}. The color dependent part
of the BCJ numerator $N^{(C)}$ is given by the usual half-ladder, with the structure
constant at the position corresponds to the gluon replaced by a Kronecker
delta, $N^{(C)}(1^{s},\dots,i^{g},\dots,n^{s})=f^{12*}f^{*3*}\dots\delta^{**}\dots f^{*n-1,n}$.
More than one Feynman graph contribute to a numerator when more gluons
are involved, but rules were developed to write down directly the
BCJ numerators from Feynman graphs \cite{Fu:2017uzt,Du:2019vzf,Tian:2021dzf,Wu:2021exa}.
In particular it was argued that gauge invariance to a large extend
fixes the kinematic part of the YMS numerators, so that ansatz can
also be introduced to simplify the calculation \cite{Fu:2017uzt,Zhou:2022orv,Hu:2023lso,Du:2024dwm}. Other powerful methods
were developed to write down the tree YMS numerators using fusion rules
\cite{Chen:2019ywi,Chen:2021chy,Chen:2022nei,Chen:2023ekh,Chen:2024gkj,Brandhuber:2022enp,Brandhuber:2021bsf}, string theory \cite{Fu:2025jpp} and Cachazo-He-Yuan formula \cite{Cachazo:2013gna,Cachazo:2013hca,Cachazo:2013iea,Cachazo:2014nsa,Cachazo:2014xea}, see e.g., \cite{Du:2017kpo,Teng:2017tbo,Du:2017gnh}.
Remarkably, it was observed by \cite{Fu:2017uzt,Brandhuber:2022enp,Brandhuber:2021bsf} that at
tree level the pure Yang-Mills numerators can be actually derived
from knowledge of YMS numerators, making them of great practical interest.\\
In view of the both rather simple numerator formulas and their
application values at tree level, we feel the YMS theory could be
a worthy investigating case for the one loop numerator problem. Especially
that the tree numerator formulas rely heavily on a language based
on the symbol $X=\sum_{\text{all momenta on the left}}k_{i}$,
defined as the momentum sum on one side of the gluon insertion position,
which is apparently a tree topology concept. One naturally wonders
how are the YMS numerator formulas generalised to one loop level.
And indeed, we shall see in section \ref{sec:one-gluon} that such
a language generalises to one loop level, with the colour-stripped
numerator for a single gluon insertion given by exactly the same formula.
\bea \boxed{C(\ell;1,\pmb\alpha)=\epsilon_p\cdot X_p(\ell;1,\pmb\alpha)} \eea
And we have 
a naturally generalised definition for the symbol $X^{\mu}$ appropriate
for the one loop topology. The two and three gluon cases are also
found to be given by the same formulas as at tree level,
but with the symbols $X^{\mu}$ replaced by the generalised ones, 
and are proved
in sections \ref{sec:two-gluons} and \ref{sec:three-gluons} respectively.\\
As for the problem concerning computing pure YM numerators using YMS
numerators, at tree level this relies on factorising the gluon numerator
into a sub-tree \cite{Brandhuber:2022enp,Brandhuber:2021bsf}, which is also a topology
dependent technique. In section \ref{sec:ym-numerators}   we use an identity
written down by \cite{Geyer:2017ela,Porkert:2022efy,Xie:2024pro,Cao:2024olg} as a substitute, which relates the colour-ordered
integrands of pure YM with YMS ones, and we explain how to construct
YM numerator at one loop using the YMS numerators. Explicitly, we obtain the pure YM numerator formulas
for the $2$-, $3$-point cases as the following\footnote{The factors ${\epsilon_i\cdot\epsilon_j\over k_i\cdot k_j}$ ($i\neq j, i,j=1,2,3$)  in the two 
 and three point numerators $N^{\text{YM}}\left(\ell;1,2\right)$ and  $N^{\text{YM}}\left(\ell;1,2,3\right)$ seem to be singular due to momentum conservation and the massless conditions of external particles. Nevertheless, these terms are of the form $\frac{0}{0}$ when the bi-ajoint scalar integrands  in the decomposition formulas are taken into account. In fact, the final result after division is finite. }.
%
\bea
N^{\text{YM}}\left(\ell;1,2\right)&=&(D-2)\left[(\epsilon_1)_{\mu}(\epsilon_2)_{\nu}-{\epsilon_1\cdot\epsilon_2\over k_1\cdot k_2}\,(k_1)_{\mu}(k_2)_{\nu}\right]\ell^{\mu}(\ell+k_1)^{\nu}+\text{Tr}\Big[F_1\cdot F_2\,\Big],\nn
N^{\text{YM}}\left(\ell;1,2,3\right)&=&(D-2)\Big[\,(\epsilon_1)_{\mu}(\epsilon_2)_{\nu}(\epsilon_3)_{\tau}-{\epsilon_1\cdot\epsilon_2\over k_1\cdot k_2}\,(k_1)_{\mu}(k_2)_{\nu}(\epsilon_3)_{\tau}\nn
&&-{\epsilon_1\cdot\epsilon_3\over k_1\cdot k_3}\,(k_1)_{\mu}(\epsilon_2)_{\nu}(k_3)_{\tau}-{\epsilon_2\cdot\epsilon_3\over k_2\cdot k_3}\,(\epsilon_1)_{\mu}(k_2)_{\nu}(k_3)_{\tau}\Big]\nn
&&~~~~~~~~~~~~~~~~~~~~~~~~~~~~~~~~~~~~~~~~~~~~~~~~~~~~~~~\times \ell^{\mu}(\ell+k_1)^{\nu}(\ell+k_1+k_2)^{\tau}\nn
&&+\text{Tr}\Big[F_2\cdot F_3\,\Big]\epsilon_1\cdot \ell+\text{Tr}\Big[F_1\cdot F_3\,\Big]\epsilon_2\cdot (\ell+k_1)\nn
&&+\text{Tr}\Big[F_1\cdot F_2\,\Big]\epsilon_3\cdot (\ell+k_1+k_2)\nn
&&-\text{Tr}\Big[F_1\cdot F_2\cdot F_3\,\Big]\epsilon_1\cdot \ell\,\epsilon_2\cdot(\ell+k_1)\,\epsilon_3\cdot(\ell+k_1+k_2),
\eea
where $\epsilon^{\mu}_i$, $k^{\mu}_i$ respectively denote the polarisation vector and momentum of $i$. The $F_{i}^{\mu\nu}=k_i^{\mu}\epsilon_i^{\nu}-\epsilon_i^{\mu}k_i^{\mu}$ refers to the strength tensor of $i$, while  $\text{Tr}\Big[F_{1}\cdot F_{2}\cdot...\cdot F_{j}\,\Big]$ denote the cyclic contraction of strength tensors $F_{1}^{\mu_1\mu_2}F_{2}^{\mu_2\mu_3}...F_{j}^{\mu_j\mu_1}$ .
The other numerator $N\left(\ell;1,3,2\right)$ with three gluons is directly obtained from $N\left(\ell;1,2,3\right)$ via exchanging the roles of $2$ and $3$. Moreover,  numerators with more than three gluons can also be decomposed by splitting the gluon set into two ordered sets corresponding to scalars and gluons. In this situation, our YMS results with 1 to 3 gluons provide the exact expressions of the corresponding terms in this decomposition.

In order to answer the aforementioned problems for YMS theory, in
this paper we present a set of organisation principles that re-groups
Feynman graphs contributing to the colour-ordered integrand. Entries
of the propagator matrix factorises in each of these groups, so that
the remaining coefficient is identified as the unique numerator solution
to this colour-ordered integrand.
\begin{equation}
\mathcal{I}^{\text{YMS}}(\ell,123\dots n)=\sum_{\sigma\in S_{n-1}}N(1\sigma)\,\tilde{M}(1\sigma|123\dots n).\label{eq:yms-bs-integrand}
\end{equation}
\item  The next problem runs parallel to the issue (ii) in the case of
YMS. We determine the one loop numerator solutions for the one, two
and three gluon cases for the YMS theory in sections \ref{sec:yms-numerators}
to \ref{sec:three-gluons}, which because of the propagator matrix
being full-rank, is unique for the colour-ordered integrand defined
by the Feynman rules. On the other hand tree numerator solutions are
known to be non-unique \cite{Bern:2008qj}. A natural question is whether the numerator
solution (\ref{eq:yms-4pt-solution}), and its multiple-gluon formula
given by the graphic rules derived in \cite{Fu:2017uzt,Du:2019vzf,Tian:2021dzf,Hou:2018bwm,Du:2017kpo,Teng:2017tbo,Du:2017gnh}
for tree level numerators correspond to this unique solution. Namely,
suppose if we write down the half-ladder numerators based on the rules
given by \cite{Fu:2017uzt,Du:2019vzf,Tian:2021dzf,Wu:2021exa,Hou:2018bwm,Du:2017kpo,Teng:2017tbo,Du:2017gnh}, taking the forward
limit and then have their both sides welded, would the resulting numerators
be the same as the $n$-gons obtained in (a)? In section \ref{sec:welding-yms-trees}
we see that the answer is affirmative. The welded tree numerators
naturally define a loop dependent formula compatible with the consistency
condition (\ref{eq:consistency-1}) and are therefore the unique solution.
Alternatively, one can see the tree numerator formulas when welded are identical to the 
loop ones computed in section \ref{sec:yms-numerators}
to \ref{sec:three-gluons}. Finally we briefly remark in section \ref{sec:gr-eym-numberators}
on the gravity and EYM cases where we do not have a colour-ordered
integrand nor Feynman rules.
\end{enumerate}

\section{Effective numerators and propagator matrix at one loop}
\label{sec:shifting-used-explained}
\subsection{Existence of a solution and a $2$-point example}
\label{sec:2pt-example}
Strictly speaking, the problem of solving kinematic numerators at
one loop is mathematically defined by the infinite matrix relation
(\ref{eq:infinite-matrix-eqn}) explained in the introduction. The
technically challenging problem however, can be circumvented by solving
for a physically equivalent set of kinematic numerators instead. A way
out is to note that adding arbitrary difference terms related by a
constant shift in the loop momentum effectively give the same $1$-loop
amplitude, a technique frequently employed in earlier study of the
integrand behaviour under the influence of a BCJ formulation \cite{Boels:2011tp,Du:2012mt}.
\begin{align}
A_{1\text{-loop}} & =\int d^{D}\ell\,\mathcal{I}_{n}(\ell)\nonumber \\
 & =\int d^{D}\ell\,\mathcal{I}_{n}(\ell)+f(\ell)-f(\ell+c).\label{eq:integrand-shift-general}
\end{align}
Assuming that our integrand is colour-ordered with respect to some
loop momentum independent colour factors (this assumption will change
when we discuss gravity integrands in section \ref{sec:double-copy}).
In the $2$-point example we add difference terms and consider the
following shifted integrand.
\begin{eqnarray}
& & \tilde{\mathcal{I}}_{2}(\ell,k_{1},k_{2}) \label{eq:2pt-shifted-integrand}\\
&& =  \left[\frac{1}{\ell^{2}(\ell+k_{1})^{2}}+\frac{1}{\ell^{2}}\,\frac{1}{i\epsilon}+\frac{1}{(\ell+k_{1})^{2}}\,\frac{1}{i\epsilon}\right]N(\ell,1,2,-\ell) \nonumber \\
 && -\left[\frac{1}{\ell^{2}}\,\frac{1}{i\epsilon}\right]N(\ell-k_{1},1,2,-(\ell-k_{1})) \nonumber \\
 && +\left(\underset{f(\ell)}{\underbrace{\left[\frac{1}{\ell^{2}}\,\frac{1}{i\epsilon}\right]N(\ell-k_{1},1,2,-(\ell-k_{1}))}}-\underset{f(\ell+k_{1})}{\underbrace{\left[\frac{1}{(\ell+k_{1})^{2}}\,\frac{1}{i\epsilon}\right]N(\ell,1,2,-\ell)}}\right)\nonumber \\
 && -\left[\frac{1}{(\ell+k_{1})^{2}}\,\frac{1}{i\epsilon}\right]N(\ell+k_{1},1,2,-(\ell+k_{1})) \nonumber \\
 && +\left(\underset{g(\ell)}{\underbrace{\left[\frac{1}{(\ell+k_{1})^{2}}\,\frac{1}{i\epsilon}\right]N(\ell+k_{1},1,2,-(\ell+k_{1}))}}-\underset{g(\ell-k_{1})}{\underbrace{\left[\frac{1}{\ell^{2}}\,\frac{1}{i\epsilon}\right]N(\ell,1,2,-\ell)}}\right)\nonumber 
\end{eqnarray}
The difference terms were added deliberately to cancel the ``misaligned"
numerators appearing in the equation, leaving
\begin{equation}
\tilde{\mathcal{I}}_{2}(\ell,k_{1},k_{2})=\frac{1}{\ell^{2}(\ell+k_{1})^{2}}N(\ell,1,2,-\ell).\label{eq:2pt-matrix-relation}
\end{equation}
Comparing with the tree level argument, we see that for this effective
integrand at $2$-point the number of independent numerator is just
$1$, and the propagator matrix is just the factor $1/\ell^{2}(\ell+k_{1})^{2}$,
which inverts to give the generalisation of the momentum kernel in
in this case, $\ell^{2}(\ell+k_{1})^{2}$. 

\textbf{Addressing (i):} The derivation above allows us to answer the
questions raised in (i) in the introduction, specifically for this
$2$-point case. We see from equation (\ref{eq:2pt-matrix-relation})
in the sense of writing an effective integrand (up to terms that vanishes
upon integration) in terms of the BCJ formulation (\ref{eq:bcj-integrand}),
the method discussed in this section always permits a solution for
the basis numerator (\ref{eq:bubble-numerator}), regardless of whether
the same theory gives BCJ satisfying amplitudes at tree level. Note
that equation (\ref{eq:2pt-matrix-relation}) is enough to determine
all numerators: Here we have assumed the integrand $\mathcal{I}_{2}(\ell,k_{1},k_{2})$,
given by Feynman rules, is a known function of the polarisations,
external and loop momenta, so that numerator $N(\ell,1,2,-\ell)$
at every $\ell$ is solved from the integrand $\mathcal{I}_{2}(\ell,k_{1},k_{2})$
at that specific value of $\ell$. Additionally, the set of all numerators
written symbolically with the other order, $N(\ell,2,1,-\ell)$, are
actually the same set with the original order due to the nature of
this notation. Namely, from (\ref{eq:consistency-1}) we see that
\begin{equation}
N(\ell,2,1,-\ell)=N(\ell-k_{1},1,2,-(\ell-k_{1})).
\end{equation}
Since these together determine all basis numerators and nothing more,
all numerators appearing at $2$-point are solved.


\subsection{Solving higher-point numerators}
\label{sec:higher-point-numerators}

The method explained through a $2$-point example discussed in section
\ref{sec:2pt-example} generalises to higher points in a straightforward
manner. At $n$-point the basis numerators are $n$-gons with $n$
external legs attached, and our nomenclature is likewise redundant.
\begin{eqnarray}
&& N(\ell,\sigma(1),\sigma(2),\sigma(3),\dots,\sigma(n),-\ell)  \label{eq:consistency-2} \\
&& =N(\ell+k_{\sigma(1)},\sigma(2),\sigma(3),\dots,\sigma(n),\sigma(1),-(\ell+k_{\sigma(1)})) \nonumber \\
 && =N(\ell+k_{\sigma(1)}+k_{\sigma(2)},\sigma(3),\dots,\sigma(n),\sigma(1),\sigma(2),-(\ell+k_{\sigma(1)}+k_{\sigma(2)}))\nonumber \\
 &&  =\dots\nonumber 
\end{eqnarray}
where $\sigma\in S_{n}$ is understood to be any of the permutations.
Starting from the knowledge of all colour-ordered integrands, which
we assume to be already computed from Feynman rules, one solves numerators
using BCJ's formula (\ref{eq:bcj-integrand}). More specifically,
one uses the redundancy of the nomenclature to fix, for example, $\sigma(1)=1$,
leaving $(n-1)!$ independent numerators for each $\ell$. In addition
one note that BCJ's formula (\ref{eq:bcj-integrand}) for integrands
also satisfy the same symmetry as our nomenclature for numerators,
\begin{eqnarray}
&& \mathcal{I}_{n}(\ell,\sigma(1),\sigma(2),\sigma(3),\dots,\sigma(n)) \\
&& =\mathcal{I}_{n}(\ell+k_{\sigma(1)},\sigma(2),\sigma(3),\dots,\sigma(n),\sigma(1)) \nonumber \\
 && =\dots\nonumber 
\end{eqnarray}
as well as the explicit integrands obtained by (not necessarily cubic)
Feynman rules. (We neglect the proof of this, as it can be readily
checked in the lower-point cases.) Therefore we have the same number
of independent equations and unknowns. Shifting terms in the integrand
properly as in the $2$-point example shown in section \ref{sec:2pt-example}
and we obtain an $(n-1)!\times(n-1)!$ matrix equation, where the
basis numerators can be solved from.

For concreteness we list the result for $3$-point. The calculation
details can be found in appendix \ref{sec:3pt-example}. At $3$-point
we have the following two independent numerators and the corresponding
consistency conditions for their notations.
\begin{equation}
\begin{minipage}{3cm} 
  \includegraphics[width=2.5cm]{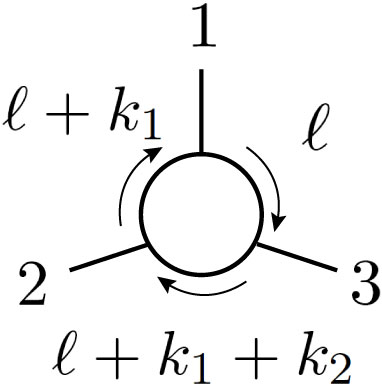}
\end{minipage}  
\begin{minipage}{4cm} 
\begin{eqnarray}
   && =N(\ell,123,-\ell)   \nonumber   \\
 && =N(\ell+k_{1},231,-(\ell+k_{1}))  \nonumber  \\
&& =N(\ell+k_{1}+k_{2},312,-(\ell+k_{1}+k_{2}))  \nonumber 
\end{eqnarray}
\end{minipage} 
\label{eq:3pt-consistency-1}
\end{equation}
and
\begin{equation}
\begin{minipage}{3cm} 
\includegraphics[width=2.5cm]{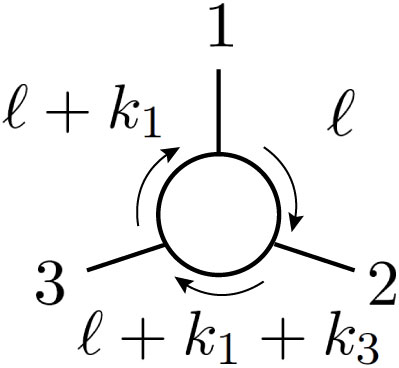}
\end{minipage}  
\begin{minipage}{4cm} 
\begin{eqnarray}
  && =N(\ell,132,-\ell) \nonumber  \\
 && =N(\ell+k_{1},321,-(\ell+k_{1}))  \nonumber  \\
&& =N(\ell+k_{1}+k_{3},213,-(\ell+k_{1}+k_{3})) \nonumber 
\end{eqnarray}
\end{minipage} 
\label{eq:3pt-consistency-2} 
\end{equation}
The above two sets are related by the following relabeling transformations,
respectively.
\begin{equation}
\begin{array}{ccccccc}
OP_{1}: & \ell\rightarrow & \ell+k_{1} & \text{ and } & OP_{2}: & \ell\rightarrow & \ell+k_{1}\\
 & k_{1}\rightarrow & k_{2} &  &  & k_{1}\rightarrow & k_{3}\\
 & k_{2}\rightarrow & k_{3} &  &  & k_{2}\rightarrow & k_{1}\\
 & k_{3}\rightarrow & k_{1} &  &  & k_{3}\rightarrow & k_{2}
\end{array}\label{eq:3pt-symmetry}
\end{equation}

Using the method just explained leads to a linear relation between
shifted integrands and the numerators,
\begin{equation}
\left[\begin{array}{c}
\tilde{\mathcal{I}}_{3}(\ell,123)\\
\tilde{\mathcal{I}}_{3}(\ell,132)
\end{array}\right]=\left[\begin{array}{cc}
\tilde{M}(23|23) & \tilde{M}(23|32)\\
\tilde{M}(32|23) & \tilde{M}(32|32)
\end{array}\right]\left[\begin{array}{c}
N(\ell,123,-\ell)\\
N(\ell,132,-\ell)
\end{array}\right],
\end{equation}
where the propagator matrix $\tilde{M}(\alpha|\beta)$ is given by
(\ref{eq:3pt-propagator-matrix}) in the appendix. We invert the propagator
matrix to solve $N(\ell,123,-\ell)$ and $N(\ell,132,-\ell)$ for
all values of $\ell$. Other orderings actually refer the same set
of numerators through (\ref{eq:3pt-consistency-1}) and (\ref{eq:3pt-consistency-2}),
and we have obtained all the basis numerators. We see again in this
example the numerator solution exists regardless of whether the same
theory has tree level solutions.

\subsection{$n$-point propagator matrix and its relation to forward limit tree propagator matrices}
\label{sec:forward-limit-propagators}

At higher points the method discussed in sections \ref{sec:2pt-example}
and \ref{sec:higher-point-numerators} would be rather complicated,
as it require considerable graphical manipulations on BCJ's formula
(\ref{eq:bcj-integrand}). In this section we present a formally different,
but more easily generalised, method to obtain the same numerator solution.
The argument presented in this section also explain the relation between
the one loop generalised propagator matrix described in sections \ref{sec:2pt-example},
\ref{sec:higher-point-numerators} and the forward limit of the conventional
tree propagator matrix $m(\alpha|\beta)$ defined in \cite{Cachazo:2013iea}. 

\textbf{Note a subtlety here}: We will not try to identify forward
limit amplitudes defined by Feynman rules, $A_{\text{Feyn}}^{\text{tree}}$,
with any object in the current section. Such identification is the main issue
raised by (ii) in the introduction and will be discussed in section
\ref{sec:forward-limit-numerators}. The method presented here is
only due to the knowledge of the cubic structure in $A_{\text{BCJ}}^{\text{tree}}$,
and the fact that integrands can be expanded in terms of $A_{\text{BCJ}}^{\text{tree}}$,
which was explained in the introduction.

The following derivation will be rather straightforward. But to avoid
unnecessarily complicated notations, let us first use the $3$-point
case to demonstrate the idea. Instead of the graphical decomposition
(\ref{eq:bcj-integrand}), we consider the forward limit decomposition
of the integrand explained in the introduction. At $3$-point, this
is given by
\begin{align}
\mathcal{I}_{3}(\ell,123)= & \frac{1}{\ell^{2}}\,A_{\text{BCJ}}^{\text{tree}}(\ell,1,2,3,-\ell)+\frac{1}{(\ell+k_{1})^{2}}\,A_{\text{BCJ}}^{\text{tree}}(\ell+k_{1},2,3,1,-(\ell+k_{1}))\label{eq:3pt-integrand}\\
 & +\frac{1}{(\ell+k_{1}+k_{2})^{2}}\,A_{\text{BCJ}}^{\text{tree}}(\ell+k_{1}+k_{2},3,1,2,-(\ell+k_{1}+k_{2})).\nonumber 
\end{align}

Recall that the amplitudes $A_{\text{BCJ}}^{\text{tree}}$ should
be understood as the colour-ordered tree amplitudes formally written
down using the tree cubic formula (\ref{eq:bcj-amplitude}), with
the two parallel legs of every numerator welded together, and then
replaced by their corresponding counterparts appear in the one loop
integrand (the $n$-gons), according to the nomenclature suggestion,
$N^{\text{tree}}(\ell,\sigma(1),\sigma(2),\dots,\sigma(n),-\ell)\rightarrow N(\ell,\sigma(1),\sigma(2),\dots,\sigma(n),-\ell)$.
This replacement implies a natural correspondence between the actual
tree basis numerators (the half-ladders) $N^{\text{tree}}(\ell,\sigma(1),\sigma(2),\dots,\sigma(n),-\ell)$
and the $n$-gons $N(\ell,\sigma(1),\sigma(2),\dots,\sigma(n),-\ell)$,
so that the amplitudes $A_{\text{BCJ}}^{\text{tree}}$ can be expanded
in exactly the same way as real tree amplitudes, such as (\ref{eq:4pt-matrix-eqn}).
At $3$-point this allows us to expand all $A_{\text{BCJ}}^{\text{tree}}$
in equation (\ref{eq:3pt-integrand}) into $3!$ unknown variables
$N(\ell,\sigma(1),\sigma(2),\sigma(3),-\ell)$. Schematically, we
have the following linear relation.
\begin{equation}
\scalebox{0.8}{
\begin{minipage}{18cm}
\begin{eqnarray}
 \underset{3!=6\text{ permutations}}{\underbrace{\left[\begin{array}{c}
\mathcal{I}_{3}(\ell,123)\\
\mathcal{I}_{3}(\ell,132)\\
\mathcal{I}_{3}(\ell,213)\\
\text{\fbox{\ensuremath{\mathcal{I}_{3}}(\ensuremath{\ell},231)}}\\
\mathcal{I}_{3}(\ell,312)\\
\mathcal{I}_{3}(\ell,321)
\end{array}\right]}} 
& = & 
\left[\begin{array}{cccccc}
\frac{1}{\ell^{2}}m(\ell123|\ell123) & \dots\\
\vdots\\
\\
\text{\fbox{}} & \text{\fbox{}} & \text{\fbox{}} & \text{\fbox{}} & \text{\fbox{}} & \text{\fbox{}}\\
\\
\\
\end{array}\right]\left[\begin{array}{c}
N(\ell,123,-\ell)\\
N(\ell,132,-\ell)\\
N(\ell,213,-\ell)\\
\text{\fbox{N(\ensuremath{\ell},231,-\ensuremath{\ell})}}\\
N(\ell,312,-\ell)\\
N(\ell,321,-\ell)
\end{array}\right]
\nonumber \\
& & +\left[\begin{array}{cccccc}
\frac{1}{(\ell+k_{1})^{2}}m(\ell+k_{1},231|\ell+k_{1},231) & \dots\\
\vdots\\
\\
\text{\fbox{}} & \text{\fbox{}} & \text{\fbox{}} & \text{\fbox{}} & \text{\fbox{}} & \text{\fbox{}}\\
\\
\\
\end{array}\right]\left[\begin{array}{c}
N(\ell+k_{1},123,-(\ell+k_{1}))\\
N(\ell+k_{1},132,-(\ell+k_{1}))\\
\vdots\\
\text{\fbox{N(\ensuremath{\ell}+\ensuremath{k_{1}},231,-(\ensuremath{\ell}+\ensuremath{k_{1}}))}}\\
\\
N(\ell+k_{1},321,-(\ell+k_{1}))
\end{array}\right]\nonumber \\
 && +\dots\nonumber 
\end{eqnarray}
 \end{minipage}
 }
 \label{eq:3pt-matrix-eqn}
\end{equation}
The coefficient $m(\alpha|\beta)$ appeared are apparently the same
tree propagator matrix entries \cite{Cachazo:2013iea}
from our discussion above.

For the same reason mentioned in (i) in the introduction, the solution
to such a matrix equation alone can not be determined, as it involves
more unknown variables related by shiftings. Solving them requires
more matrix equations involving shifted integrands, which in turn
leads to more unknown variables. The problem cannot be rigorously
solved unless we introduce boundary conditions and take infinite limits.
However let us focus on the row of the matrix equation index by permutation
$(231)$ (highlighted in (\ref{eq:3pt-matrix-eqn})). Observe that
when loop momentum takes the value $\ell+k_{1}$, this equation is
actually formally identical to the first row indexed by $\text{id}=(123)$.
A quick way to see this is to note that it came from the numerator
expansion of the following equation, the right hand side of which
is by itself invariant under the symmetry transformation $\text{OP}_{1}$
of (\ref{eq:3pt-symmetry}).
\begin{align}
\mathcal{I}_{3}(\ell,231)= & \frac{1}{\ell^{2}}\,A_{\text{BCJ}}^{\text{tree}}(\ell,2,3,1,-\ell)+\frac{1}{(\ell+k_{2})^{2}}\,A_{\text{BCJ}}^{\text{tree}}(\ell+k_{2},3,1,2,-(\ell+k_{2}))\label{eq:cubic-expansion-3pt}\\
 & +\frac{1}{(\ell+k_{2}+k_{3})^{2}}\,A_{\text{BCJ}}^{\text{tree}}(\ell+k_{2}+k_{3},1,2,3,-(\ell+k_{2}+k_{3}))\nonumber 
\end{align}
The left hand side of the equation, $\mathcal{I}_{3}(\ell,123)$,
on the other hand, belong to the known variables where we solve the
numerators from, presumably given by Feynman rules of the theory of
interest, and satisfies the same symmetry $\mathcal{I}_{3}(\ell,123)=\mathcal{I}_{3}(\ell+k_{1},231)$
by construction. This identification, together with the symmetry assumption
(\ref{eq:3pt-consistency-1}) of the nomenclature of the numerators
allows us to completely remove the redundant line. A similar argument
also identifies the line index by $(312)$ with $\text{id}=(123)$.
Similarly, lines indexed by $(321)$ and $(213)$ with $(132)$, reducing
the problem to a $2\times2$ matrix equation.

More explicitly, let us focus on just the  first row of the
matrix equation (\ref{eq:3pt-matrix-eqn}) that come from the $\frac{1}{\ell^{2}}\,A_{\text{BCJ}}^{\text{tree}}(\ell,1,2,3,-\ell)$
term in (\ref{eq:3pt-integrand}). There are three terms related by
cyclic permutation of the external legs, 
\begin{align}
 \frac{1}{\ell^{2}}\,A_{\text{BCJ}}^{\text{tree}}(\ell,1,2,3,-\ell) 
& =  \frac{1}{\ell^{2}}m(\ell,123,-\ell|\ell,123,-\ell)\,N(\ell,123,-\ell) \\
& +\frac{1}{\ell^{2}}m(\ell,123,-\ell|\ell,231,-\ell)\,N(\ell,231,-\ell)+\dots \nonumber \\
 & +\frac{1}{\ell^{2}}m(\ell,123,-\ell|\ell,312,-\ell)\,N(\ell,312,-\ell)+\dots\nonumber 
\end{align}
Equation (\ref{eq:integrand-shift-general}) allows us to trade individual
terms in the integrand with shifted ones plus terms that vanished
after the integration, by which we shift these three individual terms
by the appropriate amount so that they can be related by the symmetry
operation $\text{OP}_{1}$ of (\ref{eq:3pt-symmetry}). Consistency
of the nomenclature then allows us to identify all three numerators
symbolically as $N(\ell,123,-\ell)$.
\begin{align}
& \mathcal{\tilde{I}}_{3}(\ell,123) \\
&=  \Bigl(\frac{1}{\ell^{2}}m(\ell123|\ell123)\,N(\ell,123,-\ell)\Bigr)+\dots \nonumber \\
 & \Bigl(\frac{1}{(\ell+k_{1})^{2}}m(\ell+k_{1},123,-(\ell+k_{1})|\ell+k_{1},231,-(\ell+k_{1}))\,\underset{=N(\ell,123,-\ell)}{\underbrace{N(\ell+k_{1},231,-(\ell+k_{1}))}}\Bigr)+\dots\nonumber \\
 & +\Bigl(\frac{1}{(\ell+k_{1}+k_{2})^{2}}m(\ell+k_{1}+k_{2},123,-(\ell+k_{1}+k_{2})|\ell+k_{1}+k_{2},312,-(\ell+k_{1}+k_{2}))\, 
 \nonumber \\ & \hspace{1cm}\times 
 \underset{=N(\ell,123,-\ell)}{\underbrace{N(\ell+k_{1}+k_{2},312,-(\ell+k_{1}+k_{2}))}}\Bigr)+\dots\nonumber 
\end{align}
Repeating the same manipulation reduces the unknowns in all rows of
the matrix equation (\ref{eq:3pt-matrix-eqn}) to $3!/3=2$.

Since the derivation above is straightforward, we write the integrand-numerator
relation at generic $n$-point directly. At $n$-point we have the
following symmetry transformation that associates cyclic permutations
of the external legs with specific shift of loop momentum $\ell$,
for all ordering $(\rho(1),\rho(2),\dots,\rho(n))$ of the external
legs.
\begin{equation}
\begin{array}{cc}
\ell\rightarrow & \ell+k_{\rho(1)}\\
k_{\rho(1)}\rightarrow & k_{\rho(2)}\\
\vdots & \vdots\\
k_{\rho(n)}\rightarrow & k_{\rho(1)}
\end{array}\label{eq:npt-symmetry}
\end{equation}
The numerators are assumed to be invariant under such symmetry,
\bea
N(\ell,\rho(1),\rho(2),\dots,\rho(n),-\ell)=N(\ell+k_{\rho(1)},\rho(2),\dots,\rho(n),\rho(1),-(\ell+k_{\rho(1)})).
\eea
This allows us to reduce the basis to $n!/n=(n-1)!$ numerators. We
may choose to fix $\rho(1)=1$. Together with the same symmetry of
the integrands yields an $(n-1)!\times(n-1)!$ matrix equation relating
the integrands and the numerators.
\begin{equation}
\mathcal{I}_{n}(\ell,1,\rho)=\sum_{\sigma\in S_{n-1}}\tilde{M}(\rho|\sigma)\,N(\ell,1,\sigma,-\ell).\label{eq:npt-prop-matrix}
\end{equation}
The manipulation explained above shows that the contribution from
a single forward limit tree amplitude is
\begin{equation}    
\scalebox{0.75}{
\begin{minipage}{19cm}
\begin{align}
 & \frac{1}{\ell^{2}}A(\ell,1,\rho(2),\rho(3),\dots,\rho(n),-\ell)  \nonumber \\
 & \sim\sum_{\sigma\in S_{n-1}}\Biggl(\frac{1}{\ell^{2}}m(\ell,1,\rho(2),\rho(3),\dots,\rho(n),-\ell|\ell,1,\sigma(2),\sigma(3),\dots,\sigma(n),-\ell)\nonumber \\
 & +\frac{1}{(\ell+k_{1})^{2}}m(\ell+k_{1},1,\rho(2),\rho(3),\dots,\rho(n),-(\ell+k_{1})|\ell+k_{1},\sigma(2),\sigma(3),\dots,\sigma(n),1,-(\ell+k_{1}))\nonumber \\
 & +\frac{1}{(\ell+k_{1}+k_{\sigma(2)})^{2}}m(\ell+k_{1}+k_{\sigma(2)},1,\rho(2),\rho(3),\dots,\rho(n),-(\ell+k_{1}+k_{\sigma(2)})|\ell+k_{1}+k_{\sigma(2)},\sigma(3),\dots,\sigma(n),1,\sigma(2),-(\ell+k_{1}+k_{\sigma(2)}))\nonumber \\
 & \dots\Biggr)\times N(\ell,1,\sigma(2),\sigma(3),\dots,\sigma(n),-\ell).\nonumber 
\end{align}
\end{minipage}
}
\label{eq:npt-amplitude-numerator-relation}
\end{equation}
Reading off the coefficient, and we conclude that the $(\rho,\sigma)$-entry
of the (effective) propagator matrix $\tilde{M}(\rho|\sigma)$ at
$n$-point is
\begin{equation}    
\scalebox{0.75}{
\begin{minipage}{19cm}
\begin{align}
 & \tilde{M}(\rho|\sigma) \nonumber \\
 & =\sum_{(\bar{1},\bar{\rho})\in\text{cyclic perm. of }(1,\rho)}\Biggl(\frac{1}{\ell^{2}}m(\ell,\bar{1},\bar{\rho}(2),\bar{\rho}(3),\dots,\bar{\rho}(n),-\ell|\ell,1,\sigma(2),\sigma(3),\dots,\sigma(n),-\ell)\nonumber \\
 & +\frac{1}{(\ell+k_{1})^{2}}m(\ell+k_{1},\bar{1},\bar{\rho}(2),\bar{\rho}(3),\dots,\bar{\rho}(n),-(\ell+k_{1})|\ell+k_{1},\sigma(2),\sigma(3),\dots,\sigma(n),1,-(\ell+k_{1}))\nonumber \\
 & +\frac{1}{(\ell+k_{1}+k_{\sigma(2)})^{2}}m(\ell+k_{1}+k_{\sigma(2)},\bar{1},\bar{\rho}(2),\bar{\rho}(3),\dots,\bar{\rho}(n),-(\ell+k_{1}+k_{\sigma(2)})|\ell+k_{1}+k_{\sigma(2)},\sigma(3),\dots,\sigma(n),1,\sigma(2),-(\ell+k_{1}+k_{\sigma(2)}))\nonumber \\
 & \dots\Biggr)\nonumber 
\end{align}
\end{minipage}
}
\end{equation}
where the cyclic sum $(\bar{1},\bar{\rho})$ comes from the fact that
integrand is a cyclic sum of the forward limit terms of the form (\ref{eq:npt-amplitude-numerator-relation}).

\subsection{Momentum kernel and the Sherman-Morrison formula}
\label{sec:momentum-kernel}

As already mentioned in the introduction, the Jacobi identities impose
increasingly stringent algebraic constraints on kinematic numerators
as the loop order grows. At tree level, there exist infinitely many
possible sets of numerators related by generalised gauge transformations,
while beyond two loops these constraints can no longer be satisfied
diagram by diagram without introducing contact-term contributions
that integrate to zero\cite{Bern:2017yxu,Ochirov:2017jby}. At one loop, however, we are in the fortunate
situation where the corresponding propagator matrix happens to be
non-singular, allowing us to determine the numerators uniquely from
the integrand. We note that, when regarded as a function of loop
momentum~$\ell$, the propagator matrices at $2$- and $3$-points
given in equations~(\ref{eq:2pt-matrix-relation}) and
(\ref{eq:3pt-propagator-matrix}) respectively, are non-singular except
at a few poles and therefore can be regularised using the method
described by~\cite{Baadsgaard:2015twa}.
It is then interesting to see if
the inverse propagator matrix present any general features, considering
the fact that at tree level, momentum kernels are actually simpler
than propagator matrices in the sense that one can write down an analytic
formula at arbitrary $n$-point. Regarding this problem, we find the
propagator matrices derived from the shifting technique explained
in section \ref{sec:shifting-used-explained} however, are rather
complicated beyond $3$-point, even though a few general features
can indeed be identified. We use the Sherman-Morrison formula \cite{ref-Sherman-Morrison}
(which is a special case of the Woodbury matrix identity) to analyse
the problem, which we briefly explain in the following. 

Starting with a matrix $A$ with known inverse matrix $A^{-1}$, let
us consider a perturbation of the original matrix $A$ given by
\begin{equation}
A+\alpha uv^{T}
\end{equation}
where $\alpha$ is a numerical parameter. $u$, $v$ are column vectors.
The Sherman-Morrison formula is an explicit formula for the inverse
of the perturbed matrix $A+\alpha uv^{T}$.
\begin{equation}
\left(A+\alpha uv^{T}\right)^{-1}=A^{-1}-\frac{\alpha A^{-1}uv^{T}A^{-1}}{1+\alpha v^{T}A^{-1}u}\label{eq:sherman-morrison}
\end{equation}
To see how the formula works, let us consider the special case where
$A$ is a diagonal matrix $\text{diag}(\lambda_{1},\lambda_{2},\dots,\lambda_{n})$
with an apparent inverse, and that $u$, $v$ are happen to be unit
vectors with the non-zero component being the $i$-th and $j$-th
entry respectively, $u=e_{i}$, $v=e_{j}$, so that $u\,v^{T}$ is
a matrix with the $(i,j)$-th element in the matrix being $1$ and
others zero. The Sherman-Morrison formula (\ref{eq:sherman-morrison})
then allows us to write down the inverse of any matrix of the following
form
\begin{equation}
A+\alpha_{(i,j)}u\,v^{T}=\left(\begin{array}{ccccc}
\lambda_{1}\\
 & \lambda_{2} &  & \alpha_{(i,j)}\\
 &  & \ddots\\
 &  &  & \ddots\\
 &  &  &  & \lambda_{n}
\end{array}\right)
\end{equation}
for arbitrary values of $\lambda_{1}$, $\lambda_{2}$, $\dots$,
$\lambda_{n}$ and $\alpha$. For a generic matrix, we iterate the
above algorithm, adding non-zero entries one at a time. Namely we
regard all the off-diagonal terms as perturbations and calculate the
inverse using Sherman-Morrison formula (\ref{eq:sherman-morrison})
until all entries are included.

For a propagator matrix at one loop, we can do slightly better 
than
the brute force method explained above. The $3$-point propagator matrix written down in appendix \ref{sec:3pt-example} is $2\times2$,
its inverse can then be readily written down using the formula exclusively
for $2\times2$ matrices. However note that the $3$-point propagator
matrix $M_{3}$ can be written as a perturbation to a matrix $M_{3,0}$
that resembles the tree propagator matrix (\ref{eq:4pt-matrix-eqn}).
\begin{equation}
M_{3}=M_{3,0}+\left(\begin{array}{cc}
\alpha_{1} & \alpha_{2}\\
-\alpha_{1} & -\alpha_{2}
\end{array}\right),
\end{equation}
where
\begin{equation}
M_{3,0}=\frac{1}{D_{\ell}D_{\ell1}}\left(\begin{array}{cc}
\frac{1}{D_{\ell12}}+\frac{1}{s_{23}} & -\frac{1}{s_{23}}\\
-\frac{1}{s_{23}} & \frac{1}{D_{\ell13}}+\frac{1}{s_{23}}
\end{array}\right)
\end{equation}
and the parameters $\alpha_{1}$, $\alpha_{2}$ are given by
\begin{align}
\alpha_{1} & =\frac{1}{D_{\ell}D_{\ell12}}\left(\frac{1}{s_{12}}+\frac{1}{s_{13}}\right),\\
\alpha_{2} & =\frac{1}{D_{\ell}D_{\ell13}}\left(\frac{1}{s_{12}}+\frac{1}{s_{13}}\right), \nonumber
\end{align}
respectively.

The fact that the $3$-point propagator matrix $M_{3}$ can be written
as such a perturbation is not a coincidence. The extra pieces $\alpha_{1}$,
$\alpha_{2}$ are due to shiftings, which is the feature that distinguishes
the tree and one loop level propagators. From this viewpoint the one
loop level propagator matrix can be understood as a tree propagator
matrix perturbed by the shifting terms.

The tree level-like piece $M_{3,0}$ has an inverse that we learned
from the analytic formula for momentum kernel.
\begin{align}
S_{3,0}=M_{3,0}^{-1} & =\frac{D_{\ell}D_{\ell1}}{D_{\ell12}+D_{\ell13}+s_{23}} \\
 & \times\left(\begin{array}{cc}
D_{\ell12}\left(D_{\ell13}+s_{23}\right) & D_{\ell12}D_{\ell13}\\
D_{\ell12}D_{\ell13} & D_{\ell13}\left(D_{\ell12}+s_{23}\right)
\end{array}\right)  \nonumber
\end{align}
In viewing the Sherman-Morrison formula, we write the $3$-point propagator
matrix as the following two-step perturbations,
\begin{equation}
M_{3}=M_{3,0}+\alpha_{1}uv_{1}^{T}+\alpha_{2}uv_{2}^{T},
\end{equation}
where the vectors $u$, $v_{1}$ and $v_{2}$ are given by
\begin{equation}
u=\left(\begin{array}{c}
1\\
-1
\end{array}\right),\quad v_{1}=\left(\begin{array}{c}
1\\
0
\end{array}\right),\quad v_{2}=\left(\begin{array}{c}
0\\
1
\end{array}\right),
\end{equation}
respectively. We find the feature illustrate through the $3$-point
example here also resides in the higher-point case, even though the
complexity involved in applying the Sherman-Morrison formula increases
rapidly.

\section{Conditions required by identifying $n$-gons with welded trees}
\label{sec:forward-limit-numerators}

In this section we discuss a few problems relating to loop and tree structures.
In particular, under what condition can the welded tree numerators be used directly
as one loop numerators. Recall that we explained in (ii) in the introduction
that the BCJ formula of integrand (\ref{eq:bcj-integrand}) can be
regarded as a sum over polygons attached with various sub-trees, so
that if we use the partial fraction identity (\ref{eq:partial-fraction-npt})
to decompose the loop dependent propagators of the polygons and re-group,
the original BCJ integrand formula can be re-expressed as
a sum over forward limit amplitudes $A_{\text{BCJ}}^{\text{tree}}$.
At $2$- and $3$-points these are given by (\ref{eq:tree-decomposition-2pt})
and (\ref{eq:cubic-expansion-3pt}) respectively, and generically
we have\footnote{From equation (\ref{eq:forward-limit-decomp-npt}) one sees that the
tadpole numerators do not contribute to the (integrated) amplitude
at one loop level, regardless of the actual analytic behaviour of
the numerator solutions. See appendix \ref{sec:tadpoles} for discussion.}
\begin{align}
\mathcal{I}_{n}(\ell,k_{1},k_{2},\dots,k_{n})= & \frac{1}{\ell^{2}}\,A_{\text{BCJ}}^{\text{tree}}(\ell,1,2,3,\dots,n,-\ell) \label{eq:forward-limit-decomp-npt} \\
& +\frac{1}{(\ell+k_{1})^{2}}\,A_{\text{BCJ}}^{\text{tree}}(\ell+k_{1},2,3,\dots,n,1,-(\ell+k_{1}))
\nonumber \\
 & +\frac{1}{(\ell+k_{1}+k_{2})^{2}}\,A_{\text{BCJ}}^{\text{tree}}(\ell+k_{1}+k_{2},3,\dots,n,1,2,-(\ell+k_{1}+k_{2})) \nonumber \\
 & +\dots\nonumber 
\end{align}
where by $A_{\text{BCJ}}^{\text{tree}}$ we mean the BCJ tree formula
(\ref{eq:bcj-amplitude}) with the two parallel legs of every numerator
welded, and then swapped by the corresponding polygon numerator at
one loop. We mentioned there that $A_{\text{BCJ}}^{\text{tree}}$
formally resembles welded forward limit tree amplitudes, therefore
motivates the question about whether we may simply identify the value
obtained from Feynman rules, $A_{\text{Feyn}}^{\text{tree}}$ , with
these amplitudes. 

\begin{figure}[h!]
\centering
\begin{equation}
\mathcal{I}_{2}(\ell,k_{1},k_{2}) = \begin{minipage}{1.5cm}\includegraphics[width=1.5cm]{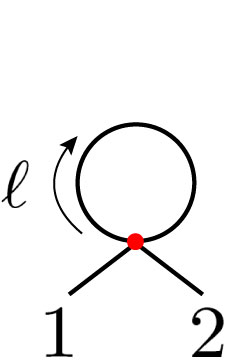}\end{minipage}+\begin{minipage}{2.1cm}\includegraphics[width=2.1cm]{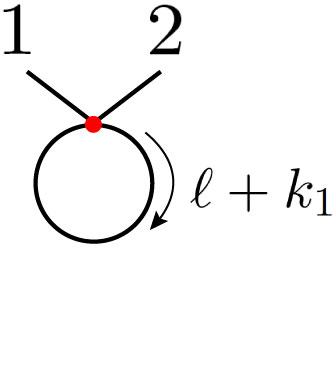}\end{minipage}
\nonumber
\end{equation}
\caption{$2$-point colour-ordered integrand of a hypothetical theory that only has $4$-vertex}
\label{fig:4-vertex-integrand}
\end{figure}

This resemblance actually goes further. The above forward limit decomposition
(\ref{eq:forward-limit-decomp-npt}) is actually not limited to cubic
structure. The fact that it generalises can be seen by considering
a hypothetical theory that contains only a $4$-vertex. The resulting
integrand in this theory at $n$-point will also be a sum over polygons,
and with vertices lying on the loop, from which sub-trees are attached.
(The polygon structure is illustrated in Fig. \ref{fig:4-vertex-integrand}.
Terms in this figure are all Feynman graphs, not numerators.) 
Following the same argument to replace loop propagators by partial
fractions and then re-group, we see a similar decomposition to (\ref{eq:forward-limit-decomp-npt}).
Further including more types of vertices to the theory and repeat,
we see that integrands defined by general colour-ordered Feynman rules
also satisfy a similar forward limit decomposition.
\begin{align}
\mathcal{I}_{n}(\ell,k_{1},k_{2},\dots,k_{n})= & \sum_{h}\frac{1}{\ell^{2}}\,A_{\text{Feyn}}^{\text{tree}}(\ell^{h},1,2,3,\dots,n,-\ell^{h})  \label{eq:feyn-forward-limit-decomp} \\
&+\frac{1}{(\ell+k_{1})^{2}}\,A_{\text{Feyn}}^{\text{tree}}((\ell+k_{1})^{h},2,3,\dots,n,1,-(\ell+k_{1})^{h})
\nonumber \\
 & +\frac{1}{(\ell+k_{1}+k_{2})^{2}}\,A_{\text{Feyn}}^{\text{tree}}((\ell+k_{1}+k_{2})^{h},3,\dots,n,1,2,-(\ell+k_{1}+k_{2})^{h}) \nonumber \\
 & +\dots\nonumber 
\end{align}
where in the above we have assumed the propagator of this theory can
be ``cut'' by inserting complete states $h$, such as the Yang-Mills
propagator in light-cone gauge $\eta^{\mu\nu}/\ell^{2}=\sum_{\pm}\epsilon^{\pm,\mu}\epsilon^{\mp,\nu}/\ell^{2}$.

The remarkable resemblance between the two decompositions (\ref{eq:forward-limit-decomp-npt})
and (\ref{eq:feyn-forward-limit-decomp}) again suggests that we identify
the values of $A_{\text{BCJ}}^{\text{tree}}$ with $A_{\text{Feyn}}^{\text{tree}}$.
Furthermore, from (\ref{eq:forward-limit-decomp-npt}) and (\ref{eq:feyn-forward-limit-decomp})
we see that the numerator solutions obtained from identifying $A_{\text{BCJ}}^{\text{tree}}$
with $A_{\text{Feyn}}^{\text{tree}}$ is actually imposing a stronger
condition than the method explained in section \ref{sec:shifting-used-explained}:
In the following we discuss solutions obtained by identifying the
two, so that the $A_{\text{BCJ}}^{\text{tree}}$ and $A_{\text{Feyn}}^{\text{tree}}$
appear in equations (\ref{eq:forward-limit-decomp-npt}) and (\ref{eq:feyn-forward-limit-decomp})
must be equal term-by-term, while in section \ref{sec:shifting-used-explained}
the numerators were solved from linear equations between the whole
integrands and numerators. Therefore the solutions discussed here,
if exists, automatically satisfies the linear equations of section
\ref{sec:shifting-used-explained}, but the reverse is not necessarily
true.

To determine the numerators from $A_{\text{Feyn}}^{\text{tree}}$,
recall that we argued in section \ref{sec:forward-limit-propagators}
that the expansion coefficients for $A_{\text{BCJ}}^{\text{tree}}$
with respect to basis $n$-gons $N(\ell,\sigma(1),\sigma(2),\dots,\sigma(n),-\ell)$
are just the tree propagator matrix $m(\alpha|\beta)$, with forward
limit taken. Therefore the basis numerators can be solved from linear
equations similar to (\ref{eq:4pt-matrix-eqn}), with the amplitudes
and numerators appear in the equation identified as $A_{\text{Feyn}}^{\text{tree}}$
and $N(\ell,\sigma(1),\sigma(2),\dots,\sigma(n),-\ell)$ respectively. 

Note however, during the partial fraction derivation that leads to
(\ref{eq:forward-limit-decomp-npt}), the propagator of one $n$-gon
is decomposed into several terms, and re-grouped into several separate
$A_{\text{BCJ}}^{\text{tree}}$. So that the $n$-gon $N(\ell,1,2,3,\dots,n,-\ell)$
appear in $A_{\text{BCJ}}^{\text{tree}}(\ell,1,2,3,\dots,n,-\ell)$
needs to be the same $n$-gon appear in $A_{\text{BCJ}}^{\text{tree}}(\ell+k_{1},2,3,\dots,n,1,-(\ell+k_{1}))$,
equivalently denoted as $N(\ell,1,2,3,\dots,n,-\ell)$ or as $N(\ell+k_{1},2,3,\dots,n,1,-(\ell+k_{1}))$
according to the nomenclature. If we were to identify $A_{\text{BCJ}}^{\text{tree}}$
with $A_{\text{Feyn}}^{\text{tree}}$, this requires that the numerator
solved from the first terms $A_{\text{Feyn}}^{\text{tree}}(\ell^{h},1,2,3,\dots,n,-\ell^{h})$
in the forward limit decomposition (\ref{eq:feyn-forward-limit-decomp})
needs to have the same value with those solved from the second terms
$A_{\text{Feyn}}^{\text{tree}}((\ell+k_{1})^{h},2,3,\dots,n,1,-(\ell+k_{1})^{h})$
and so on. These appear as separate matrix equations when solving
the numerators using the method just described. Explicitly, we need
the following extra condition.
\begin{eqnarray}
&&
\sum_{h}N^{\text{tree}}(\ell^{h},\sigma(1),\sigma(2),\dots,\sigma(n),-\ell^{h}) 
\label{eq:extra-condition-2} \\
&&
=\sum_{h}N^{\text{tree}}((\ell+k_{\sigma(1)})^{h},\sigma(2),\dots,\sigma(n),\sigma(1),-(\ell+k_{\sigma(1)})^{h})
\nonumber
\end{eqnarray}
Namely, suppose if we would like to identify an $n$-gon with the tree numerators solved from $A_{\text{Feyn}}^{\text{tree}}$, then the identifications corresponding to all possible incision locations of the $n$-gon need to be all equal. Generically, tree numerators solved from different $A_{\text{Feyn}}^{\text{tree}}$'s can take different generalised gauge. For example, the extra condition may read
\begin{align}
 & \frac{1}{\ell^{2}}\sum_{\rho\in S_{n}}\sum_{h}\mathcal{S}[\sigma(1),\sigma(2),\dots,\sigma(n)|\rho]_{\ell}\,A_{\text{Feyn}}^{\text{tree}}(\ell^{h},\rho(1),\rho(2),\dots,\rho(n),-\ell^{h})\label{eq:extra-condition}\\
 & =\frac{1}{(\ell+k_{\sigma(1)})^{2}}\sum_{\rho\in S_{n}}\sum_{h}\mathcal{S}[\sigma(2),\dots,\sigma(n),\sigma(1)|\rho]_{\ell+k_{\sigma(1)}}\, \nonumber \\
 & \hspace{4cm} \times A_{\text{Feyn}}^{\text{tree}}((\ell+k_{\sigma(1)})^{h},\rho(1),\rho(2),\dots,\rho(n),-(\ell+k_{\sigma(1)})^{h})\nonumber 
\end{align}
or some numerators may come from $(n-3)$-basis ($(n-3)!$-basis) and so on.

Additionally note that it was discovered by \cite{Ochirov:2017jby}
that the rank of the propagator matrix becomes further reduced compared
with tree level ones when forward limit is taken. In view of (\ref{eq:4pt-matrix-eqn}),
the reduced rank suggests more identities must be satisfied by the
forward limit amplitude $A_{\text{Feyn}}^{\text{tree}}$ in order
to yield numerator solutions. Here we assumed these conditions are
also met and the numerators can be solved by inverting the propagator
matrix.

\textbf{Addressing (ii):} When forward limit amplitudes $A_{\text{Feyn}}^{\text{tree}}$
calculated from Feynman rules satisfy the conditions due to reduced
rank just mentioned, and that the welded tree numerators solved by
inverting the propagator matrix are compatible with the loop numerator
nomenclature (\ref{eq:npt-symmetry}), which is explicitly given by
(\ref{eq:extra-condition}), we have a consistent set of basis numerators
(the $n$-gons). Other numerators are generated by this set through
the same manipulations as in section \ref{sec:shifting-used-explained}.
Such a set is indeed a solution can be seen from that when multiplied
with propagators, they yield $A_{\text{Feyn}}^{\text{tree}}$, so
that the whole integrand is correctly reconstructed term-by-term through
(\ref{eq:feyn-forward-limit-decomp}). When the extra conditions are
violated, the numerators obtained conflicts with the nomenclature.
Also those supposedly be the same $n$-gons appear in $A_{\text{Feyn}}^{\text{tree}}(\ell^{h},1,2,3,\dots,n,-\ell^{h})$
and $A_{\text{Feyn}}^{\text{tree}}((\ell+k_{1})^{h},2,3,\dots,n,1,-(\ell+k_{1})^{h})$
disagree with each other, so that we cannot identify the welded tree
numerators as the solutions of the one loop numerators.

From the discussion at the beginning of this section one also sees
that in the extreme case, where we have a Jacobi satisfying cubic
theory, and additionally when there is a natural operation that welds
tree numerator to give exactly what is supposed to be the loop numerator,
then equations (\ref{eq:forward-limit-decomp-npt}) and (\ref{eq:feyn-forward-limit-decomp})
are identical graph-by-graph, instead of just amplitude-by-amplitude.
The welded tree numerator therefore naturally is the loop numerator.
Such condition is apparently satisfied by the self-dual Yang-Mills
numerators, as they are given by products of structure constants of
diffeomorphism algebras, contracted to form a loop, $N_{\text{SDYM}}^{\text{tree}}(\ell^{h},\sigma(1),\sigma(2),\dots,\sigma(n),-\ell^{h})=\sum_{\pm}f^{\ell^{\pm},\sigma(1)}{}_{*}f^{*,\sigma(2)}{}_{*}\dots f^{*,\sigma(n)}{}_{\ell^{\mp}}$.
The same can be said for NLSM numerators, which we demonstrate in
appendix \ref{sec:nlsm-numerators}. Note that unlike the bypassing
method mentioned in (i) in the introduction and explained in section
\ref{sec:shifting-used-explained}, not all theories satisfy the stronger
condition discussed in this section. A quick way to see this is by
considering a counter example where we have a hypothetical theory
that has only a $4$-vertex in its Feynman rules. For concreteness
we may choose the vertex to be a polynomial $V(1,2,3,4)=c+c_{\mu}^{(i)}k_{i}^{\mu}+c_{\mu,\nu}^{(i,j)}k_{i}^{\mu}k_{j}^{\nu}+\dots$,
assuming only KK and BCJ relations for tree amplitudes, which leave
undetermined parameters that may be adjusted to violate the condition
(\ref{eq:extra-condition}). The fact that both NLSM and SDYM satisfy
this extra condition seems to suggest that this is the more physically
interesting scenario that we should consider.

\textbf{Remark:} The fact that numerator solution always exist when
solving from integrands $\mathcal{I}_{n}(\ell,\sigma(1),\dots,\sigma(n))$
(using the method explained in section \ref{sec:shifting-used-explained})
but not from forward limit tree amplitudes $A_{\text{Feyn}}^{\text{tree}}(\ell,\sigma(1),\dots,\sigma(n),-\ell)$
might appear puzzling at first sight. Roughly speaking, the distinct
feature that affects the solution is that $A_{\text{Feyn}}^{\text{tree}}$
is not in general invariant under the symmetry transformation (\ref{eq:npt-symmetry})
(cyclic permutation together with a shifting in the loop momentum
$\ell$) while integrand can be made so with proper loop momentum
assignment. This is because integrand constructed from Feynman rules
is cyclic, and that it is possible to choose a loop momentum convention
to make it compatible with this symmetry. On the other hand forward
limit tree amplitudes $A_{\text{Feyn}}^{\text{tree}}(\ell,\sigma(1),\dots,\sigma(n),-\ell)$
constructed from the same Feynman rules has a distinctive ``cut''.
$4$-vertices and higher can not be placed across this cut. It is
only when cyclically summed (\ref{eq:feyn-forward-limit-decomp})
that the symmetry is restored.
\subsection{Relation between tree and loop numerators}
\label{sec:loop-and-tree-relation}

In the event when the extra conditions are not satisfied, as we have
just argued in section \ref{sec:forward-limit-numerators}, $A_{\text{BCJ}}^{\text{tree}}$
and $A_{\text{Feyn}}^{\text{tree}}$ cannot be identified, and that
the one loop numerators are not simply the welded tree numerators.
In this scenario the one loop numerators, which can still be determined
using the shifting method explained in section \ref{sec:shifting-used-explained},
and the welded tree numerators, by which we mean the solutions to
linear equations such as (\ref{eq:4pt-matrix-eqn}) determined by
$A_{\text{Feyn}}^{\text{tree}}$, are still related, as the two must
give the same colour-ordered integrand.

We use the $2$-point case to illustrate the relation, even though
generalisation to $n$-point should be straightforward. At $2$-point
we have only one independent colour-ordered integrand $\mathcal{I}_{2}(\ell,k_{1},k_{2})$,
and its decomposition (\ref{eq:feyn-forward-limit-decomp}) contains
only two terms, $\frac{1}{\ell^{2}}\,A_{\text{Feyn}}^{\text{tree}}(\ell,1,2,-\ell)$
and $\frac{1}{(\ell+k_{1})^{2}}\,A_{\text{Feyn}}^{\text{tree}}(\ell+k_{1},2,1,-(\ell+k_{1}))$.
(This is not to be confused with (\ref{eq:forward-limit-decomp-npt}).)
The first term has the following relation with forward limit tree
numerators.
\begin{align}
& \frac{1}{\ell^{2}}\sum_{h}A_{\text{Feyn}}^{\text{tree}}(\ell^{h},1,2,-\ell^{h}) \\
& =  \sum_{h}\frac{1}{\ell^{2}}m(\ell,1,2,-\ell|\ell,1,2,-\ell)\,N^{\text{tree}}(\ell^{h},1,2,-\ell^{h})\\
 & +\frac{1}{\ell^{2}}m(\ell,1,2,-\ell|\ell,2,1,-\ell)\,N^{\text{tree}}(\ell^{h},2,1-\ell^{h})\nonumber \\
\sim & \sum_{h}\frac{1}{\ell^{2}}m(\ell,1,2,-\ell|\ell,1,2,-\ell)\,N^{\text{tree}}(\ell^{h},1,2,-\ell^{h})\\
 & +\frac{1}{(\ell+k_{1})^{2}}m(\ell+k_{1},1,2,-(\ell+k_{1})|\ell+k_{1},2,1,-(\ell+k_{1}))\,N^{\text{tree}}((\ell+k_{1})^{h},2,1-(\ell+k_{1})^{h})\nonumber 
\end{align}
where we have used the fact that $\frac{1}{\ell^{2}}\,A_{\text{Feyn}}^{\text{tree}}(\ell,1,2,-\ell)$
is a term in the integrand so that adding difference terms related
by a constant shift does not alter the integral result. Similar relation
can be written down for term $\frac{1}{(\ell+k_{1})^{2}}\,A_{\text{Feyn}}^{\text{tree}}(\ell+k_{1},2,1,-(\ell+k_{1}))$.
Note for tree numerators, now we are not assuming $N^{\text{tree}}(\ell^{h},1,2,-\ell^{h})=N^{\text{tree}}((\ell+k_{1})^{h},2,1-(\ell+k_{1})^{h})$,
so that these two terms do not combine.

On the other hand the same integrand is assumed to be expressible
as (\ref{eq:forward-limit-decomp-npt}). For the $2$-point case this
is (\ref{eq:tree-decomposition-2pt}), or more explicitly given by
(\ref{eq:2pt-integrand}). We have argued in great length that this
equals $\frac{1}{\ell^{2}(\ell+k_{1})^{2}}N(\ell,1,2,-\ell)$ in section
\ref{sec:2pt-example}. This time using $N(\ell,1,2,-\ell)=N(\ell+k_{1},2,1-(\ell+k_{1}))$,
because both refer to the same bubble numerator.

The two expressions above must be equal, as they should both give
the same integrand that is given by the colour-ordered Feynman rules,
therefore we have
\begin{align}
& N(\ell,1,2,-\ell) \\
&=  \ell^{2}(\ell+k_{1})^{2}\times\Bigl(\sum_{h}\frac{1}{\ell^{2}}m(\ell,1,2,-\ell|\ell,1,2,-\ell)\,N^{\text{tree}}(\ell^{h},1,2,-\ell^{h})
\nonumber \\
 & +\frac{1}{(\ell+k_{1})^{2}}m(\ell+k_{1},1,2,-(\ell+k_{1})|\ell+k_{1},2,1,-(\ell+k_{1}))\,N^{\text{tree}}((\ell+k_{1})^{h},2,1-(\ell+k_{1})^{h}) \nonumber \\
 & +\dots\Bigr)
 \nonumber 
\end{align}
where the $\dots$ refers to the similar terms that come from $\frac{1}{(\ell+k_{1})^{2}}\,A_{\text{Feyn}}^{\text{tree}}(\ell+k_{1},2,1,-(\ell+k_{1}))$.
At higher-point, the above becomes a matrix equation.

\section{KLT and double copy formulation at one loop}
\label{sec:double-copy}

Recall that at tree level, the KLT \cite{Kawai:1985xq,Mafra:2011nw,Bjerrum-Bohr:2010pnr},
the Del Duca-Dixon-Maltoni (DDM) \cite{DelDuca:1999rs,Bern:2010yg}, and the double-copy
formulation \cite{Bern:2010ue} are all known to be equivalent
\cite{Bern:2010yg}. Starting from double copies, the other
two formulations can be readily obtained by expanding with respect
to half-ladder basis and using the fact that the momentum kernel is
the inverse of the propagator matrix.
\begin{align}
A_{GR}^{\text{tree}}(123\dots n) & =\sum_{\text{all cubic graphs }\Gamma}\frac{N_{\Gamma}\,\bar{N}_{\Gamma}}{\prod_{i\in\Gamma}D_{i}}\label{eq:tree-db-copy}\\
 & =\sum_{\sigma\in S_{n-2}}\,N(1\sigma n)\,\bar{A}_{YM}(1\sigma n)\label{eq:half-ladder}\\
 & =\frac{1}{k_{n}^{2}} \sum_{\alpha , \beta \in S_{n-2}} 
 A_{YM}(1\alpha n)\,S[\alpha^{T}|\beta]\,\bar{A}_{YM}(1\beta n)\label{eq:tree-klt-relation}
\end{align}
The relation between their generalizations at one loop level following
the reasoning explained in section \ref{sec:shifting-used-explained},
however, is less straightforward. Generically one finds a relative
shift between the loop momentum dependence of each component. In
this section we illustrate this issue through a $2$-point example,
even though the same issue also resides in the $3$-point amplitudes.

We start from a double-copy gravity integrand $\mathcal{I}_{GR}$
with the assumption that it can be defined by bi-adjoint $\phi^{3}$-like
structure, with the loop momentum dependence of the two sets of kinematic
numerators aligned.
\begin{equation}
\mathcal{I}_{GR}(\ell,1,2)
= \frac{1}{\ell^{2}(\ell+k_{1})^{2}}
\times\begin{minipage}{1.8cm}\includegraphics[width=1.8cm]{2ptEq1a}\end{minipage} \times\begin{minipage}{1.8cm}\includegraphics[width=1.8cm]{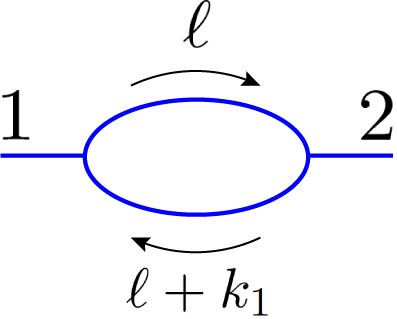}\end{minipage} 
+ \frac{1}{\ell^{2}}\,\frac{1}{s_{12}} \times \begin{minipage}{1cm}\includegraphics[width=1cm]{2ptEq1b}\end{minipage}\times\begin{minipage}{1cm}\includegraphics[width=1cm]{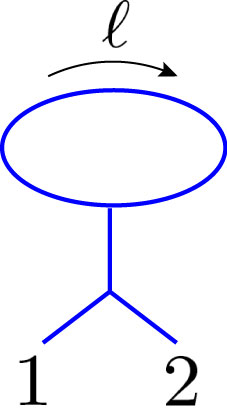}\end{minipage}.
\label{eq:2pt-db-copy} 
\end{equation}
We leave the first term as it is. For the second term, we expand one
manually chosen copy (the unbarred numerators represented by numerator graphs shown in black) in the following with respect
to the corresponding basis. This requires using Jacobi identities
and then shift $\ell$ by the appropriate amount to allow the un-barred
numerator be traded by the consistency condition (\ref{eq:consistency-1}),
creating a difference term in the process.
\begin{align}
 & \frac{\biggl(N(\ell,12,-\ell)-N(\ell,21,-\ell)\biggr)\biggl(\bar{N}(\ell,12,-\ell)-\bar{N}(\ell,21,-\ell)\biggr)}{D_{\ell}\,s_{12}}\\
 & =\frac{N(\ell,12,-\ell)}{D_{\ell}\,s_{12}}\biggl(\bar{N}(\ell,12,-\ell)-\bar{N}(\ell,21,-\ell)\biggr)\label{eq:shifted-tadpole}\\
 & -\frac{N(\ell,12,-\ell)}{D_{\ell1}\,s_{12}}\biggl(\bar{N}(\ell+k_{1},12,-(\ell+k_{1}))-\bar{N}(\ell+k_{1},21,-(\ell+k_{1}))\biggr)+\text{difference terms}\nonumber 
\end{align}
Collecting terms with respect to the unbarred basis, we see there
are now two (barred/blue) tadpoles appearing as its coefficients, as required
by colour-ordered integrand.
\begin{eqnarray}
&& \tilde{\mathcal{I}}_{GR}(\ell,12) \nn
&& =   \Biggl(
 \frac{1}{D_{\ell}D_{\ell,1}} \times\begin{minipage}{1.8cm}\includegraphics[width=1.8cm]{2ptEq1a}\end{minipage} \times\begin{minipage}{1.8cm}\includegraphics[width=1.8cm]{2ptEq1a2}\end{minipage} 
\nn
&&
~~~~~~~~~~~~~~~~~~~~~~~~~~~~~~~~~~~~~~+ \frac{1}{D_{\ell}\,s_{12}} \times \begin{minipage}{1.8cm}\includegraphics[width=1.8cm]{2ptEq1a}\end{minipage}\times\begin{minipage}{1cm}\includegraphics[width=1cm]{2ptEq1b3}\end{minipage}
+ \frac{1}{D_{\ell,1}\,s_{12}}\times \begin{minipage}{1.8cm}\includegraphics[width=1.8cm]{2ptEq1a}\end{minipage}\times\begin{minipage}{1cm}\includegraphics[width=1cm]{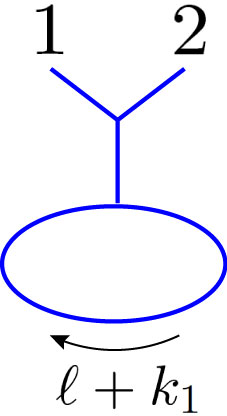}\end{minipage}
\Biggr) \nn
&& =  N(\ell,12,-\ell) \times \bar{\mathcal{I}}_{YM}(\ell,12).
\label{2pt-half-ladder}
\end{eqnarray}
Note the barred integrand has not acquired any shifting in this derivation.

To obtain the analogue of KLT expression (\ref{eq:tree-klt-relation})
we need to further translate the unbarred numerator into integrand.
However recall the manipulation explained in section \ref{sec:2pt-example}
requires shifting the integrand to allow identification using the
consistency condition (\ref{eq:consistency-1}), otherwise one faces
the infinite matrix problem. Therefore the reasoning we use in this
paper relates the $2$-point numerator with shifted integrand instead
of the original one, where loop momentum $\ell$ follows the convention
inherited from string integrand \cite{Geyer:2015bja}.
\begin{equation}
N(\ell,12,-\ell)=(D_{\ell}D_{\ell,1})\tilde{\mathcal{I}}_{YM}(\ell,12)=(D_{\ell}D_{\ell,1})\,(\mathcal{I}_{YM}(\ell,12)+\Delta_{\ell})
\end{equation}
Plugging the above back into (\ref{2pt-half-ladder}) and we see that
one either choose manually an unequal-footing expression in the KLT
formula, where one of the integrand acquires an extra difference term,
or we start with a double copy formulation that has loop momentum
dependence that violates the convention of \cite{Geyer:2015bja}.
\begin{align}
\mathcal{\tilde{I}}_{GR}(\ell,12) & =\left(\mathcal{I}_{YM}(\ell,12)+\Delta_{\ell}\right)\,S[1|1]_{\ell}\,\bar{\mathcal{I}}_{YM}(\ell,12).
\end{align}
Here the difference term $\Delta_{\ell}=f(\ell)-f(\ell+c)$ acts as
a factor in the integrand so that it generically has non-vanishing contribution
to the loop integral. We find the same issue exists also at $3$-point.

Note that this issue was identified early in the development of the colour-kinematics duality by \cite{feng:2012, Stieberger:2023nol,Stieberger:2022lss} and various research groups. We revisit them here for the sake of conceptual completeness and to adapt the discussion to the specific assumptions and settings used in this paper.

While this paper was being completed, we became aware of ref. \cite{Cao:2025ygu}, which also discusses one-loop KLT-type relations. In that reference, the resulting expression for the gravitational integrand appears manifestly symmetric between the two YM copies. This difference can be traced back to the distinct starting assumptions adopted in the two approaches. In \cite{Cao:2025ygu}, the symmetry arises from assuming a one-loop generalised DDM relation from the outset, together with the use of a shifted colour-ordered YM integrand $\tilde{\mathcal{I}}$ (see equation (9) of \cite{Cao:2025ygu}). The subsequent integrand $\mathcal{I}(1,\alpha)$ is then obtained by collecting terms related by symmetry. By contrast, in the present work we assume the validity of a symmetric double-copy formula (\ref{eq:2pt-db-copy}) and show that this assumption is equivalent to a one-loop generalised DDM representation formulated in terms of the unshifted YM integrand (\ref{2pt-half-ladder}), which in turn is equivalent to a non-symmetric KLT relation. In terms of the conventions used in this paper, the two YM integrands appearing in \cite{Cao:2025ygu} are therefore both shifted, and is equivalent to a non-symmetric double copy formula.


\section{A Two-Step Expansion Strategy for One-Loop Yang-Mills Numerators}
\label{sec:ym-numerators}

In previous sections, we addressed the question of whether BCJ-compatible numerators exist for a given one-loop integrand, and, if so, whether they are unique. By exploring the consistency conditions of the 
nomenclature of the numerators and integrands, we established a full-rank linear transformation between the two, showing that for a properly shifted integrand, the numerators satisfying Jacobi identities are uniquely determined.

Although this establishes the conceptual existence and uniqueness of one-loop numerators, the explicit inversion of the associated transformation matrix becomes rapidly intractable beyond three points. To circumvent this difficulty, we develop in this and the next few sections a constructive, algebraically motivated approach that yields explicit numerator expressions without resorting to matrix inversion.

Our key idea is to express the Yang–Mills integrand as its unique linear expansion of propagator matrix elements, allowing us to directly read off the numerators as the corresponding expansion coefficients. In fact, writing the Yang–Mills integrand in the form of the propagator-matrix expression~(\ref{eq:npt-prop-matrix}) is precisely equivalent to expanding it as a linear combination of BS integrands. This two-step procedure proceeds as follows:

\begin{itemize}
  \item In the first step, the Yang-Mills integrand is expanded in terms of scalar-loop integrands in the so-called YMS (Yang-Mills-scalar) theory, where external legs are gluons and scalars while the loop contains only scalar propagators.
  
  \item In the second step, each YMS integrand is recursively expanded in terms of bi-adjoint scalar (BS) integrands, resulting in an overall factorisation of the original Yang-Mills integrand in terms of known BS integrands.
\end{itemize}

To make this expansion precise, we introduce a consistent notation system to describe the Yang-Mills integrands, the YMS integrands (with scalar internal lines), and the BS integrands. We then formulate the expansion in terms of these structures, and extract the kinematic coefficients (which we later identify as the colour-stripped numerators) in such a way that they manifestly satisfy the one-loop consistency conditions upon shifting and relabeling.


\subsection{Notation for Scalar-loop YMS Integrands}

We begin by introducing a convenient notation to describe one-loop integrands where the internal loop consists purely of scalars, and external particles can be scalars and gluons. We will be referring such integrand as the scalar-loop Yang–Mills-scalar (YMS) integrand, which is actually only part of the YMS integrand, as the full YMS integrand defined by Feynman rules should include gluon loops (more detail can be found in \cite{Porkert:2022efy}). The absence of gluon self-interactions in the loop allows for a simpler structure of integrands, making them suitable intermediate bases for constructing Yang–Mills numerators.

We denote such YMS integrand by $\mathcal{I}^{\,\text{YMS}}(\pmb{j} || \pmb{g}\,|\,\pmb{\rho})$, where the arguments are defined as follows (This is the notation used by \cite{Xie:2024pro,Xie:2025utp}):
\begin{itemize}
\item $\pmb{\rho}$ denotes the global cyclic ordering of all external particles
(i.e., both scalars and gluons), which determines the arrangement of propagators 
in the full loop integrand.

\item $\pmb{j} = \{j_1,\dots,j_m\}$ denotes the cyclically ordered external scalars. 
These scalars determine the second colour-order index. 
We use both $\pmb{\rho}$ and $\pmb{j}$ to label the integrand 
$\mathcal{I}^{\,\text{YMS}}$.  
Both colour factors corresponding to $\pmb{\rho}$ and $\pmb{j}$ are stripped.

\item $\pmb{g} = \{g_1,\dots,g_{n-m}\}$ denotes the set of external gluons.
\end{itemize}

Each integrand $\mathcal{I}^{\,\text{YMS}}(\pmb{j} || \pmb{g} \,|\, \pmb{\rho})$ represents a sum over Feynman graphs, in which tree level currents with gluons and/or scalars are attached to the scalar loop, via three-point scalar vertex, two-scalar-one-gluon vertex or two-scalar-two-gluon vertex, as demonstrated by \cite{Xie:2024pro}.

This notation cleanly separates the contributions from different loop insertion patterns and will prove instrumental in formulating the Yang–Mills integrand as a linear combination of YMS building blocks in the next subsection.

\subsection{Decomposition of Yang-Mills Integrands into YMS Building Blocks}

Having introduced the notation for scalar-loop YMS integrands, we now present a decomposition of the one-loop Yang–Mills integrand $\mathcal{I}^{\,\text{YM}}(\pmb{\rho})$ into a sum over YMS integrands of the form $\mathcal{I}^{\,\text{YMS}}(\pmb{j} || \mathsf{G} \setminus \pmb{j} \,|\, \pmb{\rho})$. This expansion separates out contributions according to the number and ordering of gluons inserted on the scalar loop and reflects the algebraic structure of color-dressed integrands.

The decomposition reads \cite{Geyer:2017ela,Porkert:2022efy,Xie:2024pro,Cao:2024olg}:
\begin{align}
\mathcal{I}^{\,\text{YM}}(\pmb{\rho}) &= (D - 2)\, \mathcal{I}^{\,\text{YMS}}(\emptyset || \mathsf{G} \mid \pmb{\rho}) \nonumber\\
&\quad + \sum_{l=2}^{n} (-1)^l \sum_{\substack{\pmb{j} \subset \mathsf{G} \\ |\pmb{j}| = l}} 
\mathrm{Tr}(F_{j_1} \cdots F_{j_l})\, \mathcal{I}^{\,\text{YMS}}(j_1,\dots,j_l || \mathsf{G} \setminus \{j_1,\dots,j_l\}\,|\,\pmb{\rho}) \, .
\label{Eq:YMbyYMS}
\end{align}
In this expression:\\
- $\mathsf{G} = \{1,2,\dots,n\}$ denotes the full set of external gluon labels; \\
- the sum runs over all subsets of fixed length $l$,  $\pmb{j} \subset \mathsf{G}$ of size $l$, modulo cyclic permutations;\\
- $\mathrm{Tr}(F_{j_1} \cdots F_{j_l})$ denotes the color-ordered trace of field-strength tensors; \\
- each $\mathcal{I}^{\,\text{YMS}}(\pmb{j} || \mathsf{G} \setminus \pmb{j} \,|\, \pmb{\rho})$ is the one loop integrand defined by the YMS colour-ordered Feynman rules, with   $\pmb{j}$ scalars and  $\mathsf{G} \setminus \pmb{j}$ gluons inserted on the scalar loop.


 This decomposition formula sets the stage for directly reading off numerators as expansion coefficients once the YMS integrands are further expressed in terms of bi-adjoint scalar integrands.

By rewriting the Yang–Mills integrand as a sum over trace-labelled YMS building blocks, we effectively translate the problem of constructing BCJ-compatible numerators into a more tractable task: determining the expansion of each YMS integrand in terms of BS integrands. We will carry out this step in the following section.


\section{Expanding Scalar-loop Integrands into Bi-adjoint Scalar Integrands}
\label{sec:yms-to-bs}

Having expressed the Yang–Mills integrand as a sum over scalar-loop YMS integrands, our next task is to expand each YMS integrand into a linear combination of bi-adjoint scalar (BS) integrands. This second step is grounded in a direct comparison of the Feynman rules of the two theories: the scalar-loop YMS theory, which includes scalar, gluon, and mixed vertices, and the purely scalar BS theory. By matching loop graphs and their associated kinematic factors, we systematically determine the expansion coefficients. In doing so, we complete the two-stage reduction of the original problem of constructing one-loop BCJ numerators into the determination of kinematic coefficients appearing in a well-structured expansion.

As discussed earlier in our analysis of color-ordered one-loop integrands, scalar-loop integrands in the YMS theory naturally admit a representation in terms of the propagator matrix structure  (\ref{eq:npt-prop-matrix}). Furthermore, each YMS integrand $\mathcal{I}^{\text{YMS}}$ can be further stripped of its scalar-loop color factor, yielding a \textbf{doubly color stripped} integrand $\mathcal{I}^{\text{YMS}}_{\text{DCS}}(l;1,\pmb{\sigma}||\mathsf{G}\,|\, \pmb{\rho})$. With the second set of color factors stripped from \ref{eq:npt-prop-matrix}, this object can be written as the following expansion.
\begin{equation}
\mathcal{I}_{\text{DCS}}^{\,\text{YMS}}(\ell;1,\pmb{\sigma}||\mathsf{G}\,|\, \pmb{\rho}) = \sum_{\pmb{\gamma} \in \mathrm{Perms(\mathsf{G}),\ \pmb{\alpha}\in \pmb{\gamma} \shuffle \pmb{\sigma}  }} N_{\text{DCS}}(\ell;1,\pmb{\alpha}\,|\, \pmb{\rho}){I}^{\,\text{BS}}(\ell;1,\pmb{\alpha} \,|\,\pmb{\rho}) ,
\label{Eq:Exp0}
\end{equation}
By comparing the Feynman-graph expansions of the YMS and BS integrands, we find that the coefficient $N_{\text{DCS}}(\ell;1,\pmb{\alpha}\,|\, \pmb{\rho})$ depends only on the ordering $\pmb{\alpha}$. We therefore denote it by $C(l;1,\pmb{\alpha})$ and rewrite~(\ref{Eq:Exp0}) as
\begin{equation}
\mathcal{I}_{\text{DCS}}^{\,\text{YMS}}(\ell;1,\pmb{\sigma}||\mathsf{G} \mid \pmb{\rho}) = \sum_{\pmb{\gamma} \in \mathrm{Perms(\mathsf{G}),\ \pmb{\alpha}\in \pmb{\gamma} \shuffle \pmb{\sigma}  }} C(\ell;1,\pmb{\alpha}\, ) {I}^{\,\text{BS}}(\ell;1,\pmb{\alpha} \mid \pmb{\rho}) \,,
\label{Eq:Exp1}
\end{equation}
where:\\
- The symbol $\shuffle$ denotes the shuffle between the scalar index $\pmb{\sigma}$  and the gluons $\pmb{\gamma}$, preserving the scalar orderings while summing over all gluon orderings;\\
- $\mathcal{I}_{\text{DCS}}^{\,\text{BS}}(\ell;1,\pmb{\alpha}\,|\, \pmb{\rho})$ denotes the doubly color ordered bi-adjoint scalar integrand with color ordering $1,\pmb{\alpha}$ and  $\pmb{\rho}$, constructed from cubic graphs with quadratic propagators;\\
- $C(\ell;1,\pmb{\alpha})$ are the expansion coefficients (loop-momentum-dependent in general) associated with the ordering $1,\pmb{\alpha}$ around the loop. The coefficients  $C(\ell;1,\pmb{\alpha})$  are purely kinematic and will be determined in the next subsection.

This expansion inherits a crucial property from the YMS integrands: all coefficients $C(\ell;1,\pmb{\alpha})$ are independent of loop momentum routing ambiguities and can be chosen to satisfy a one-loop version of the kinematic consistency conditions. In particular, they obey:
\begin{equation}
C(\ell;1,\alpha_1,\dots,\alpha_r) = C(\ell + k_1;\alpha_1,\dots,\alpha_r,1) = \cdots = C\bigg(\ell + \sum_{i=1}^{r} k_{\alpha_i}; \alpha_r,1,\alpha_1,\dots,\alpha_{r-1}\bigg)\,,
\label{Eq:ConstraintC}
\end{equation}
ensuring compatibility with the cyclic invariance and shift identities of the loop integrand.

As a result, combining the YMS decomposition of the YM integrand in \eqref{Eq:YMbyYMS} with the expansion~\eqref{Eq:Exp1} of each YMS integrand into BS building blocks leads directly to the desired propagator-matrix representation:
\begin{equation}
\mathcal{I}^{\,\text{YM}}(\pmb{\rho}) = \sum_{\pmb{\sigma} \in S_{n-1}} N(l;1,\pmb\sigma)\, \mathcal{I}^{\,\text{BS}}(\ell;1,\pmb{\sigma}\,|\, \pmb{\rho}) \,,
\end{equation}
with the coefficients $ N(l;1,\pmb\sigma)$ being linear combinations of the field-strength traces $\mathrm{Tr}(F_{j_1} \cdots F_{j_l})$ weighted by the expansion coefficients $C(\ell;1,\pmb{\alpha})$. 

A key observation is that the field-strength traces $\mathrm{Tr}(F_{j_1} \cdots F_{j_l})$ are independent of the loop momentum, so that cyclic symmetry, which is apparently satisfied by traces, in this special case becomes identical to the 
consistency condition of the nomenclature (\ref{eq:consistency-2}). On the other hand, we require the factors $C(\ell;1,\pmb{\alpha})$ satisfy the same condition (\ref{Eq:ConstraintC}) separately, therefore the 
weighted sum 
 $N(\ell;1,\pmb{\sigma})$ automatically satisfy the loop-shift consistency conditions
\begin{equation}
 N(\ell;1,\sigma_1,\dots,\sigma_{n-1}) = N(\ell+k_{\sigma_1};\sigma_1,\dots,\sigma_{n-1},1) = \cdots = N(\ell + \sum_{i=1}^{n-1} k_{\sigma_i}; \sigma_{n-1},1,\sigma_1,\dots,\sigma_{n-2})\,,
\label{Eq:ConsistencyRepeat}
\end{equation}
making them suitable as BCJ numerators for use in the double-copy construction at one loop. We write the $i$-th element $\sigma(i)$ of the ordered set $\pmb\sigma$ as $\sigma_i$ for short.   

In the next subsection, we will provide an explicit construction of the coefficients $C(\ell;1,\pmb{\alpha})$ in \eqref{Eq:Exp1} from a diagrammatic algorithm based on scalar loop topologies and insertion rules.
  Final formulae for the $2$- and $3$-point YM numerators $N(l;1,\pmb{\sigma})$  calculated using the two-step method explained above is shown in equation (\ref{pure-ym-numerators})

 
\section{Constructive prescription for the YMS expansion coefficients}
\label{sec:yms-numerators}

To construct the coefficients in~(\ref{Eq:Exp1}) that satisfy the constraint~(\ref{Eq:ConstraintC}), 
 we note that the YMS integrand $\mathcal{I}_{\text{DCS}}^{\text{YMS}}(\ell;1,\pmb{\sigma}||\mathsf{G}\,|\, \pmb{\rho})$ can be arranged as a sum over Feynman graphs, each of which is expressed by attaching tree level currents to the loop as demonstrated in \cite{Xie:2024pro}. Each tree level current is further expanded in terms of BS currents with proper kinematic coefficients. The YMS integrand is therefore expressed by a combination of Feynman graphs with proper kinematic coefficients. After this manipulation, one can collect all contributions of such graphs with respect to a given cyclic ordering ${1,\pmb{\alpha}}$. We denote such terms by $\mathcal{I}^{\text{YMS}}(l;1,\pmb{\alpha} \,|\,\pmb{\rho})$, the YMS integrand is then expressed as a sum over such terms with varying $\pmb{\alpha}$, where $\pmb{\alpha}$ runs over all shuffles of the fixed scalar sequence $\pmb{\sigma}$ with all permutations of the gluon set $\mathsf{G}$.
\bea
\mathcal{I}_{\text{DCS}}^{\,\text{YMS}}(\ell;1,\pmb{\sigma}||\mathsf{G}\,|\,\pmb\rho)=\Sl_{\pmb\alpha\,\in\,\pmb{\sigma}\shuffle\text{\,perms\,}\mathsf{G}}\mathcal{I}^{\text{YMS}}(\ell;1,\pmb\alpha\,|\,\pmb\rho).
\eea
 Thus, we can formally define the coefficient for a given permutation $\pmb\alpha\in\pmb\gamma\shuffle\pmb\sigma$ (where $\pmb\gamma\in \text{perms}\,\mathsf{G}$ are permutations of gluons) as follows.
\bea
C(\ell;1,\pmb\alpha)\equiv \frac{\mathcal{I}^{\text{YMS}}(\ell;1,\pmb\alpha\,|\,\pmb\rho)}{\mathcal{I}^{\,\text{BS}}(\ell;1,\pmb\alpha\,|\,\pmb\rho)}.\label{Eq:Coefficient1}
\eea
Apparently, these coefficients can be considered as the expansion coefficients in (\ref{Eq:Exp1}) and produce the correct YMS integrand. {\it Although the coefficient itself does not satisfy the consistency condition (\ref{Eq:ConstraintC}), we will show that they can induce compact expressions of coefficients satisfying the consistency condition (\ref{Eq:ConstraintC}).}

In the following, we answer the questions: What are the expressions of $\mathcal{I}^{\text{BS/YMS}}(\ell;1,\pmb\alpha\,|\,\pmb\rho)$? How to induce the coefficients satisfying the consistency condition \eqref{Eq:ConstraintC}, starting from \eqref{Eq:Coefficient1}?  
Recall that both $\mathcal{I}^{\text{BS}}$ and $\mathcal{I}^{\text{YMS}}$ are defined as sums over one-loop Feynman graphs, which in these scalar-loop theories take the form of polygons built from scalar propagators. External particles are attached to the loop via tree-level currents emanating from the vertices of the polygon.
A natural way to systematically enumerate such Feynman graphs is as follows:\\
• First, enumerate all possible cyclic divisions $\mathcal{D}$  of the external ordering ${1,\pmb\alpha}$. (A cyclic division of  ${1,\pmb\alpha}$ is defined as follows.  Consider ${1,\pmb\alpha}$ as cyclic order and divide this cyclic order into ordered subsets.);\\
• Then, for each ordered subset in $\mathcal{D}$, enumerate all possible  structures that are expressed by attaching currents to the loop via possible vertices.
   

We group together all such Feynman graphs that correspond to a fixed cyclic division $\mathcal{D}$, and denote their sum as
$\mathcal{I}_{\mathcal{D}}^{\,\text{YMS}}(l;1,\pmb\alpha\,|\,\pmb\rho) \,$.
The full integrand $\mathcal{I}^{\text{YMS}}(l;1,\pmb\alpha\,|\,\pmb\rho)$ is then the sum over all such division-based contributions.
\bea
\mathcal{I}^{\text{YMS/BS}}(\ell;1,\pmb\alpha\,|\,\pmb\rho)=\Sl_{\mathcal{D}\in\text{CycDiv}\,\{1,\pmb\alpha\}}\mathcal{I}_{\,\mathcal{D}}^{\text{YMS/BS}}(\ell;1,\pmb\alpha\,|\,\pmb\rho).
\eea

We now turn to the detailed construction of $\mathcal{I}_{\,\mathcal{D}}^{\text{YMS/BS}}(\ell;1,\pmb\alpha\,|\,\pmb\rho)$ associated with a given cyclic partition $\mathcal{D}$.

\noindent~{\bf(i)} For {\it BS integrand}, recall that $\mathcal{I}_{\,\mathcal{D}}^{\text{BS}}(\ell;1,\pmb\alpha\,|\,\pmb\rho)$ denotes the sum of BS Feynman graphs associated with a fixed cyclic division $\mathcal{D}$ of the ordering $1,\pmb\alpha$.  More explicitly, if the division takes the form $\{1,\pmb\alpha\}\to\pmb\alpha_1=\pmb\alpha_{1L}1\pmb\alpha_{1R},\pmb\alpha_2,...,\pmb\alpha_I$, then $\mathcal{I}_{\,\mathcal{D}}^{\text{BS}}(\ell;1,\pmb\alpha\,|\,\pmb\rho)$ is obtained by summing all BS Feynman graphs whose loop topology matches the cyclic polygon defined by the parts
${\pmb\alpha_1, \dots, \pmb\alpha_I}$, with each part attached to a tree corresponding to the double ordering $(\pmb\alpha_i | \pmb\rho_i)$\footnote{ Graphs in this section represent Feynman graphs. In addition, the direction of loop momentum $\ell$ is supposed to be reflected. These adjustments are introduced in order to agree with the conventions in \cite{Xie:2024pro}.}.

\bea
\mathcal{I}_{\,\mathcal{D}}^{\text{BS}}(\ell;1,\pmb\alpha\,|\,\pmb\rho)&=&\Sl_{\substack{\text{CycDiv}\,\pmb\rho\\ |\pmb\alpha_i|=|\pmb\rho_i|} }\begin{minipage}{4.5cm}  \includegraphics[width=4.5cm]{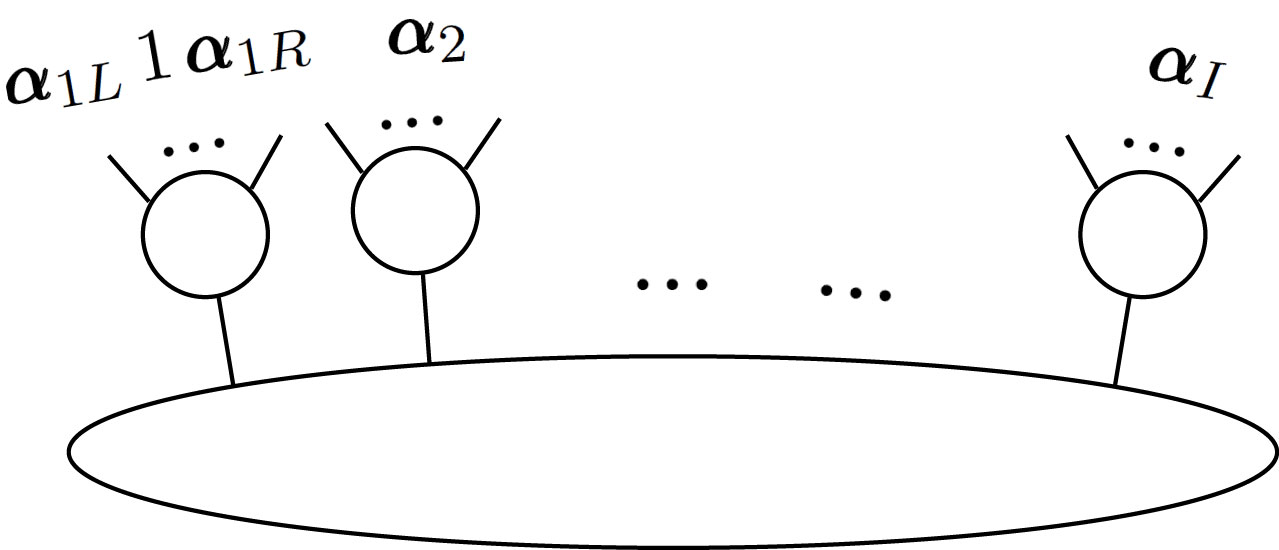} \end{minipage}\nn
&\equiv&\Sl_{\substack{\text{CycDiv}\,\pmb\rho\\ |\pmb\alpha_i|=|\pmb\rho_i|} }{1\over (\ell-k_{\pmb\alpha_{1L}})^2}\,\phi_{\pmb\alpha_{1L}1\pmb\alpha_{1R}|\pmb\rho_1}\,{1\over (\ell+k_1+k_{\pmb\alpha_{1R}})^2}\,\phi_{\pmb\alpha_{2}|\pmb\rho_2}\nn
&&~~~~~~~~~\times...\times{1\over (\ell+k_1+k_{\pmb\alpha_{1R}}+k_{\pmb\alpha_2}+...+k_{\pmb\alpha_I})^2}\,\phi_{\pmb\alpha_{I}|\pmb\rho_I},\label{Eq:IDBS}
\eea
in which, we summed over 
 all cyclic divisions of $\pmb\rho$ such that  each ordered subset $\pmb\rho_i$ has the same number of elements with the corresponding $\pmb\alpha_i$. The right permutations in the graph on the first line have been hidden for convenience, while each $\phi_{\pmb\alpha_{i}|\pmb\rho_i}$ ($i=1,...,I$) on the second line denotes a tree-level Berends-Giele current of BS (see \appref{app:BScurrents}).

\noindent~{\bf(ii)} The {\it YMS integrand} $\mathcal{I}_{\,\mathcal{D}}^{\text{YMS}}(\ell;1,\pmb\alpha\,|\,\pmb\rho)$ can be given by
\bea
\mathcal{I}_{\,\mathcal{D}}^{\text{YMS}}(\ell;1,\pmb\alpha\,|\,\pmb\rho)&=&C(\pmb\alpha_{1L}1\pmb\alpha_{1R})\,C(\pmb\alpha_{2})\,...\,C(\pmb\alpha_{I})\,\times\,\Sl_{\substack{\text{CycDiv}\,\pmb\rho\\ |\pmb\alpha_i|=|\pmb\rho_i|} }\begin{minipage}{4.5cm}  \includegraphics[width=4.5cm]{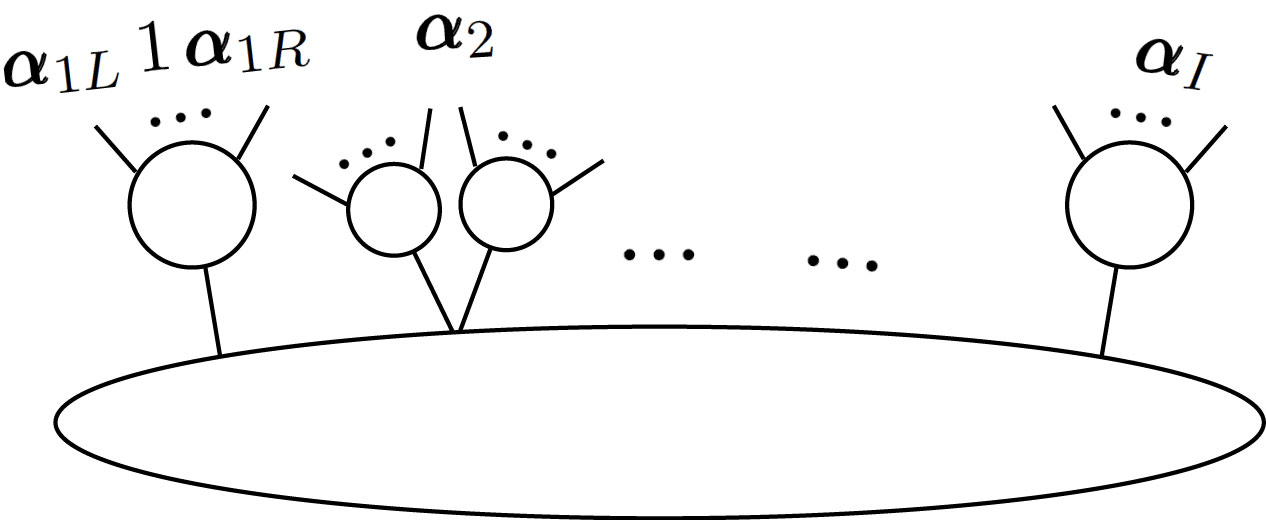} \end{minipage}\label{Eq:IDYMS}\\
&\equiv&\Sl_{\substack{\text{CycDiv}\,\pmb\rho\\ |\pmb\alpha_i|=|\pmb\rho_i|} } C(\pmb\alpha_{1L}1\pmb\alpha_{1R})\,C(\pmb\alpha_{2})\,...\,C(\pmb\alpha_{I})\,{1\over (\ell-k_{\pmb\alpha_{1L}})^2}\,\phi_{\pmb\alpha_{1L}1\pmb\alpha_{1R}|\pmb\rho_1}\nn
&&~~~~\times\,{1\over (\ell+k_1+k_{\pmb\alpha_{1R}})^2}\,\phi_{\pmb\alpha_{2}|\pmb\rho_2}\,...\,{1\over (\ell+k_1+k_{\pmb\alpha_{1R}}+k_{\pmb\alpha_2}+...+k_{\pmb\alpha_I})^2}\,\phi_{\pmb\alpha_{I}|\pmb\rho_I}.\nonumber
\eea
The above expression originates from the local expression of YMS integrand derived in \cite{Xie:2024pro}. In particular, we express the full YMS integrand according to the formula given in \cite{Xie:2024pro}, expand the subcurrents atthached to the loop in terms of BS ones and collect all contributions (i.e., all cyclic divisions) for a given ordering $\{1,\pmb\alpha\}$. For a given cyclic division $\{1,\pmb\alpha\}\to\pmb\alpha_1=\pmb\alpha_{1L}1\pmb\alpha_{1R},\pmb\alpha_2,...,\pmb\alpha_I$, the coefficient
\bea
C(\pmb\alpha_{1L}1\pmb\alpha_{1R})\,C(\pmb\alpha_{2})\,...\,C(\pmb\alpha_{I})
\eea
is just the product of kinematic factors coming from the corresponding subcurrents. If there is a four-point vertex structure for a given subset $\pmb\alpha_{i}$ ($|\pmb\alpha_{i}|\geq 2$, $\pmb\alpha_{i}$ only contains gluons) in the division, the coefficient $C(\pmb\alpha_{i})$ must be given by Lorentz contraction of two coefficients corresponding to the two subcurrents attached to this four-point vertex structure. More detail can be found in \appref{app:kinematicC}.

In the following subsections, we evaluate (\ref{Eq:Coefficient1}) and simplify it into a more compact form without loop-momentum denominators, focusing on integrands with up to three gluons.

\subsection{YMS integrand with one gluon}
\label{sec:one-gluon}
We first investigate the simplest case of the expansion formula \eqref{Eq:Exp1} with coefficients (\ref{Eq:Coefficient1}), i.e., the YMS integrand with only one gluon. Without loss of generality, we take the integrand with three scalars $1$, $2$, $3$ and one gluon $p$ as an example. According to (\ref{Eq:Coefficient1}), the coefficient reads
\bea
C(\ell;1,\pmb\alpha)=\frac{\mathcal{I}^{\text{YMS}}(\ell;1,\pmb\alpha\,|\,\pmb\rho)}{\mathcal{I}^{\,\text{BS}}(\ell;1,\pmb\alpha\,|\,\pmb\rho)}.\quad \pmb\alpha\in \{p\}\shuffle\{2,3\}\label{Eq:CoefficientEG1}
\eea
To show the explicit expressions of $\mathcal{I}^{\text{YMS}}$ and $\mathcal{I}^{\,\text{BS}}$, we take the permutation $\{1,\pmb{\alpha}\}=\{1,p,2,3\}\in \{1,\{p\}\shuffle\{2,3\}\}$ as an example. All possible cyclic partitions of this ordering are 
\bea
&&~~\{1,p,2,3\},~~~\,\{3,1,p,2\},~~~\,\{2,3,1,p\},~~~\underline{\{p,2,3,1\}},~~~\underline{\{1\}\,\{p,2,3\}},~~~\{1,p,2\}\,\{3\},\nn
&&~~\{2,3,1\}\,\{p\},~~\{3,1,p\}\,\{2\},~~\{1,p\}\,\{2,3\},~~\underline{\{3,1\}\,\{p,2\}},~~\{1\}\,\{p\}\,\{2,3\},~\{1,p\}\,\{2\}\,\{3\},\nn
&&~~\underline{\{1\}\,\{p,2\}\,\{3\}},~\{3,1\}\,\{p\}\,\{2\},\,\,\{1\}\,\{p\}\,\{2\}\,\{3\},
\eea
where the divisions with underlines are those  exist only in $\mathcal{I}^{\,\text{BS}}$ but not in $\mathcal{I}^{\,\text{YMS}}$ (note that in the definition of YMS  current with both scalars and gluons, the first element in the permutation must be scalar, see \appref{app:kinematicC}). Let us write down explicit expressions for specific divisions as illustrative examples, according to \appref{app:kinematicC}. For division  $\mathcal{D}\,\to\,\{1,p\}\,\{2\}\,\{3\}$, the $\mathcal{I}_{\,\mathcal{P}}^{\text{BS}}$ and $\mathcal{I}_{\,\mathcal{P}}^{\text{YMS}}$ are  respectively given by
\bea
\mathcal{I}_{\,\mathcal{D}}^{\text{BS}}(\ell;1,p,2,3\,|\,\pmb\rho)=\begin{minipage}{1.5cm}  \includegraphics[width=1.5cm]{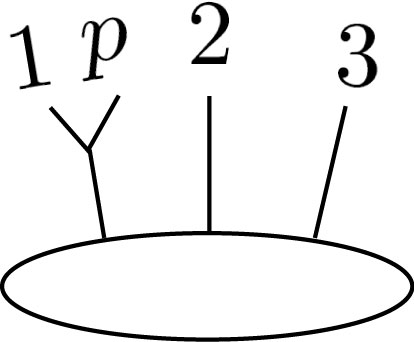} \end{minipage},~~~~\mathcal{I}_{\,\mathcal{D}}^{\text{YMS}}(\ell;1,p,2,3\,|\,\pmb\rho)=\epsilon_{p}\cdot k_1\times\begin{minipage}{1.5cm}\includegraphics[width=1.5cm]{EG1FD1} \end{minipage}.
\eea
For division $\mathcal{D}\,\to\,\{3,1\}\,\{p\}\,\{2\}$, the explicit expressions of $\mathcal{I}_{\,\mathcal{D}}^{\text{BS}}$ and $\mathcal{I}_{\,\mathcal{D}}^{\text{YMS}}$ are displayed as
\bea
\mathcal{I}_{\,\mathcal{D}}^{\text{BS}}(\ell;1,p,2,3\,|\,\pmb\rho)=\begin{minipage}{1.5cm}  \includegraphics[width=1.5cm]{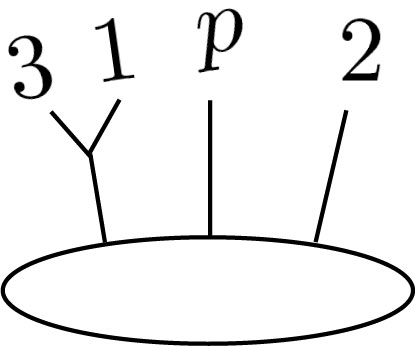} \end{minipage},~~\mathcal{I}_{\,\mathcal{D}}^{\text{YMS}}(\ell;1,p,2,3\,|\,\pmb\rho)=\epsilon_{p}\cdot (\ell+k_1)\times\begin{minipage}{1.5cm}\includegraphics[width=1.5cm]{EG1FD2} \end{minipage}.
\eea
Following a similar way, the $\mathcal{I}_{\,\mathcal{D}}^{\text{BS/YMS}}$ with other divisions can also be provided explicitly.

The coefficient (\ref{Eq:CoefficientEG1}) itself does not obey the consistency condition (\ref{Eq:ConstraintC}). In addition, there is loop momentum in the denominator of each coefficient (\ref{Eq:CoefficientEG1}). In the following, we try to reduce the expansion coefficient (\ref{Eq:CoefficientEG1}) into an appropriate form such that they obey the condition (\ref{Eq:CoefficientEG1}) and the loop momentum is only involved in the numerator. To achieve this, we classify the cyclic divisions into the following two types.
\begin{itemize}

\item [\bf(i)] Gluon $p$ is a single-particle subcurrent attached to the loop, e.g., the division $\mathcal{D}\,\to\,\{3,1\}\,\{p\}\,\{2\}$.

\item [\bf(ii)] Gluon $p$ is involved in a subcurrent with at least one scalar, e.g., the division $\mathcal{D}\,\to\,\{1,p\}\,\{2\}\,\{3\}$.

\end{itemize}
For a {\it type (i) division}, e.g.,  $\mathcal{D}\,\to\,\{3,1\}\,\{p\}\,\{2\}$, the kinematic coefficient is presented as
\bea
C(3,1)C(p)C(2)=\epsilon_p\cdot (\ell+k_1),
\eea 
where the fact that the coefficient associating to a {\it pure scalar subcurrent} is $1$ has been applied. This coefficient can be generalised to the pattern
\bea
\epsilon_p\cdot X_p(\ell;1,\pmb\alpha),\label{Eq:CoefficientEG12}
\eea
where $X_p^{\mu}(\ell;1,\pmb\alpha)$ is defined by the sum of loop momentum $l^{\mu}$ and the total momentum of elements between $\ell^{\mu}$ and the gluon $p$
\bea
X_p^{\mu}(\ell;\pmb\alpha)=\ell^{\mu}+\Sl_{i\prec p}k_i^{\mu},
\eea
in which, the notation $i\prec p$ means the position of $i$ is less than the position of $p$, when $1$ is considered as the first element in the permutation $\{1,\pmb\alpha\}$.
For instance, if the permutation is $\{1,\pmb{\alpha}\}=\{1,p,2,3\}$, the $X_p^{\mu}$ is $X_p^{\mu}=\ell^{\mu}+k_1^{\mu}$, while if the permutation is $\{1,\pmb\alpha\}=\{1,2,p,3\}$, we have $X_p^{\mu}=\ell^{\mu}+k_1^{\mu}+k_2^{\mu}$.

The {\it type (ii) divisions}, e.g., the division $\mathcal{D}\,\to\,\{1,p\}\,\{2\}\,\{3\}$, would contribute to  $\mathcal{I}^{\text{YMS}}(l;1,\pmb\alpha \,|\,\pmb\rho)$ with different orderings $\pmb\alpha$:
The kinematic coefficient from the sub-tree current is given by 
\bea
C(1,p)=\epsilon_p\cdot k_1.
\eea
%
%
With the $U(1)$-decoupling identity
\bea
\begin{minipage}{1.8cm}\includegraphics[width=1.8cm]{EG1FD1}\end{minipage}\,+\,\begin{minipage}{1.8cm}\includegraphics[width=1.8cm]{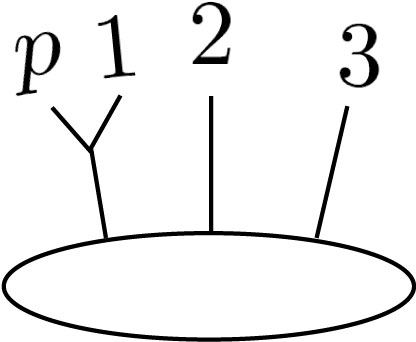}\end{minipage}=0,
\eea
the expression for $\mathcal{I}_{\{\{1,p\},\{2\},\{3\}\}}^{\text{YMS}}(\ell;1,p,2,3\,|\,\pmb\rho)$ can be equivalently written as 
\bea
(\epsilon_p\cdot k_1)\,\times\,\begin{minipage}{1.8cm}\includegraphics[width=1.8cm]{EG1FD1}\end{minipage}\cong\epsilon_p\cdot (\ell+k_1)\,\times\,\begin{minipage}{1.8cm}\includegraphics[width=1.8cm]{EG1FD1}\end{minipage}\,+\epsilon_p\cdot \ell\,\times\,\begin{minipage}{1.8cm}\includegraphics[width=1.8cm]{EG1FD1b}\end{minipage}\label{eq:1g-id2}
\eea
 The first term can combine with $\epsilon_p \cdot (\ell+k_1) \times \mathcal{I}_{\{\{1\},\{p\},\{2\},\{3\}\}}^{\text{BS}}(\ell;1,p,2,3\,|\,\pmb\rho)$ and other partitions in the ordering $\{1,p,2,3\}$ to get 
$\mathcal{I}^{\text{YMS}}(\ell;1,p,2,3\,|\,\pmb\rho)=\epsilon_p \cdot (\ell+k_1) \times \mathcal{I}^{\text{BS}}(\ell;1,p,2,3\,|\,\pmb\rho)$ and the second term in (\ref{eq:1g-id2}) will combine with other terms in the ordering $\{1,2,3,p\}$ to get 
$\mathcal{I}^{\text{YMS}}(\ell;1,2,3,p\,|\,\pmb\rho)=\epsilon_p \cdot (\ell+k_1+k_2+k_3) \times \mathcal{I}^{\text{BS}}(\ell;1,2,3,p\,|\,\pmb\rho)$.

%
%
With this manipulation, the contribution from partitions of type (ii) also agrees with the general pattern (\ref{Eq:CoefficientEG12}). When the above discussions are applied to all permutations $\pmb\alpha$ and all possible cyclic divisions, the $\mathcal{I}^{\text{YMS}}(\ell;1,\pmb\alpha),\ \forall\pmb\alpha\in\{p\}\shuffle\{2,3\}$ turns into  a multiple of $\mathcal{I}^{\text{BS}}(\ell;1,\pmb\alpha)$, with coefficient $\epsilon_p\cdot X_p(\ell;1,\pmb\alpha)$. Therefore, we can redefine the coefficient (\ref{Eq:CoefficientEG1}) as
\bea
C(\ell;1,\pmb\alpha)\cong\epsilon_p\cdot X_p(\ell;1,\pmb\alpha)\frac{\mathcal{I}^{\text{BS}}(\ell;1,\pmb\alpha)}{\mathcal{I}^{\,\text{BS}}(\ell;1,\pmb\alpha)}=\epsilon_p\cdot X(\ell;1,\pmb\alpha).\quad \forall\pmb\alpha\in\{p\}\shuffle\{2,3\}
\eea

The above example is straightforwardly generalised to YMS integrands with one gluon and an arbitrary number of scalars:
\bea
\boxed{C(\ell;1,\pmb\alpha)\cong\epsilon_p\cdot X_p(\ell;1,\pmb\alpha)}\,.\label{Eq:CoefficientEG1a}
\eea
{\it Apparently, this coefficient satisfies the consistency condition (\ref{Eq:ConstraintC}) and it is a polynomial function of the loop momentum. Another important feature of (\ref{Eq:CoefficientEG1a})  is that $C(l;1,\pmb\alpha)$ is independent of the choice of the right permutation $\pmb\rho$.}

\subsection{YMS integrand with two gluons}
\label{sec:two-gluons}
For integrands with two gluons, we use the example with two scalars $1$, $2$ and two gluons $p$, $q$ to demonstrate the general construction of expansion coefficients.  The  coefficient (\ref{Eq:Coefficient1}) is now explicitly given by
\bea
C(\ell;1,\pmb\alpha)=\frac{\mathcal{I}^{\text{YMS}}(\ell;1,\pmb\alpha\,|\,\pmb\rho)}{\mathcal{I}^{\,\text{BS}}\,(\ell;1,\pmb\alpha\,|\,\pmb\rho)},\quad\pmb\alpha\in \{p\}\shuffle\{q\}\shuffle\{2\}. \label{Eq:CoefficientEG2}
\eea
The concrete expression of (\ref{Eq:CoefficientEG2}) for a given $\pmb\alpha$ can be obtained by expressing the YMS and BS integrands by summations over cyclic divisions, as we have done in the case of only one gluon.
Particularly, the denominator $\mathcal{I}^{\,\text{BS}}\,(\ell;1,\pmb\alpha\,|\,\pmb\rho)$ in (\ref{Eq:CoefficientEG2}) for a given $\pmb\alpha\in\{p\}\shuffle\{q\}\shuffle\{2\}$, is just given by summing all possible BS Feynman graphs corresponding to the cyclic divisions for $\{1,\pmb\alpha\}$. For example, all cyclic divisions contributing to $\mathcal{I}^{\,\text{BS}}\,(l;1,\pmb\alpha\,|\,\pmb\rho)$ are following 
\bea
&&~\{1,p,q,2\},~~~~\,\{2,1,p,q\},~~~~\,\{q,2,1,p\},~~~~\,\{p,q,2,1\},~~~~\,\{1,p,q\}\,\{2\},~~~~\,\{2,1,p\}\,\{q\},\nn
&&\,\{q,2,1\}\,\{p\},~~\,\{1,p\}\,\{q,2\},~~\{2,1\}\,\{p,q\},~~\,\{1,p\}\,\{q\}\,\{2\},~\{2,1\}\,\{p\}\,\{q\},~~\{1\}\,\{p,q\}\,\{2\},\nn
&&\{1\}\,\{p\}\,\{q,2\},~\{1\}\,\{p\},\{q\},\{2\}.\label{Eq:DivisionsBSEG2}
\eea
Thus the denominator is given by summing all possible BS diagrams (with cubic vertices) corresponding to these cyclic divisions
\bea
\mathcal{I}^{\text{BS}}(\ell;1,p,q,2\,|\,\pmb\rho)&=&\begin{minipage}{1.5cm}\includegraphics[width=1.5cm]{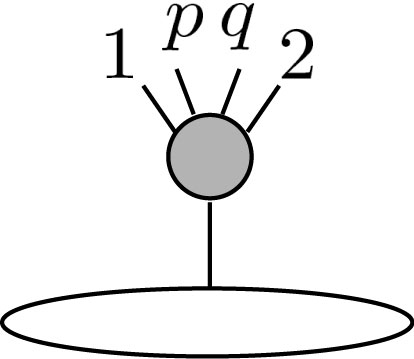}\end{minipage}+\begin{minipage}{1.5cm}\includegraphics[width=1.5cm]{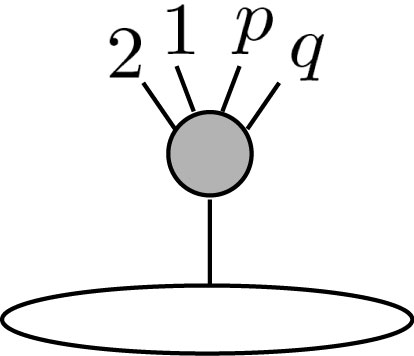}\end{minipage}+\begin{minipage}{1.5cm}\includegraphics[width=1.5cm]{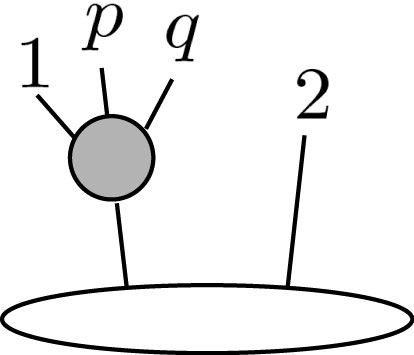}\end{minipage}+\begin{minipage}{1.5cm}\includegraphics[width=1.5cm]{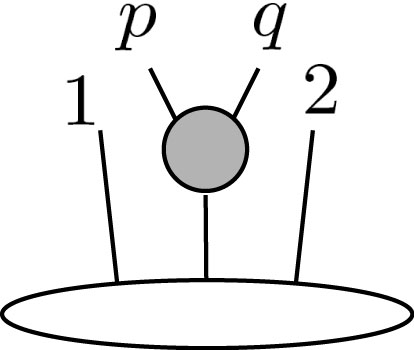}\end{minipage}+\begin{minipage}{1.5cm}\includegraphics[width=1.5cm]{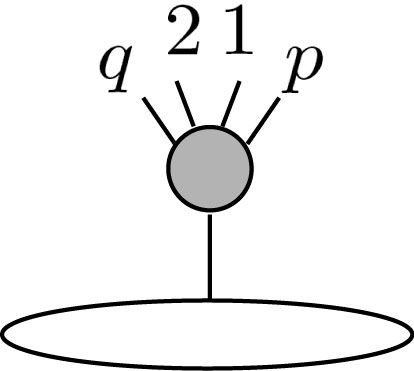}\end{minipage}+\begin{minipage}{1.5cm}\includegraphics[width=1.5cm]{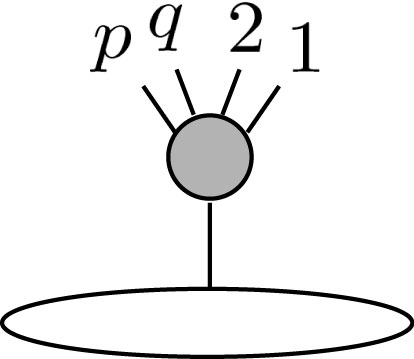}\end{minipage}+\,...,\nn
\eea
 The numerator $\mathcal{I}^{\text{YMS}}(\ell;1,\pmb\alpha\,|\,\pmb\rho)$ can also be provided by summing over all cyclic divisions with correct kinematic coefficients according to \appref{app:kinematicC}.

In the following, we classify the coefficients $C(\ell;1,\pmb\alpha)$ according to different powers of $\epsilon_p\cdot\epsilon_q$ in the numerator, and then reduce each coefficient into a proper form that satisfies the condition (\ref{Eq:ConstraintC}) and does not involve loop momentum in the denominator.

\subsubsection{The part with $(\epsilon_p\cdot\epsilon_q)^0$}

Now let us investigate the terms in $C(\ell;1,\pmb\alpha)$  without powers of  $(\epsilon_p\cdot\epsilon_q)$. According to (\ref{Eq:IDYMS}), the numerator of (\ref{Eq:CoefficientEG2})  , i.e., $\mathcal{I}^{\text{YMS}}$ in this part, for a given permutation $\{1,\pmb\alpha\}\in\{1,\{p\}\shuffle\{q\}\shuffle\{2\}\}$ is obtained by summing over all Feynman  graphs (or equivalently all cyclic divisions) that are constructed by attaching subcurrents (with proper kinematic coefficients) to the scalar loop. Each diagram is characterized by a cyclic division of this permutation. For example, for the doubly colour-ordered integrand $\mathcal{I}^{\,\text{YMS}}\,(\ell;1,\pmb\alpha\,|\,\pmb\rho)$,  if $\pmb{\alpha}=\{p,q,2\}$, the cyclic divisions $\mathcal{D}$ of $\{1,\pmb\alpha\}$ with terms containing $(\epsilon_p\cdot\epsilon_q)^0$ factor are
\bea
&&~\{1,p,q,2\},~~\,\{2,1,p,q\},~~\,\{1,p,q\}\,\{2\},~~\,\{2,1,p\}\,\{q\},\,~~\{2,1\}\,\{p,q\},~~\,\{1,p\}\,\{q\}\,\{2\},\nn
&&~\{2,1\}\,\{p\}\,\{q\},~\{1\}\,\{p,q\}\,\{2\},~\{1\}\,\{p\},\{q\},\{2\}.\label{Eq:DivisionsYMSPart1EG2}
\eea
Comparing to (\ref{Eq:DivisionsBSEG2}), we have just deleted those divisions with parts satisfying the following two conditions: (i) it contains both gluons and scalars,  (ii) the first element is a gluon. 

In general, subcurrents containing gluons can be classified into the following two types:
\begin{itemize}
 \item {\bf (i)} The subcurrent consists  of only  gluons e.g., the subcurrent $\{p,q\}$ in the division $\{1\}\,\{p,q\}\,\{2\}$, and  the single particle subcurrent  $\{p\}$ in the division $\{1,p\},\{q\},\{2\}$.
 
 \item {\bf (ii)} The subcurrent containing both gluon and scalar , e.g., the subcurrent  $\{2,1,p\}$ in the division $\{2,1,p\}\,\{q\}$.
\end{itemize} 
The contribution of division with these two types of subcurrents is similar to those introduced in the case with only one gluon. After adding proper terms vanishing due to the generalised $U(1)$-decoupling identity
\bea
\Sl_{\shuffle}\begin{minipage}{1.8cm}\includegraphics[width=1.8cm]{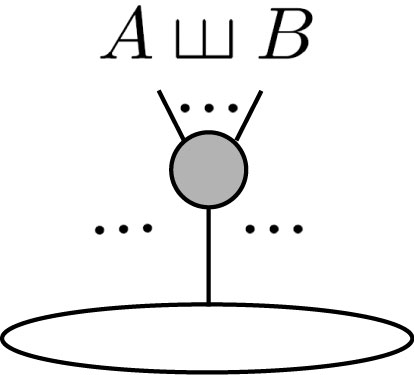}\end{minipage}=0,\label{Eq:GenU1}
\eea
where $A$, $B$ are two ordered sets, we can arrange them into contributions to multiple terms  in the doubly color-ordered {\it full YMS integrand} with a compact formula of  $C(l;1,\pmb\alpha)$. In particular, these vanishing terms used for the current example are 
\bea
&&~~(\epsilon_p\cdot \ell)(\epsilon_q\cdot \ell)\times\Sl_{\shuffle}\begin{minipage}{2cm}\includegraphics[width=2cm]{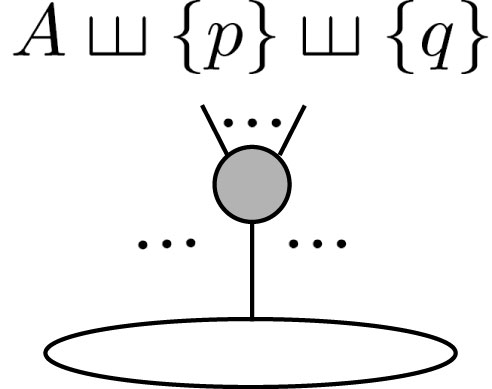}\end{minipage},~~(\epsilon_p\cdot \ell)\times\Sl_{\shuffle}\begin{minipage}{2cm}\includegraphics[width=2cm]{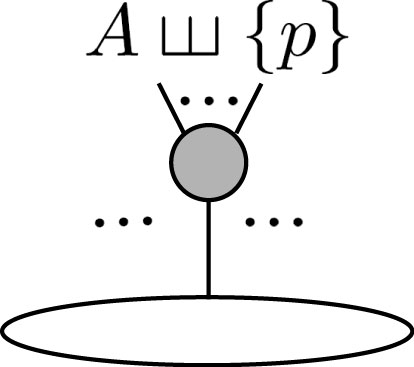}\end{minipage},~~(\epsilon_q\cdot \ell)\times\Sl_{\shuffle}\begin{minipage}{2cm}\includegraphics[width=2cm]{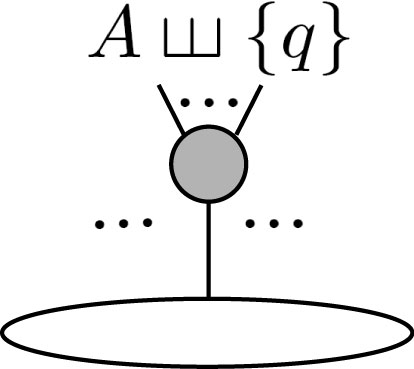}\end{minipage},\nn
&&~~(\epsilon_p\cdot \ell)\times\Sl_{\shuffle}\begin{minipage}{2cm}\includegraphics[width=2cm]{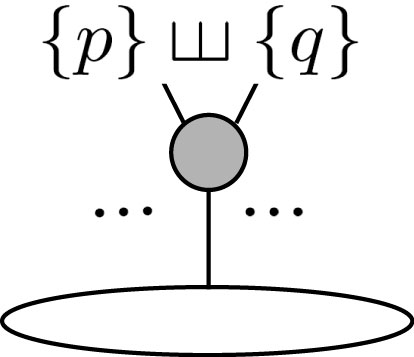}\end{minipage},~~(\epsilon_q\cdot \ell)\times\Sl_{\shuffle}\begin{minipage}{2cm}\includegraphics[width=2cm]{EG2U1d}\end{minipage}.
\eea
After this step, the new defined integrand will introduce the following coefficients associating with both  $p$ and $q$
\bea
\epsilon_{p,q}\cdot X_{p,q}(1,\pmb\alpha),
\eea
regardless of the type of the subcurrent containing $p$ and/or $q$. In addition, the possible cyclic divisions of elements in the new defined YMS integrand now agree with those for BS ones. For example, if $\{1,\pmb\alpha\}=\{1,p,q,2\}$, now the cyclic divisions are given by (\ref{Eq:DivisionsBSEG2}) instead of (\ref{Eq:DivisionsYMSPart1EG2}). This implies that we have rearranged the  YM integrand properly so that each numerator in \eqref{Eq:CoefficientEG2} is a multiple of the corresponding denominator. As a consequence, the kinematic coefficients in {\it the $(\epsilon_p\cdot\epsilon_q)^0$ sector} turn into  
\bea
C^{(0)}(\ell;1,\pmb\alpha)\cong\epsilon_p\cdot X_p(\ell;1,\pmb\alpha)\,\epsilon_q\cdot X_q(\ell;1,\pmb\alpha)\,.\label{Eq:CoefficientEG2a}
\eea
The above discussions can be generalised to the $(\epsilon\cdot\epsilon)^0$ sector of YMS integrand with an arbitrary number of gluons
\bea
C(\ell;1,\pmb\alpha)\cong\prod\limits_{i\in \mathsf{G} }\epsilon_i\cdot X_i(\ell;1,\pmb\alpha),
\eea
which has already been observed in \cite{Xie:2024pro,Dong:2023stt}.

\subsubsection{Terms with $(\epsilon_p\cdot\epsilon_q)^1$}

Next we move on to investigate the  $(\epsilon_p\cdot\epsilon_q)^1$ terms of the coefficient (\ref{Eq:CoefficientEG2}), by collecting all partitions of  permutation $\{1,\pmb\alpha\}$ that could lead to terms with a $(\epsilon_p\cdot\epsilon_q)^1$ factor. {\it In this case, only the divisions where $p$, $q$ live in the same part would contain terms with the factor $(\epsilon_p\cdot\epsilon_q)^1$  }. We take $\pmb\alpha=\{p,q,2\}$ as an example. All such cyclic divisions of $\{1,\pmb\alpha\}=\{1,p,q,2\}$ that contribute to the YMS integrand in the denominator of (\ref{Eq:CoefficientEG2}) are the following
\bea
\{1,p,q,2\},\,\{2,1,p,q\},\,\{1,p,q\}\,\{2\},\,\{1\}\,\{p,q\}\,\{2\},\,\{1\}\,\{p\text{-}q\}\,\{2\},\,\{2,1\}\,\{p,q\},\,\{2,1\}\,\{p\text{-}q\},\nn
\eea
where the notation $\{p\text{-}q\}$ is introduced to stand for a four-point vertex structure see \appref{app:kinematicC}. The full expression of $\mathcal{I}^{\text{YMS}}(\ell;1,p,q,2\,|\,\pmb\rho)$ is then given by the sum of all terms with respect to the above partitions
\bea
\mathcal{I}^{\text{YMS}}(\ell;1,p,q,2\,|\,\pmb\rho)&=&(\epsilon_p\cdot\epsilon_q)\Biggl\{\,(-k_p\cdot k_1)\times\begin{minipage}{1.5cm}\includegraphics[width=1.5cm]{EG2FD1}\end{minipage} +[-k_p\cdot (k_1+k_2)]\times\begin{minipage}{1.5cm}\includegraphics[width=1.5cm]{EG2FD2}\end{minipage}\nn
&&~~~~~~~~+(-k_p\cdot k_1)\times\begin{minipage}{1.45cm}\includegraphics[width=1.45cm]{EG2FD3}\end{minipage}+[-k_p\cdot (k_1+\ell)]\times\begin{minipage}{1.45cm}\includegraphics[width=1.45cm]{EG2FD4}\end{minipage}\nn
&&~~~~~~~~+\left(-{1\over 2}\right)\times\begin{minipage}{1.5cm}\includegraphics[width=1.5cm]{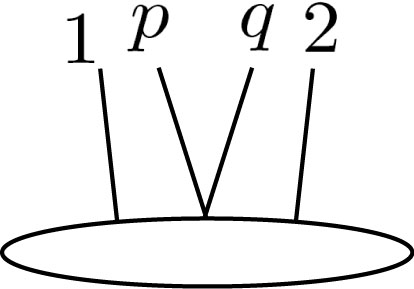}\end{minipage}+[-k_p\cdot (k_1+\ell)]\times\begin{minipage}{1.45cm}\includegraphics[width=1.45cm]{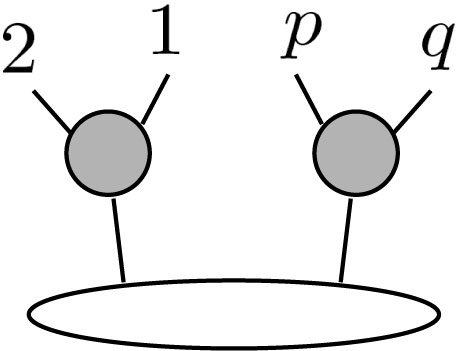}\end{minipage}\nn
&&~~~~~~~~+~\,\left(-{1\over 2}\right)\times\begin{minipage}{1.5cm}\includegraphics[width=1.5cm]{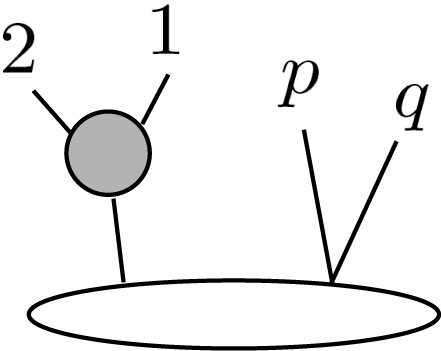}\end{minipage}
\,\Biggr\}.
\eea
Thus the coefficient $C(\ell;1,p,q,2)$ is obtained by substituting this expression into (\ref{Eq:CoefficientEG2}).
 Coefficients for other permutations can be obtained analogously.

Now we transform the coefficients (\ref{Eq:CoefficientEG2}) into equivalent ones which do not contain loop momentum in the denominator. The crucial point of this is to add terms which integrate to zero to the YMS integrand $\mathcal{I}^{\text{YMS}}$, so that
{\bf(i)} Each new $\mathcal{I}^{\text{YMS}}(\ell;1,\pmb\alpha\,|\,\pmb\rho)$ is proportional to $\mathcal{I}^{\text{BS}}(\ell;1,\pmb\alpha\,|\,\pmb\rho)$, and the proportionality factor (in other words the new defined kinematic coefficient) is a rational function of Lorentz contractions of external polarizations, momenta and the loop momentum.
{\bf(ii)} The new coefficient does not involve loop momentum $\ell^{\mu}$ in the denominators.
We now derive such coefficients by the following steps.

 \noindent\,{\bf\text{Step-1}} Reversing the localization process in (\cite{Xie:2024pro}) to transform the local YMS integrand
\bea
\mathcal{I}^{\,\text{YMS}}(\ell;1,2||\{p,q\}\,|\,1,\pmb\rho)
  \eea
as a proper combination of BS Feynman graphs with {\it nonlocal coefficients (i.e. Lorentz contraction separated by loop propagators)}. There are three key techniques for this step:
\begin{itemize}
 \item {\it First}, the generalized $U(1)$-decoupling identity of a BS subcurrent (\ref{Eq:GenU1}), helps one to introduce a factor $(-k_p\cdot \ell_i)$ into a subcurrent by adding the following zeros to the full integrand: 
\bea
&&(\epsilon_p\cdot\epsilon_q)(-k_p\cdot \ell_i)\times\Sl_{\shuffle}\begin{minipage}{1.8cm}\includegraphics[width=1.8cm]{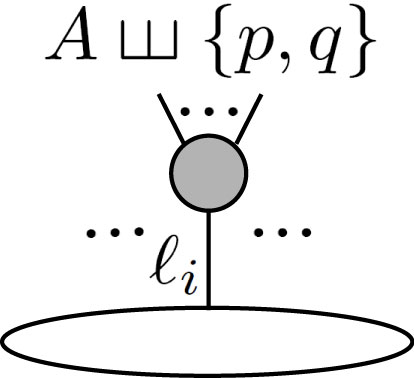}\end{minipage},~~~~~~~~(\epsilon_p\cdot\epsilon_q)(-k_p\cdot \ell_i)\times\Sl_{\shuffle}\begin{minipage}{2.3cm}\includegraphics[width=2.3cm]{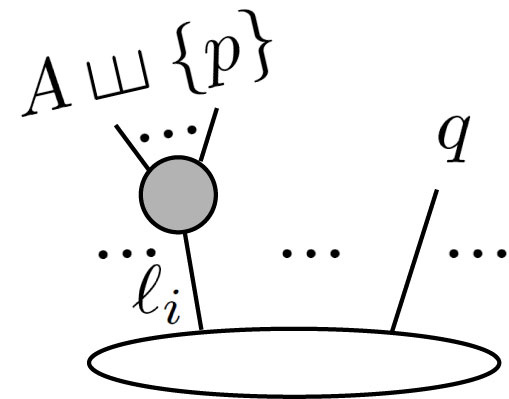}\end{minipage}\,,\nn
&&
(\epsilon_p\cdot\epsilon_q)(-k_p\cdot \ell_i)\times\Sl_{\shuffle}\begin{minipage}{2.4cm}\includegraphics[width=2.4cm]{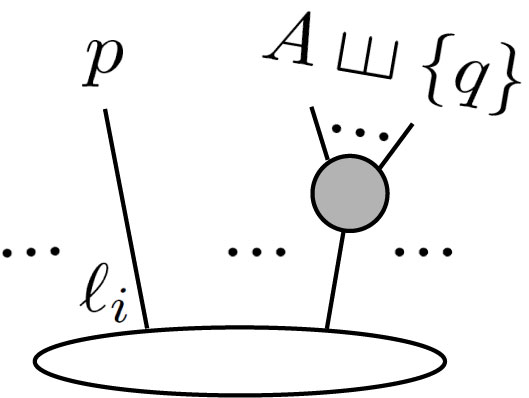}\end{minipage},\nn
&&
\Sl_{\shuffle_1}(\epsilon_p\cdot\epsilon_q)(-k_p\cdot (\ell_i+X_p(\ell;\shuffle_1)))\times\Sl_{\shuffle_2}\begin{minipage}{2.6cm}\includegraphics[width=2.6cm]{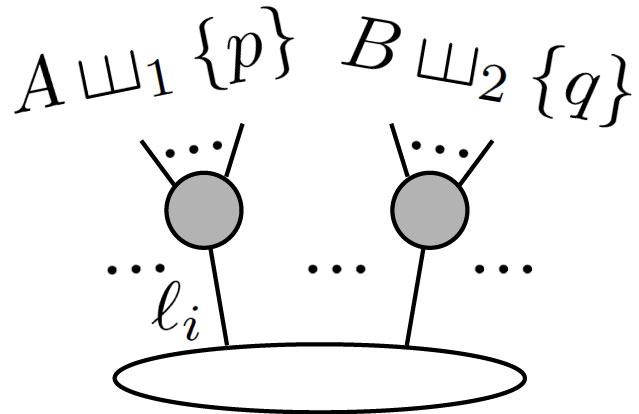}\end{minipage},
\eea
where $A$, $B$ are ordered scalar subsets, $\ell_i$ denotes the momentum of the loop propagator that is connected to the corresponding subcurrent from left (as shown in the Feynman graphs).

\item {\it Second,} if there are  subcurrents $\phi_{A|\W A}$, $\phi_{p|p}$ and $\phi_{B|\W B}$ attached to the loop in turn, where $\{x_1,A\}$, $\{y_1,B\}$ refer to ordered scalar subsets, the elements of $\{x_1,A\}$ ($\{y_1,B\}$) are identical to those of $\W A$ ($\W B$) so that these subcurrent do not vanish, we have the following identity related to such structure: 
\bea
0&=&\Sl_{\shuffle}(-k_p\cdot X_p(\ell;\shuffle))\times\begin{minipage}{2cm}\includegraphics[width=2cm]{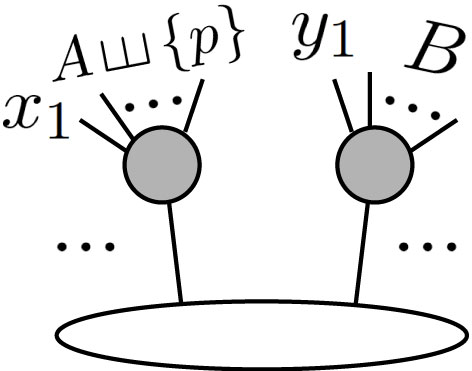}\end{minipage}+(-k_p\cdot \ell_i)\times\begin{minipage}{1.9cm}\includegraphics[width=1.9cm]{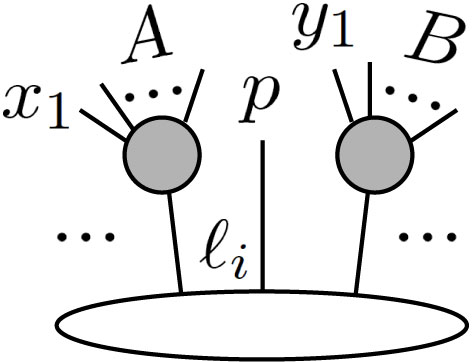}\end{minipage}\nn
&&~~~~~~~~~~~~~~~~~~~~~~~~~~~~~~~~~~~~~~~~~+\Sl_{\shuffle}(-k_p\cdot X_p(\ell;\shuffle))\begin{minipage}{2.2cm}\includegraphics[width=2.3cm]{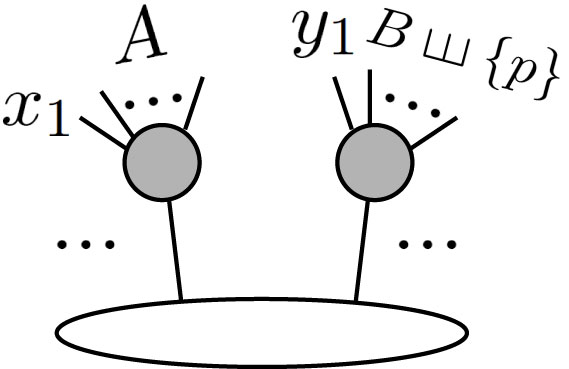}\end{minipage},\nn
\label{Eq:EG2Cancel}
\eea
where the right permutation of the subcurrent in the first and the last terms are respectively $\{x_1,A,p\}$ and $\{p,y_1,B\}$.

\item {\it Third,} the following relation, by which a four-point vertex structure is transformed into BS Feynman graphs (with only three-point vertices), as 
\bea
\begin{minipage}{2.2cm}\includegraphics[width=2.2cm]{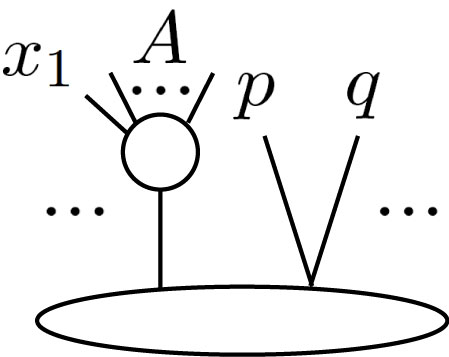}\end{minipage}=\Sl_{\shuffle}(-k_p\cdot X_p(\ell,\shuffle))\times\begin{minipage}{2.3cm}\includegraphics[width=2.3cm]{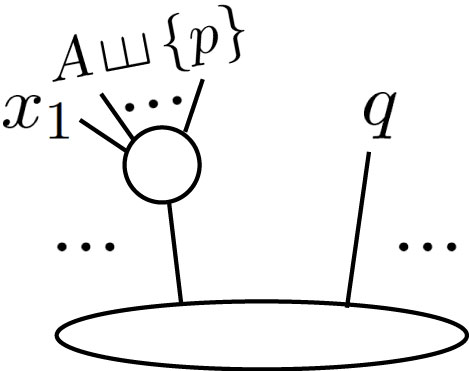}\end{minipage}+(-k_p\cdot \ell_i)\times\begin{minipage}{2.2cm}\includegraphics[width=2.2cm]{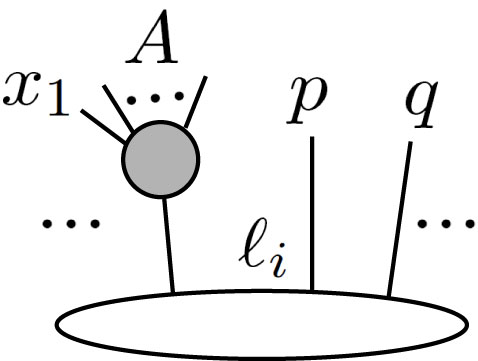}\end{minipage}\nn
\label{Eq:4PtTrans}
\eea
where $A$ is an ordered set of scalars, $X_p(\ell;\shuffle)$ in the first term denotes the total momentum of the particles to the left of $p$ in the permutation $\{x_1,A\shuffle\{p\}\}$, $\ell_i$ in the second term denotes the momentum of the propagator attached to $p$ from left. 
\end{itemize}
After this step, the full integrand $\mathcal{I}^{\text{YMS}}(\ell;1,2||\{p,q\}\,|\,\pmb\rho)$ is transformed into a proper combination of BS Feynman graphs
\bea
\mathcal{I}^{\text{YMS}}(\ell;1,2||\{p,q\}\,|\,\pmb\rho)&\cong& (\epsilon_p\cdot\epsilon_q)\Biggl[\,\Sl_{\pmb\alpha}(-k_p\cdot X_p(\ell;1,\pmb\alpha))\,\mathcal{I}_1^{\text{BS}}(\ell;1,\pmb\alpha\in\{2\}\shuffle\{p,q\}\,|\,\pmb\rho)\nn
&&+\Sl_{\pmb\beta\,\text{s.t}\,\beta_1\neq 1}(-k_p\cdot X_p(\ell';\pmb\beta))\,\mathcal{I}_2^{\text{BS}}(\ell';\pmb\beta\in\text{cyc}\,\{1,2\}\shuffle\{p,q\}\,|\,\pmb\rho)\,\Biggr].\nn\label{Eq:CoefficientEG2Sector21}
\eea
In the above expression, $\mathcal{I}_1^{\text{YMS}}(\ell;1,\pmb\alpha\,|\,\pmb\rho)$ and $\mathcal{I}_2^{\text{YMS}}(\ell';\pmb\beta\,|\,\pmb\rho)$ respectively refer to the sum of all possible diagrams (in other words all cyclic divisions) of permutations $\{1,\pmb\alpha\}$ and $\pmb\beta$, where the first element in the subcurent containing the scalar $1$ is $1$ (for $\mathcal{I}_1^{\text{YMS}}$) and $\beta_1(\neq1)$ (for $\mathcal{I}_2^{\text{YMS}}$). In  $\mathcal{I}_1^{\text{YMS}}$, the momentum of the loop propagator, which is attached to the sucurrent containing $1$, is $\ell$, while in  $\mathcal{I}_2^{\text{YMS}}$, this loop momentum is $\ell'=\ell-k_{\pmb\beta_L}$ ($\pmb\beta_L$ denotes the elements between $\ell'$ and  $1$ in the permutation $\pmb\beta\equiv\{\pmb\beta_L,1,\pmb\beta_R\}$). The $X_p$ always denotes the sum of $\ell$ (or $\ell'$) and the total external momenta between  $\ell$ (or $\ell'$), $p$. To clarify terms in (\ref{Eq:CoefficientEG2Sector21}), we write down the coefficients associating with the following diagrams (which come from $\mathcal{I}_1^{\text{YMS}}(\ell;1,p,q,2\,|\,\pmb\rho)$ and $\mathcal{I}_2^{\text{YMS}}(\ell';2,p,1,q\,|\,\pmb\rho)$)
\bea
&&\mathcal{I}_1^{\text{YMS}}(\ell;1,p,q,2\,|\,\pmb\rho):~~\begin{minipage}{1.45cm}\includegraphics[width=1.45cm]{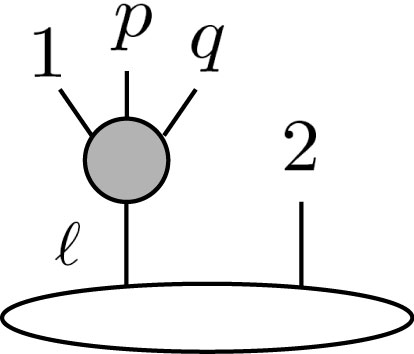}\end{minipage}\to (\epsilon_p\cdot\epsilon_q)[-k_p\cdot (\ell+k_1)],\nn
&&\mathcal{I}_2^{\text{YMS}}(\ell';2,p,1,q\,|\,\pmb\rho):~~\begin{minipage}{1.45cm}\includegraphics[width=1.45cm]{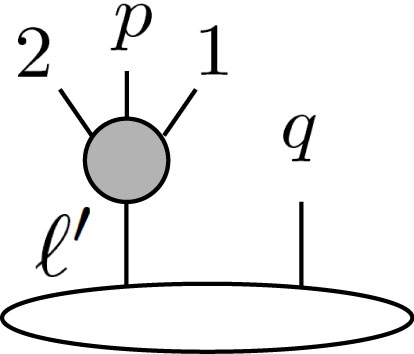}\end{minipage}\to (\epsilon_p\cdot\epsilon_q)[-k_p\cdot (\ell'+k_2)]=(\epsilon_p\cdot\epsilon_q)[-k_p\cdot (\ell-k_p)].\nn
&&\mathcal{I}_2^{\text{YMS}}(\ell';p,2,1,q\,|\,\pmb\rho):~~\begin{minipage}{1.45cm}\includegraphics[width=1.45cm]{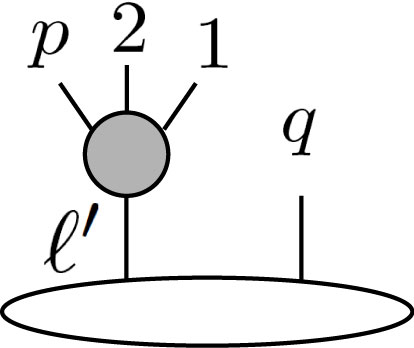}\end{minipage}\to (\epsilon_p\cdot\epsilon_q)(-k_p\cdot \ell')=(\epsilon_p\cdot\epsilon_q)[-k_p\cdot (\ell-k_p-k_2)].\nn
\eea

{\it An important fact we emphasise here is the expression (\ref{Eq:CoefficientEG2Sector21}) now is nonlocal }(i.e. there are contractions between polarizations and momenta living in distinct subcurrents that are separated by loop propagators) and does not satisfy the consistency condition (\ref{Eq:ConstraintC}). To 
remedy the consistency and to construct expressions proportional to the $\mathcal{I}_1^{\text{BS}}$ in the denominator of (\ref{Eq:CoefficientEG2}), we need a further step.

\noindent\,{\bf \text{Step-2}} Adding to the full integrand (\ref{Eq:CoefficientEG2Sector21}) of YMS the following {\it vanishing term}:
\bea
0&=&{\epsilon_p\cdot\epsilon_q\over k_p\cdot k_q}\Biggl[\Sl_{\pmb\alpha}(k_q\cdot Y_q(\ell;1,\pmb\alpha))(-k_p\cdot X_p(\ell;1,\pmb\alpha))\,\mathcal{I}_1^{\text{BS}}(\ell;1,\pmb\alpha\in\{2\}\shuffle\{p\}\shuffle\{q\}\,|\,\pmb\rho)\nn
&&+\Sl_{\pmb\beta}(k_q\cdot Y_q(\ell';\pmb\beta))(-k_p\cdot X_p(\ell';\pmb\beta))\mathcal{I}_2^{\text{BS}}(\ell';\pmb\beta\in\text{cyc}\{2,1\}\shuffle\{p\}\shuffle\{q\}\,|\,\pmb\rho)\Biggr],\label{Eq:CoefficientEG2Sector22}
\eea
where $Y_q(\pmb\alpha)$ ($Y_q(\pmb\beta)$) denotes the sum of $\ell$ (${\ell'}$) and the total momenta of {\it scalars} (i.e. the gluons $p$, $q$ are excluded) between $\ell$ (${\ell'}$) and $q$. In fact, each term of the identity (\ref{Eq:CoefficientEG2Sector22}) has to vanish. This is a consequence of the fundamental BCJ relation \cite{Boels:2011tp,Du:2012mt} for gluon $q$ when the position of $p$ is fixed. Such BCJ relation can be verified via expressing the $\mathcal{I}_1^{\text{BS}}$ or $\mathcal{I}_2^{\text{BS}}$ in terms of Feynman graphs and then grouping graphs with the substructures in (\ref{Eq:EG2Cancel}), or with generalized $U(1)$-decoupling identity (\ref{Eq:GenU1}) together.

Now we sum  (\ref{Eq:CoefficientEG2Sector21}) and (\ref{Eq:CoefficientEG2Sector22}) together, we get a newly defined $\mathcal{I}^{\text{YMS}}$:
\bea
\mathcal{I}^{\text{YMS}}&\cong&{\epsilon_p\cdot\epsilon_q\over k_p\cdot k_q}\Biggl[\Sl_{\pmb\alpha}(k_q\cdot X_q(\ell;1,\pmb\alpha))(-k_p\cdot X_p(\ell;1,\pmb\alpha))\,\mathcal{I}_1^{\text{BS}}(\ell;1,\pmb\alpha\in\{2\}\shuffle\{p\}\shuffle\{q\}\,|\,\pmb\rho)\nn
&&+\Sl_{\pmb\beta}(k_q\cdot X_q(\ell';\pmb\beta))(-k_p\cdot X_p(\ell';\pmb\beta))\mathcal{I}_2^{\text{BS}}(\ell';\pmb\beta\in\{2,1\}\shuffle\{p\}\shuffle\{q\}\,|\,\pmb\rho)\Biggr],\label{Eq:CoefficientEG2Sector23}
\eea
noting that in a permutation if $p\prec q$, $X^{\mu}_{q}=Y^{\mu}_q+ k^{\mu}_p$, if $q\prec p$, $X^{\mu}_{q}=Y^{\mu}_q$.

A further observation is that the $X_p(\ell';\pmb\beta)$ ($X_q(\ell';\pmb\beta)$) of the second term in fact is equivalent to those defined by the sum of $\ell$ (instead of ${\ell'}$) and the total external momentum between $\ell$ and $p$ ($q$). For example, for $\mathcal{I}_2^{\text{BS}}(\ell';p,2,1,q|\pmb\rho)$, we have
\bea
X^{\mu}_p&=&{\ell'}^{\mu}=\ell^{\mu}-k^{\mu}_p-k^{\mu}_2=\ell^{\mu}+k_1^{\mu}+k_q^{\mu},\nn
X^{\mu}_q&=&{\ell'}^{\mu}+k_p^{\mu}+k_2^{\mu}+k_1^{\mu}=(\ell^{\mu}-k^{\mu}_p-k^{\mu}_2)+k^{\mu}_p+k^{\mu}_2+k_1=\ell^{\mu}+k_1^{\mu}.
\eea
Therefore, the $\mathcal{I}^{\text{YMS}}$ in (\ref{Eq:CoefficientEG2Sector23}) is finally converted into 
\bea
\mathcal{I}^{\text{YMS}}&\cong&\Sl_{\pmb\alpha}\left(-{\epsilon_p\cdot\epsilon_q\over k_p\cdot k_q}\right)\,\left(k_p\cdot X_p(\ell,1,\pmb\alpha)\right)\left(k_q\cdot X_q(\ell,1,\pmb\alpha)\right)\mathcal{I}^{\,\text{BS}}\,(\ell;1,\pmb\alpha\in \{2\}\shuffle\{p\}\shuffle\{q\}\,|\,\pmb\rho),\nn
\eea
where $\mathcal{I}^{\,\text{BS}}\,(\ell;1,\pmb\alpha\in \{2\}\shuffle\{p\}\shuffle\{q\}|\pmb\rho)$ is the full integrand when the scalar $1$ is fixed as the first element. It is now clear that the $(\epsilon_p\cdot\epsilon_q)^1$ part of coefficient can be defined by 
\bea
C^{(1)}(\ell;1,\pmb\alpha)=\left(-{\epsilon_p\cdot\epsilon_q\over k_p\cdot k_q}\right)\,\left(k_p\cdot X_p(\ell; 1,\pmb\alpha)\right)\left(k_q\cdot X_q(\ell;1,\pmb\alpha)\right),
\eea
which naturally satisfies the consistency condition (\ref{Eq:ConstraintC}) and is apparently independent of the choice of the right permutation $\pmb\rho$.

\subsubsection{The final result of the kinematic coefficents}
Now we generalize the calculations of the above example to integrand with two gluons $p$, $q$ and an arbitrary number of scalars $1,...,r$. The YMS integrand $\mathcal{I}^{\,\text{YMS}}(\ell;1,...,r||\{p,q\}\,|\,\pmb\rho)$ can be expanded in terms of BS ones as 
\bea
\mathcal{I}^{\,\text{YMS}}(\ell;1,...,r||\{p,q\}\,|\,\pmb\rho)=\Sl_{\pmb\alpha\in\{p\}\shuffle\{q\}}C(\ell;1,\pmb\alpha)\mathcal{I}^{\,\text{BS}}(\ell;1,\pmb\alpha||\{p,q\}\,|\,\pmb\rho),\label{Eq:YMSExp2Gluons}
\eea
in which, the kinematic coefficient $C(l;1,\pmb\alpha)$ for any given $\pmb\alpha$ can be defined by the following compact expression
\bea
\boxed{C(\ell;1,\pmb\alpha)\cong\left[(\epsilon_p)_{\mu}(\epsilon_q)_{\nu}-{\epsilon_p\cdot\epsilon_q\over k_p\cdot k_q}\,(k_p)_{\mu}(k_q)_{\nu}\right]X^{\mu}_p(\ell;1,\pmb\alpha)X^{\mu}_q(\ell;1,\pmb\alpha)}\,,\label{Eq:YMSExp2GluonsCF}
\eea
which satisfies the consistency condition (\ref{Eq:ConstraintC}), is independent of the choice of the right permutation, and does not contain loop momentum in the denominator. With these coefficients, one can write down the expansion of single-trace EYM integrand:
\bea
\mathcal{I}^{\,\text{EYM}}(\ell;1,...,r||\{p,q\})=\Sl_{\pmb\alpha\in\{p\}\shuffle\{q\}}C(\ell;1,\pmb\alpha)\,\mathcal{I}^{\,\text{EYM}}(\ell;1,\pmb\alpha||\{p,q\}),\label{Eq:EYMExp2Gravitons}
\eea
where $1,...,r$ now denote gluons with fixed colour ordering, while $p$, $q$ are gravitons.

\subsection{YMS integrand with three gluons}
\label{sec:three-gluons}
The discussions on YMS integrands with one and two gluons can further be extended to YMS integrands with three gluons. This can be derived following similar discussions on the one and two gluon cases, but the vanishing terms that are added to the original YMS integrand have a more complicated form.  We just present the main result without providing the details of the calculation. The integrand $\mathcal{I}^{\,\text{YMS}}(\ell;1,...,r||\{p,q,s\}\,|\,\pmb\rho)$ with scalars $1,...,r$ and three gluons $p$, $q$, $s$ can be expressed as
\bea
\mathcal{I}^{\,\text{YMS}}(\ell;1,...,r||\{p,q,s\}\,|\,\pmb\rho)=\Sl_{\pmb\alpha\in\{p\}\shuffle\{q\}\shuffle\{s\}}C(\ell;1,\pmb\alpha)\mathcal{I}^{\,\text{BS}}(\ell;1,\pmb\alpha||\{p,q,s\}\,|\,\pmb\rho),\label{Eq:YMSExp3Gluons}
\eea
in which, the coefficients are defined by
\bea
\boxed{
\begin{matrix}
C(\ell;1,\pmb\alpha)&=&\biggl[\,(\epsilon_p)_{\mu}(\epsilon_q)_{\nu}(\epsilon_s)_{\tau}-{\epsilon_p\cdot\epsilon_q\over k_p\cdot k_q}\,(k_p)_{\mu}(k_q)_{\nu}(\epsilon_s)_{\tau}&\\
&&-{\epsilon_p\cdot\epsilon_s\over k_p\cdot k_s}\,(k_p)_{\mu}(\epsilon_q)_{\nu}(k_s)_{\tau}-{\epsilon_q\cdot\epsilon_s\over k_q\cdot k_s}\,(\epsilon_p)_{\mu}(k_q)_{\nu}(k_s)_{\tau}&\biggr]X^{\mu}_p(\ell;1,\pmb\alpha)X^{\mu}_q(\ell;1,\pmb\alpha)X^{\tau}_{s}(\ell;1,\pmb\alpha)
\end{matrix} }\nonumber
\label{Eq:YMSExp3GluonsCF}
\\
%
%
\eea
This coefficient $C(\ell;1,\pmb\alpha)$ has a symmetric form under the exchangings of $p$, $q$, $s$. Moreover, They naturally satisfy the consistency condition
(\ref{Eq:ConstraintC}) and do not contain loop momentum in the numerators. As the results with one and two gluons (\ref{Eq:CoefficientEG2a}), these coefficients are independent of the right permutations.
One can verify the expansion formula of YMS integrand with  the coefficients defined by (\ref{Eq:CoefficientEG1a}), (\ref{Eq:YMSExp2GluonsCF}) and
(\ref{Eq:YMSExp3GluonsCF}) by combining unitary cut and the tree level relations which will be presented in the next section. 

{\bf Comments on the results with coefficients (\ref{Eq:CoefficientEG1a})
,
 (\ref{Eq:YMSExp2GluonsCF}) 
 and (\ref{Eq:YMSExp3GluonsCF}):}\\

 (i) Although the first element in the above discussion is always fixed as a scalar, say 1,
one can also generalise these results to expressions with a gluon fixed as the first element
instead, following a similar discussion. For example, if we fix gluon $p$ as the one next to the loop momentum $\ell$, we simply replace $X_i(\ell;1,\pmb\alpha)$ ($i=p,q,r$) by $X_i(\ell;p,\pmb\alpha)$ in (\ref{Eq:CoefficientEG1a}), 
(\ref{Eq:YMSExp2GluonsCF}) 
 or (\ref{Eq:YMSExp3GluonsCF}), without changing the definition of $X$.
 \\

 (ii) For YMS integrands involving two or three gluons but no external scalar, momentum
conservation and the on-shell condition result in a vanishing denominator in the $\epsilon\cdot\epsilon$ terms of  (\ref{Eq:YMSExp2GluonsCF}) 
 or (\ref{Eq:YMSExp3GluonsCF}). However, this singularity is divided out
when we consider the full integrand, and the final result is nonsingular.

Taking into account the subtleties discussed above, and plugging the results of sections \ref{sec:one-gluon}, \ref{sec:two-gluons} and \ref{sec:three-gluons} back, we see that the pure YM numerators at $2$- and $3$-point are given by
%
\bea
N^{\text{YM}}\left(\ell;1,2\right)&=&(D-2)\left[(\epsilon_1)_{\mu}(\epsilon_2)_{\nu}-{\epsilon_1\cdot\epsilon_2\over k_1\cdot k_2}\,(k_1)_{\mu}(k_2)_{\nu}\right]\ell^{\mu}(\ell+k_1)^{\nu}+\text{Tr}\Big[F_1\cdot F_2\,\Big],\nn
N^{\text{YM}}\left(\ell;1,2,3\right)&=&(D-2)\Big[\,(\epsilon_1)_{\mu}(\epsilon_2)_{\nu}(\epsilon_3)_{\tau}-{\epsilon_1\cdot\epsilon_2\over k_1\cdot k_2}\,(k_1)_{\mu}(k_2)_{\nu}(\epsilon_3)_{\tau}\nn
&&-{\epsilon_1\cdot\epsilon_3\over k_1\cdot k_3}\,(k_1)_{\mu}(\epsilon_2)_{\nu}(k_3)_{\tau}-{\epsilon_2\cdot\epsilon_3\over k_2\cdot k_3}\,(\epsilon_1)_{\mu}(k_2)_{\nu}(k_3)_{\tau}\Big]\nn
&&~~~~~~~~~~~~~~~~~~~~~~~~~~~~~~~~~~~~~~~~~~~~~~~~~~~~~~~\times \ell^{\mu}(\ell+k_1)^{\nu}(\ell+k_1+k_2)^{\tau}\nn
&&+\text{Tr}\Big[F_2\cdot F_3\,\Big]\epsilon_1\cdot \ell+\text{Tr}\Big[F_1\cdot F_3\,\Big]\epsilon_2\cdot (\ell+k_1)\nn
&&+\text{Tr}\Big[F_1\cdot F_2\,\Big]\epsilon_3\cdot (\ell+k_1+k_2)\nn
&&-\text{Tr}\Big[F_1\cdot F_2\cdot F_3\,\Big]\epsilon_1\cdot \ell\,\epsilon_2\cdot(\ell+k_1)\,\epsilon_3\cdot(\ell+k_1+k_2).
\label{pure-ym-numerators}
\eea
%

\section{Welding YMS forward limit tree numerators}
\label{sec:welding-yms-trees}
Having seen that the YMS numerators at one loop can be calculated directly using the method explained in the previous section, the next natural question then is whether this result is identical to the welded forward limit tree numerators. Namely, suppose if we instead use tree level techniques 
 to solve for half-ladder numerators from YMS tree amplitudes, then 
take the forward limit on these half-ladders, would the $n$-gon basis numerators calculated in the previous section be identical to the forward limit 
half-ladder with both sides welded? 

Note especially that the numerator solutions are known to be non-unique at tree level, so that even when given a theory such as NLSM or self-dual 
Yang-Mills with built-in kinematic algebra structure and therefore inherited a 
natural numerator solution expressible as products of structure constants,
one can still write down other equivalent numerator solutions by deliberately applying
generalised gauge shifts. The resulting sets of kinematic numerator 
yield exactly the same tree amplitudes and are therefore also legitimate numerator
solutions. On the other hand, we saw in the derivation in section \ref{sec:shifting-used-explained} that the propagator matrix at one loop level are generically full-rank, so that given an integrand, the numerator 
the solution at one loop is unique. Additionally, we argued in section
\ref{sec:forward-limit-numerators} that the forward limit tree numerator 
solution, when both sides are welded, gives exactly this unique numerator
solution at one loop if they satisfy the consistency conditions that arise 
from the nomenclature \ref{eq:extra-condition-2}. For the YMS case
actually one sees readily that the numerator solution at tree level based on
 graphical rules of \cite{Hou:2018bwm} indeed satisfy these conditions, and therefore
 corresponds to this unique numerator solution at one loop.

Recall at tree level, the YMS amplitude and the propagator matrix (equivalently, the doubly colour-ordered bi-adjoint scalar amplitudes)
are related by the following.
\begin{equation}
A_{\text{tree}}^{\text{YMS}}(1,...,r||\mathsf{G}\,|\,\pmb\rho) = \Sl_{\pmb\alpha\in\{2,...r-1\}\,\shuffle\,\text{perms}\,\mathsf{G}}\,C(1,\pmb\alpha)\,A_{\text{tree}}^{\text{BS}}(1,\pmb\alpha,r\,|\,\pmb\rho),\label{tree-yms-expansion}
\end{equation}
where  $1,...,r$ and $\mathsf{G}$ denote the scalar and gluon legs respectively. The expansion coefficients, for the one, two and three gluon cases are given by
\bea
C(1,\pmb\alpha)&\equiv&\epsilon_p\cdot X_p(1,\pmb\alpha),~~~~~~~~~~~~~~~~~~~~~~~~~~~~~~~~~~~~~~~~~~~~~~~~~~~~~~~~~~~~\,\,\text{if}~~\mathsf{G}=\{p\}\nn
C(1,\pmb\alpha)&\equiv&\left[(\epsilon_p)_{\mu}(\epsilon_q)_{\nu}-{\epsilon_p\cdot\epsilon_q\over k_p\cdot k_q}\,(k_p)_{\mu}(k_q)_{\nu}\right]X^{\mu}_p(1,\pmb\alpha)X^{\mu}_q(1,\pmb\alpha),~~~~~~~~~~~\,\text{if}~~\mathsf{G}=\{p,q\}\nn
C(1,\pmb\alpha)&\equiv&\biggl[\,(\epsilon_p)_{\mu}(\epsilon_q)_{\nu}(\epsilon_s)_{\tau}-{\epsilon_p\cdot\epsilon_q\over k_p\cdot k_q}\,(k_p)_{\mu}(k_q)_{\nu}(\epsilon_s)_{\tau}-{\epsilon_p\cdot\epsilon_s\over k_p\cdot k_s}\,(k_p)_{\mu}(\epsilon_q)_{\nu}(k_s)_{\tau}\nn
&&-{\epsilon_q\cdot\epsilon_s\over k_q\cdot k_s}\,(\epsilon_p)_{\mu}(k_q)_{\nu}(k_s)_{\tau}\biggr]X^{\mu}_p(1,\pmb\alpha)X^{\mu}_q(1,\pmb\alpha)X^{\tau}_{s}(1,\pmb\alpha),~~~~~~~~~~~\text{if}~~\mathsf{G}=\{p,q,s\}.
\nonumber \\
&& \hspace{13cm}{(\theequation)}
\label{Eq:TreeCF} 
\refstepcounter{equation}\nonumber
\eea
Take the two-scalar-one-gluon scattering event as an example, and
let $k_{1}$, $k_{2}$ be the momenta carried by the scalar and $k_{3}$
by the gluon respectively. In view of (\ref{eq:feyn-forward-limit-decomp}),
the colour-ordered integrand $\mathcal{I}_{2}(\ell,k_{1}^{s},k_{2}^{s},k_{3}^{g})$
is given by
\begin{eqnarray}
& 
\mathcal{I}_{2}(\ell,k_{1}^{s},k_{2}^{g})= & \frac{1}{\ell^{2}}A_{\text{Feyn}}^{\text{tree}}(\ell,1^{s},2^{s},3^{g},-\ell)  \\
&& +\frac{1}{(\ell+k_{1})^{2}}A_{\text{Feyn}}^{\text{tree}}(\ell+k_{1},2^{s},3^{g},1^{s},-(\ell+k_{1}))
\nonumber \\
&& +\frac{1}{(\ell+k_{1}+k_{2})^{2}}A_{\text{Feyn}}^{\text{tree}}(\ell+k_{1}+k_{2},3^{g},1^{s},2^{s},-(\ell+k_{1}+k_{2}))
\nonumber
\end{eqnarray}
Restoring the colour factor carried by the scalars on both sides of
the equation (\ref{tree-yms-expansion}), and we read off the forward
limit tree numerators. For $A_{\text{Feyn}}^{\text{tree}}(\ell,1^{s},2^{s},3^{g},-\ell)$,
we have the following half-ladder.
\begin{align}
N^{\text{tree}}(\ell,1^{s},2^{s},3^{g},-\ell) & =\sum_{e_{1},e_{2}}C(1,\alpha)f^{e_{1},1,e_{2}}f^{e_{2},2,e_{1}}\\
 & =\epsilon_{3}\cdot(\ell+k_{1}+k_{2})\sum_{e_{1},e_{2}}f^{e_{1},1,e_{2}}f^{e_{2},2,e_{1}}\label{eq:welded-tree-numerator-3pt-1}
\end{align}
The kinematic part of the numerator $C(1,\alpha)=\epsilon_{3}\cdot X$
contains a sum over momenta lying on one side of the gluon line, which
in this case includes the loop scalar $\ell$, and two external scalar
lines $k_{1}$, $k_{2}$. On the other hand, for $A_{\text{Feyn}}^{\text{tree}}(\ell+k_{1},2^{s},3^{g},1^{s},-(\ell+k_{1}))$
we have
\begin{equation}
N^{\text{tree}}(\ell+k_{1},2^{s},3^{g},1^{s},-(\ell+k_{1}))=\epsilon_{3}\cdot((\ell+k_{1})+k_{2})\sum_{e_{1},e_{2}}f^{e_{1},1,e_{2}}f^{e_{2},2,e_{1}}\label{eq:welded-tree-numerator-3pt-2}
\end{equation}
The kinematic part of the numerator this time includes only two lines:
The loop scalar, now carrying momenta $\ell+k_{1}$, and a scalar,
carrying $k_{2}$. The momentum sum is identical to that in (\ref{eq:welded-tree-numerator-3pt-1}),
and we see that 
\begin{equation}
N^{\text{tree}}(\ell,1^{s},2^{s},3^{g},-\ell)=N^{\text{tree}}(\ell+k_{1},2^{s},3^{g},1^{s},-(\ell+k_{1}))
\end{equation}
as required. Similarly for the numerator $N^{\text{tree}}(\ell+k_{1}+k_{2},3^{g},1^{s},2^{s},-(\ell+k_{1}+k_{2}))$,
and others related by permutations of the external legs.

It is straightforward to see the above example generalises to cases
with a single gluon and an arbitrary number of scalars. At $n$-point
the basis numerators are $n$-gons, with the colour part of the numerator
given by a product of $(n-1)$ structure constants $f^{*\sigma(1)*}f^{*\sigma(2)*}\dots f^{*\sigma(n)*}$
which depends only on the cyclic ordering of the scalar legs, so that
the welded tree numerators $N^{\text{tree}}(\ell,\sigma(1),\sigma(2),\dots,\sigma(n),-\ell)$,
$N^{\text{tree}}(\ell+k_{\sigma(1)},\sigma(2),\dots,\sigma(n),\sigma(1),-(\ell+k_{\sigma(1)}))$,
$\dots$, $N^{\text{tree}}(\ell+k_{\sigma(1)}+\dots+k_{\sigma(n-1)},\sigma(n),\sigma(1),\dots,\sigma(n-1),-(\ell k_{\sigma(1)}+\dots+k_{\sigma(n-1)}))$
all carry this common (loop independent) factor. The kinematic part
of the tree numerator $\epsilon_{g}\cdot X_{g}:=\epsilon_{g}\cdot\sum_{\text{all momenta on the left}}k_{i}$
is given by a sum of all momenta lying on one side of the gluon. Tree
numerators $N^{\text{tree}}(\ell,\sigma(1),\sigma(2),\dots,\sigma(n),-\ell)$,
$N^{\text{tree}}(\ell+k_{\sigma(1)},\sigma(2),\dots,\sigma(n),\sigma(1),-(\ell+k_{\sigma(1)}))$,
$\dots$, $N^{\text{tree}}(\ell+k_{\sigma(1)}+\dots+k_{\sigma(n-1)},\sigma(n),\sigma(1),\dots,\sigma(n-1),-(\ell k_{\sigma(1)}+\dots+k_{\sigma(n-1)}))$
have different set of scalar legs lying on the left of the gluon line,
however the $\ell$ appears in each tree numerator is shifted accordingly
to compensate the missing scalar leg momenta, so that the sums all
yield the same result, and the welded tree numerators therefore all
agree with each other, fulfilling the consistency requirement (\ref{eq:extra-condition-2}).
For cases involving two or three gluons and arbitrary number of scalars,
one sees that the welded forward limit tree numerators also satisfy
the consistency requirement (\ref{eq:extra-condition-2}) and are
therefore identical to the unique numerator solution at one loop.
The verification follows more or less the same calculations, as they
also expressible in terms of the factor $X_{g}:=\sum_{\text{all momenta on the left of }g}k_{i}$.
Equivalently, one can also check the welded tree numerators are in
these cases identical to the $n$-gon numerator formulae (\ref{Eq:CoefficientEG2a}),
(\ref{Eq:YMSExp2GluonsCF}) and (\ref{Eq:YMSExp3GluonsCF}) calculated
in the previous section.


\section{A brief remark on gravity and EYM numerators at one loop}
\label{sec:gr-eym-numberators}

In this section we briefly remark on the gravity and Einstein-Yang-Mills (EYM) numerators. A first observation is that tree level EYM amplitudes satisfy the same expansion formula with the YMS ones.
\begin{equation}
A_{\text{tree}}^{\text{EYM}}(1,...,r||\mathsf{H}) = \Sl_{\pmb\alpha\in\{2,...r-1\}\,\shuffle\,\text{perms}\,\mathsf{H}}\,C(1,\pmb\alpha)\,A_{\text{tree}}^{\text{YM}}(1,\pmb\alpha,r\,),\label{tree-eym-expansion}
\end{equation}
where the $1,...,r$ denotes gluons, while $\mathsf{H}$ is the graviton  set. The coefficients $C(1,\pmb\alpha)$ have the same form with those in (\ref{Eq:TreeCF}) but the polarizations therein are now ‘half gravitons’. This can be straightforwardly proved using either BCJ relations or CHY formula.

When combining the tree-level expansion formula (\ref{tree-eym-expansion}) with the forward limit expressed  one loop EYM integrand, as stated by the worldsheet approach (\cite{He:2016mzd,He:2017spx,Geyer:2017ela,Porkert:2022efy}), we get 
\bea
&&\mathcal{I}^{\,\text{EYM}}(\ell;1...,r||\mathsf{H})\nn
&=&\Sl_{\pmb\rho\in\text{perms}(\{1,...,r\}\cup\mathsf{H}) }~\Sl_{\pmb\sigma\in\{1,...,r\}\shuffle \text{perms}\mathsf{H} }{1\over (\ell-k_{\sigma_L})^2}C(+,\pmb\sigma,-)\times A^{\text{YM}}(+,\pmb\sigma,-)+\text{cyc}(1,...,r)\nn
&=&\Sl_{\pmb\rho\in\text{perms}(\{1,...,r\}\cup\mathsf{H}) }~\Sl_{\pmb\sigma\in\{1,...,r\}\shuffle \text{perms}\mathsf{H} }{1\over (\ell-k_{\sigma_L})^2}C(+,\pmb\sigma,-)\nn
&&~~~~~~~~~~~~~~~~~~~~~~~~~~~~~~~~~~~~\times A^{\text{BS}}(+,\pmb\sigma,-|\,+,\pmb\rho,-)\W N(+,\pmb\rho,-)+\text{cyc}(1,...,r),\label{EYMeg0}
\eea
where the YM tree amplitudes with forward limit is further expanded in terms of BS ones, by using the numerators $\W N(+,\pmb\rho,-)$ satisfying the consistency condition. A technical treatment \cite{Xie:2024pro,Xie:2025utp} is the following. When expanding the $A^{\text{BS}}(+,\pmb\sigma,-|\,+,\pmb\rho,-)$ in terms of Feynman graphs which have the form that tree level BS currents are planted at the loop propagator line between $\pm$, one can always find other related terms in the summations over $\pmb\rho$, $\pmb\sigma$ and the cyclic orderings of external gluons, so that these terms are related to each other by a cyclic permutation of the BS currents. When all these terms are collected together and the consistency conditions of both  $C(+,\pmb\sigma,-)$ and $\W N(+,\pmb\rho,-)$ are taken into account, the Feynman graphs related by cyclic permutations of BS currents produce one loop Feynman graphs of BS  integrand, with quadratic propagators and coefficients  $C(+,\pmb\sigma,-)$ and $\W N(+,\pmb\rho,-)$. Therefore, we arrive at the expansion formula of EYM integrand 
\bea
&&\mathcal{I}^{\,\text{EYM}}(\ell;1...,r||\mathsf{H})\nn
&=&\Sl_{\pmb\sigma\in\{1,...,r\}\shuffle \text{perms}~\mathsf{H} }C(\ell;1,\pmb\sigma)\left[\,{1\over (\ell-k_{\sigma_L})^2}\,\Sl_{\pmb\rho\in\text{perms}~(\{1,...,r\}\cup\mathsf{H}) }~\mathcal{I}^{\text{BS}}(\ell;1,\pmb\sigma|\,\pmb\rho)\,\W N(\ell';1,\pmb\rho)\right]\nn
&=&\Sl_{\pmb\sigma\in\{1,...,r\}\shuffle \text{perms}~\mathsf{H} }C(\ell;1,\pmb\sigma)~\mathcal{I}^{\text{YM}}(\ell;1,\pmb\sigma),\label{EYMeg7}
\eea
where the coefficients $C(\ell;1,\pmb\sigma)$ were given by (\ref{Eq:CoefficientEG1a}), (\ref{Eq:YMSExp2GluonsCF}) and (\ref{Eq:YMSExp3GluonsCF}) corresponding to the $1$, $2$ and $3$ graviton cases. The loop momentum $\ell'$ in the right numerator $\W N(\ell';1,\pmb\rho)$ is understood as follows. If the subcurrent containing scalar $1$ in a diagram of $\mathcal{I}^{\text{BS}}$ have the form $\phi_{\pmb\sigma_A1\pmb\sigma_B|\pmb\rho_A1\pmb\rho_B}$, $\ell'\equiv l-k_{\rho_A}+k_{\rho_B}$. More detail can be found in the paper \cite{Du:2025rty} by two of the authors of the current work.

For gravity integrand\footnote{By gravity here we mean the  Einstein-Hilbert gravity coupled to $B$ fields and dilaton.} there is an analogue to the expansion formula  (\ref{Eq:YMbyYMS}) that expanses
the gravity integrand in terms of EYM integrands with fewer graviton external lines (\cite{He:2016mzd,He:2017spx,Geyer:2017ela,Porkert:2022efy,Xie:2025utp,Cao:2025ygu}),
\bea
\mathcal{I}^{\,\text{GR}}&\cong&(D-2)\,\mathcal{I}^{\,\text{EYM}}\big(\emptyset||\mathsf{H}\big)\nn
&&+\Sl_{l=2}^{n}(-1)^l\Sl_{{{\small(j_1...j_l)}\in\text{S}_i\setminus\text{Z}_i}}\text{Tr}\big[{F}_{j_1}\cdot...\cdot {F}_{j_l}\big]\mathcal{I}^{\,\text{EYM}}\big(j_1,...,j_l||\mathsf{H}\setminus\{j_1,...,j_l\}\big).\label{Eq:GRbyEYM}
\eea
This relation, together with the formula (\ref{EYMeg7}) will result in an expansion formula for GR integrands
\begin{equation}
\mathcal{I}^{\,\text{GR}} = \Sl_{\pmb\sigma\in S_{n-1}}n_{\ell;1,\pmb\sigma}\,\mathcal{I}^{\,\text{YM}}(\ell;1,\pmb{\sigma}).\label{Eq:GRNumerators}
\end{equation}

We emphasise that the discussions in this section are based on equation (\ref{EYMeg0}), which is derived from the worldsheet approach. Establishing equation (\ref{EYMeg0}) purely within a field-theoretic framework, in particular by starting directly from the EYM and YM Feynman rules, appears to be highly non-trivial. Indeed, experience from the scattering amplitude literature indicates that demonstrating double copy relations directly at the level of Feynman integrands is often technically challenging\footnote{One possible way to do this is to adapt the argument used in \cite{Borsten:2020zgj,Borsten:2021hua,Borsten:2021gyl}  to the EYM case.}.

Alternatively, the reasoning presented here can be understood in the following conditional sense. Assuming the field theory double copy relation between EYM and YM observed at tree level \cite{Bern:1999bx,Cachazo:2014nsa} generalises to one loop (for example at $2$-point we have equation (\ref{eq:2pt-db-copy}) and similarly at higher points), the argument of section \ref{sec:double-copy} allows one to write the EYM integrand in the form of the one-loop generalised DDM representation (\ref{2pt-half-ladder}). This representation can then be recast into equation (\ref{EYMeg0}) by applying the partial-fraction decomposition described in section \ref{sec:forward-limit-numerators}, see equation (\ref{eq:feyn-forward-limit-decomp}). The calculation presented in this section then proceeds from equation (\ref{EYMeg0}).

\section{Conclusion}
\label{sec:conclusion}

In this work we have analysed the algebraic consistency conditions underlying the 
construction of one-loop kinematic numerators, clarified their relation to welded 
forward-limit tree numerators, and provided explicit solutions for Yang--Mills-scalar 
theory up to three gluons. Our findings highlight several conceptual features of 
colour--kinematics duality at one loop that are absent at tree level.

At tree level, BCJ numerators are defined only up to the kernel of propagator matrices.  
Consequently, many inequivalent sets of half-ladder numerators lead to the same 
physical amplitude. At one loop, however, the situation is markedly different: 
the nomenclature consistency conditions identify all $n$-gon representations of the 
same loop graph---that is, all expressions related by cyclic relabelling of external 
legs and corresponding loop-momentum shifts---as describing a single physical object.  
The resulting quotient removes the redundancy associated with different ``pictures'' 
of the same polygon diagram, and the effective propagator matrix derived from this 
quotient is generically of full rank. As a result, a colour-ordered integrand fixes a 
\emph{unique} numerator assignment (up to integrand-level shift symmetries).  
This gives a natural projection map from any set of tree-level numerators---with their 
usual gauge freedom---onto the unique one-loop solution compatible with the 
nomenclature consistency condition.  Our analysis of welded forward-limit 
numerators confirms that, whenever the physical conditions discussed in 
section~3 are satisfied, this projection coincides with welding the tree numerators 
and identifying the two parallel legs.The same analysis can be extended to Einstein--Yang--Mills theory, whose mixed gluon--graviton integrands follow the same algebraic pattern once the loop-momentum dependence of both sectors is taken into account.

The situation is more subtle for gravity-like theories such as Einstein--Yang--Mills and gravity.  
The discussion in section~9 suggests that, once both copies of the integrand carry 
loop-momentum dependence, consistent double-copy constructions may require  
\emph{additional} physical conditions.  
In particular, the naive expectation that the KLT kernel provides a loop-level inverse 
to the propagator matrix is obstructed: ordering the gravity integrand via the 
forward-limit decomposition necessarily introduces either misaligned loop momenta 
between the two Yang--Mills copies or a difference term of the form 
$\Delta_\ell = f(\ell) - f(\ell+c)$.  
This misalignment indicates that the standard KLT and double-copy prescriptions 
cannot both be simultaneously realised in a loop-momentum--local form, and the 
resolution may again lie in additional physical constraints on the loop integrands.

Beyond these structural observations, our explicit results for YMS and EYM theories demonstrate that the simple tree-level building blocks, organised by generalised momentum-sum symbols $X^\mu$, extend naturally to one loop.
The $n$-gon decomposition 
\[
\mathcal{I}_n^{\mathrm{YMS}}(\ell,1,\dots,n)
= \sum_{\sigma\in S_{n-1}} N(\ell;1,\sigma)\,\widetilde{M}(1\sigma \mid 1\,2\cdots n),
\]
together with the quotient over cyclic relabellings and loop-momentum shifts, provides 
a constructive framework from which numerators can be extracted directly without 
requiring inversion of large propagator matrices.  
Our one-, two-, and three-gluon results confirm that the one-loop kinematic 
coefficients follow the same algebraic pattern as at tree level, once the appropriate 
generalised $X^\mu$ are used.  
This strongly suggests that the higher-point YMS numerators admit similarly compact 
and symmetry-transparent expressions.

Looking forward, several interesting directions remain open.  
It would be valuable to extend the explicit YMS numerator formulas to higher points, 
 and to explore whether the same organisational principles extend further to  gravity amplitudes and to higher loops.  
Clarifying the precise physical requirements for a fully consistent loop-level double 
copy---and resolving the apparent incompatibility between KLT relations and 
double-copy constructions---would shed further light on the analytic structure of 
gravity integrands.  
We expect that the structural perspective developed in this work, based on 
nomenclature consistency and forward-limit factorisation, will provide a useful 
foundation for these developments.





\begin{acknowledgments}
We thank Pierre Vanhove for his inspiring comments and valuable suggestions on this work.
CF is supported by the High-level Talent 
Research Start-up Project Funding of Henan Academy of Sciences
(Project No. 241819244). YW is supported by China National Natural Science Funds for Distinguished Young Scholar (Grant No. 12105062). YD is supported by NSFC under Grant No. 11875206.
\end{acknowledgments}

\appendix
\section{Nonlinear sigma model numerators}
\label{sec:nlsm-numerators}

In this appendix we perform a simple consistency check on the NLSM
numerators. It was shown in \cite{Carrasco:2016ldy} that the tree level numerators
of NLSM can be efficiently expressed as momentum kernel $N_{NLSM}^{\text{tree}}(123\dots n)=S_{1}[23\dots n-1|23\dots n-1]$.
The corresponding forward limit numerator is therefore
\begin{align}
 & N(\ell,1,2,3,\dots,n,-\ell)\label{eq:nlsm-numerator-formula}\\
 & =S_{\ell}[123\dots n|123\dots n]\nonumber \\
 & =\left(\ell\cdot k_{1}\right)\left((\ell+k_{1})\cdot k_{2}\right)\left((\ell+k_{1}+k_{2})\cdot k_{3}\right)\dots\left((\ell+k_{1}+k_{2}\dots+k_{n-1})\cdot k_{n}\right)\nonumber 
\end{align}
At $n$-point, consistency requires that the above numerator should
be invariant under the following symmetry transformation.
\begin{equation}
\begin{array}{cc}
\ell\rightarrow & \ell+k_{1}\\
k_{1}\rightarrow & k_{2}\\
\vdots & \vdots\\
k_{n}\rightarrow & k_{1}
\end{array}
\end{equation}
Equation (\ref{eq:nlsm-numerator-formula}) is manifestly invariant
because each factor becomes the next one under such relabelling.
\begin{equation}
\begin{array}{cc}
\ell\cdot k_{1}\rightarrow & (\ell+k_{1})\cdot k_{2}\\
(\ell+k_{1})\cdot k_{2}\rightarrow & \ell+k_{1}+k_{2})\cdot k_{3}\\
\vdots & \vdots\\
(\ell+k_{1}+k_{2}\dots+k_{n-1})\cdot k_{n}\rightarrow & \ell\cdot k_{1}
\end{array}
\end{equation}
where the last line above is guaranteed by momentum conservation.

\section{$3$-point integrand numerator relation}

\label{sec:3pt-example}

At $3$-point we have two independent numerators (\ref{eq:3pt-consistency-1})
and (\ref{eq:3pt-consistency-2}), and the integrand is assumed to
have the following graphical decomposition. 
\bea
 \mathcal{I}_{3}(\ell,1,2,3)&=&\begin{minipage}{2cm}\includegraphics[width=2cm]{graph123}\end{minipage}\,+\,\begin{minipage}{2.15cm}\includegraphics[width=2.15cm]{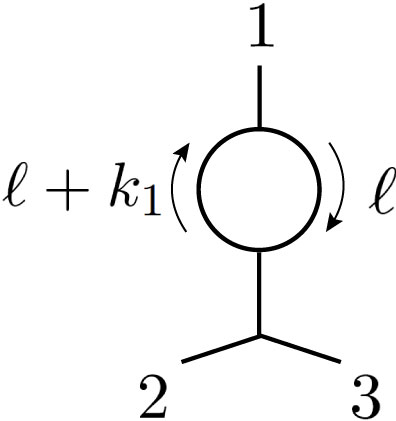}\end{minipage}\,+\,\begin{minipage}{2.1cm}\includegraphics[width=2.1cm]{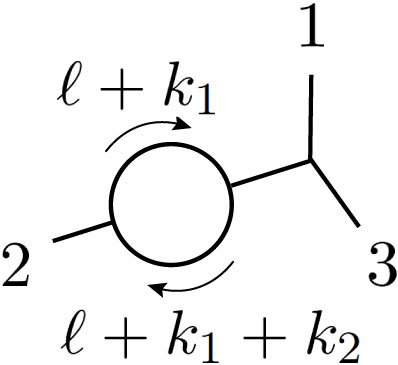}\end{minipage}\,+\,\begin{minipage}{2.0cm}\includegraphics[width=2.0cm]{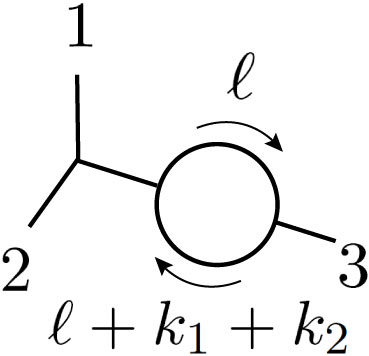}\end{minipage}\nn
&= & \frac{N(\ell,123,-\ell)}{D_{\ell}D_{\ell1}D_{\ell12}}+\frac{N(\ell,123,-\ell)-N(\ell,132,-\ell)}{D_{\ell}D_{\ell1}s_{23}}\label{eq:3pt-integrand-rel-1}\\
& &~~~~~~~~~~~~~~~~~~+\frac{N(\ell+k_{1},231,-(\ell+k_{1}))-N(\ell+k_{1},213,-(\ell+k_{1}))}{D_{\ell1}D_{\ell12}s_{13}}\nn
& &~~~~~~~~~~~~~~~~~~+\frac{N(\ell,123,-\ell)-N(\ell,213,-\ell)}{D_{\ell}D_{\ell12}s_{12}}\nonumber 
\eea
where we introduced the notation $D_{\ell,a,b,\dots,c}=(\ell+k_{a}+k_{b}+\dots+k_{c})^{2}$
frequently employed in loop amplitude calculations to denote the propagator
involving $\ell$ squared. Adding difference terms as in (\ref{eq:2pt-shifted-integrand})
effectively shifts loop momenta. In the current example this allows
us to write
\begin{align}
\tilde{\mathcal{I}}_{3}(\ell,123)= & \left(\frac{1}{D_{\ell}D_{\ell1}D_{\ell12}}+\frac{1}{D_{\ell}D_{\ell1}s_{23}}+\frac{1}{D_{\ell1}D_{\ell12}s_{13}}+\frac{1}{D_{\ell}D_{\ell12}s_{12}}\right)\,N(\ell,123,-\ell)\label{eq:integrand-123}\\
 & +\left(\frac{-1}{D_{\ell}D_{\ell1}s_{23}}+\frac{-1}{D_{\ell}D_{\ell13}s_{13}}+\frac{-1}{D_{\ell1}D_{\ell13}s_{12}}\right)\,N(\ell,132,-\ell),\nonumber 
\end{align}
where again we use $\tilde{\mathcal{I}}_{3}$ to indicate that it
is an effective integrand, related to the original one by terms that
vanish after the integration. Cyclic symmetry suggests that there
are only two independent colour-ordered integrands. The other integrand
can be re-written following the same reasoning, which together leads
to a $2\times2$ matrix equation between the colour-ordered integrands
and the master numerators, now with their loop momentum aligned to
$\ell$.
\begin{equation}
\left[\begin{array}{c}
\tilde{\mathcal{I}}_{3}(\ell,123)\\
\tilde{\mathcal{I}}_{3}(\ell,132)
\end{array}\right]=\left[\begin{array}{cc}
\tilde{M}(23|23) & \tilde{M}(23|32)\\
\tilde{M}(32|23) & \tilde{M}(32|32)
\end{array}\right]\left[\begin{array}{c}
N(\ell,123,-\ell)\\
N(\ell,132,-\ell)
\end{array}\right].
\end{equation}
The one loop generalisation of the propagator matrix is given by the
same formula obtained in the context of CHY in \cite{Feng:2022wee}.
\begin{equation}
\tilde{M}=\left[\begin{array}{cc}
\frac{1}{D_{\ell}D_{\ell1}D_{\ell12}}+\frac{1}{D_{\ell}D_{\ell1}s_{23}}+\frac{1}{D_{\ell1}D_{\ell12}s_{13}}+\frac{1}{D_{\ell}D_{\ell12}s_{12}} & \frac{-1}{D_{\ell}D_{\ell1}s_{23}}+\frac{-1}{D_{\ell}D_{\ell13}s_{13}}+\frac{-1}{D_{\ell1}D_{\ell13}s_{12}}\\
\frac{-1}{D_{\ell}D_{\ell1}s_{23}}+\frac{-1}{D_{\ell}D_{\ell12}s_{12}}+\frac{-1}{D_{\ell1}D_{\ell12}s_{13}} & \frac{1}{D_{\ell}D_{\ell1}D_{\ell13}}+\frac{1}{D_{\ell}D_{\ell1}s_{23}}+\frac{1}{D_{\ell1}D_{\ell13}s_{12}}+\frac{1}{D_{\ell}D_{\ell13}s_{13}}
\end{array}\right].\label{eq:3pt-propagator-matrix}
\end{equation}


\section{Tadpole cancellation}
\label{sec:tadpoles}

In this appendix we briefly show that tadpole numerators do not contribute
to the one loop amplitude after integration. Similar arguments were
presented in \cite{Baadsgaard:2015twa,He:2016mzd}. However, we would
like to show that tadpoles also drop out specifically for the settings
used in this paper. Note the tadpole term in the tree amplitude $A_{\text{BCJ}}^{\text{tree}}(\ell,1,2,3,\dots,n,-\ell)$
is the following.
\begin{equation}
 A_{\text{BCJ}}^{\text{tree}}(\ell,1,2,3,\dots,n,-\ell) \sim \begin{minipage}{1.7cm}\includegraphics[width=1.7cm]{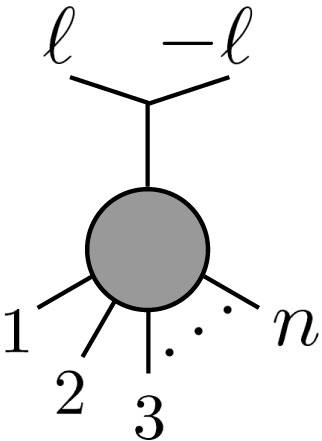}\end{minipage}
\end{equation}
Therefore if we pick out only the tadpoles in (\ref{eq:forward-limit-decomp-npt}),
up to shiftings which do not contribute to the integral, this is
\begin{eqnarray}
  && \mathcal{I}_{n}(\ell,1,2,\dots,n) \label{eq:tadpole-2} \\
  && \sim \frac{1}{\ell^{2}}\tilde{A}_{\text{BCJ}}^{\text{tree}}(\ell,1,2,3,\dots,n,-\ell)+ \frac{1}{\ell^{2}} A_{\text{BCJ}}^{\text{tree}}(\ell,2,3,\dots,n,1,-\ell)+\dots  \nonumber \\
  && \hspace{0.5cm}+ \frac{1}{\ell^{2}} A_{\text{BCJ}}^{\text{tree}}(\ell,n,1,2,\dots,n-1,-\ell) 
  \nonumber \\
  && \sim \begin{minipage}{1.7cm}\includegraphics[width=1.7cm]{Tadpole1}\end{minipage}\,+\,\begin{minipage}{1.7cm}\includegraphics[width=1.7cm]{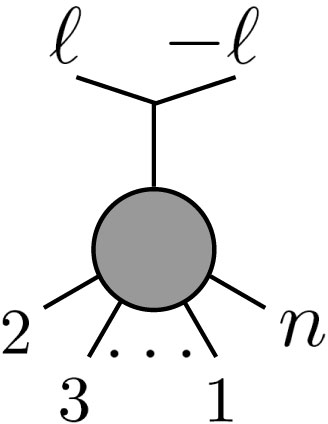}\end{minipage}\,+\dots +\,\begin{minipage}{2.6cm}\includegraphics[width=2.6cm]{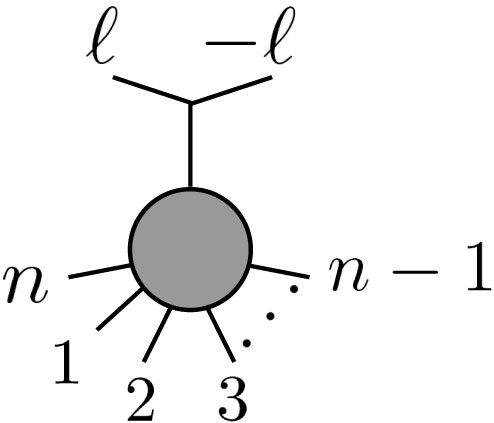}\end{minipage}
\end{eqnarray}
We would like to emphasize the above is a sum of numerators, which
are regarded as yet-determined variables whose analytic behaviour
are to a large extent unknown. However this sum still vanishes because
$U(1)$-decoupling identity requires only the knowledge of anti-symmetry
of the cubic structure, and this is an assumed property of the BCJ
numerators.

\section{Berends-Giele currents of BS}\label{app:BScurrents}

The Berends-Giele current $\phi_{A|\W A}$, where $A$ and  $\widetilde{A}$ are ordered sets with the same number of elements, is defined as follows (\cite{Mafra:2016ltu}):
\bea
\phi_{A|\widetilde{A}}={1\over s_A}\Sl_{\substack{A\to A_L,A_R\\ \widetilde{A}\to\widetilde{A}_L,\widetilde{A}_R}}\Bigl[\,\phi_{A_L|\widetilde{A}_L}\phi_{A_R|\widetilde{A}_R}-\phi_{A_R|\widetilde{A}_L}\phi_{A_L|\widetilde{A}_R}\,\Bigr],~\label{Eq:BScurrent}
\eea
where $s_A\equiv k_A^2$, $k_A^{\mu}$ denotes the total momentum of elements in $A$. We summed over divisions $A\to A_L,A_R$, $\widetilde{A}\to\widetilde{A}_L,\widetilde{A}_R$ in (\ref{Eq:BScurrent}) so that in the first term $|A_L|=|\W A_L|$, $|A_R|=|\W A_R|$,  or in the second term  $|A_R|=|\W A_L|$, $|A_L|=|\W A_R|$. The starting point of this definition is $\phi_{a|a}=1$, $\phi_{a|b}=0$ $(a\neq b)$. A significant property of the BS current (\ref{Eq:BScurrent}) is the generalized $U(1)$-decoupling identity (\ref{Eq:GenU1}). Another property is the off-shell BCJ relation, see  \cite{Du:2011js,Frost:2020eoa,Wu:2021exa,Du:2022vsw}. The off-shell BCJ relation is the key technique for deriving (\ref{Eq:EG2Cancel}) and (\ref{Eq:4PtTrans}), see \cite{Xie:2024pro,Xie:2025utp}.

\section{The  kinematic coefficient for a given subset in a cyclic division}\label{app:kinematicC}

As shown in \cite{Xie:2024pro}, a YMS integrand $\mathcal{I}^{\,\text{YMS}}(l;1,\pmb{\sigma}||\mathsf{G}\,|\,\pmb\rho)$ can be expressed as a summation over independent cyclic partitions of external particles (scalars and gluons) $\{1,\pmb\sigma\,\shuffle\text{perms}\,\mathsf{G}\}\to \{A_1,A_2,...,A_I\}$, so that the relative cyclic ordering of scalars $\{1,\pmb\sigma\}$ is preserved. The concrete expression relies on the so-called {\it reference order} of elements in the gluon set $\mathsf{G}$. Each subset $A_i$, is associated with a substructure which involves effective currents attached to the loop via a proper vertex.  When decomposed in terms of BS currents, an effective subcurrent is further expressed as a summation over permutations of elements in $A_i$, each permutation is accompanied by a kinematic coefficient. According to different particle types in $A_i$, the possible permutations, the coefficients (with upto two gluons) and the corresponding BS subcurrents are presented as follows:
\begin{itemize}
\item If $A_i$ contains only scalars in order $\{x_1,...,x_j\}$, the coefficient, permutation and BS currents are respectively presented as 
 \bea
C(\pmb\alpha_i)=1,~~~~\pmb\alpha_i=\{x_1,...,x_j\},~~~\phi_{\pmb\alpha_i|\,\pmb\rho_i},
 \eea
 where $\pmb\rho_i$ denotes the right permutation for this subcurrent.
 
\item If $A_i$ contains both scalars in order $\{x_1,...,x_j\}$ and elements in the gluon set $\mathsf{G}_i$ ($\mathsf{G}_i=\{p\},\{q\}\,\text{or}\,\{p,q\}$), we have
\begin{equation}
\scalebox{0.8}{
\hspace{-3cm}
\begin{minipage}{17cm}
\begin{eqnarray}
&&\text{For $\mathsf{G}_i=\{p\}\,(\text{or}\,\{q\})$},~\,C(\pmb\alpha_i)=\epsilon_{p\,(\text{or}\,q)}\cdot X_{p\,(\text{or}\,q)}(\pmb\alpha_i),~~~\,\pmb\alpha_i\in\{x_1,\{x_2,...,x_j\}\shuffle\{p\}\,(\text{or}\,\{q\}),~\phi_{\pmb\alpha_i|\,\pmb\rho_i},\nn
&&\text{For $\mathsf{G}_i=\{p,q\}$},~~~~~\,\,C^{(0)}(\pmb\alpha_i)=\epsilon_p\cdot X_p(\pmb\alpha_i)\,\epsilon_q\cdot X_a(\pmb\alpha_i),~\pmb\alpha_i\in\{x_1,\{x_2,...,x_j\}\shuffle\{p\}\shuffle\{q\}\},~\,\phi_{\pmb\alpha_i|\,\pmb\rho_i},\nn
&&~~~~~~~~~~~~~~~~~~~~~~~~~~\,\,C^{(1)}(\pmb\alpha_i)=(\epsilon_p\cdot\epsilon_q)\,(-k_p\cdot X_p(\pmb\alpha_i)),\,\pmb\alpha_i\in\{x_1,\{x_2,...,x_j\}\shuffle\{p,q\}\},~~~~~~\,\phi_{\pmb\alpha_i|\,\pmb\rho_i},\nonumber
\end{eqnarray}
\end{minipage}
}
\end{equation}
where $C^{(0)}$ and $C^{(1)}$ respectively denote the $(\epsilon_p\cdot\epsilon_q)^{0}$ and the $(\epsilon_p\cdot\epsilon_q)^{1}$ sectors. 

\item If $A_i$ contains one or two gluons but no scalar, the coefficients, permutations and corresponding BS subcurrents are given by

\bea
&&\text{For $\mathsf{G}_i=\{p\}\,(\text{or}\,\{q\})$},~~C(p\,(\text{or}\,q))=\epsilon_{p\,(\text{or}\,q)}\cdot \ell_i,~~\pmb\alpha_i=\{p\}\,\text{or}\,\{q\},~~~\phi_{\pmb\alpha_i|\,\pmb\rho_i}\\
&&\text{For $\mathsf{G}_i=\{p,q\}$},~~~~~~~~~~C^{(0)}(\pmb\alpha_i)=\Bigg\{\begin{matrix}\epsilon_{p}\cdot \ell_i\,\epsilon_q\cdot\ k_p,~~&\pmb\alpha_i=\{p,q\}\\ \epsilon_{q}\cdot \ell_i\,\epsilon_p\cdot\ k_q,~~&\pmb\alpha_i=\{q,p\}\end{matrix},~~~~\,\phi_{\pmb\alpha_i|\,\pmb\rho_i}\\
&&~~~~~~~~~~~~~~~~~~~~~~~~~~~~~~\,C^{(1)}(\pmb\alpha_i)=\Bigg\{\begin{matrix}(-k_p\cdot \ell_i)(\epsilon_{p}\cdot\epsilon_q),~~&\pmb\alpha_i=\{p,q\},&~\phi_{\pmb\alpha_i|\,\pmb\rho_i}\\ \left(-\frac{1}{2}\right)\epsilon_p\cdot\epsilon_q,~~&\pmb\alpha_i=\{p\text{-}q\},&~\phi_{p|\,a}\phi_{q|\,b}\end{matrix},
\eea
In the case of $\mathsf{G}_i=\{p,q\}$, the $C^{(1)}$ has nonvanishing expression only when $p$ appears to the left of $q$, due to our choice of reference order $p\prec q$. The total contribution of $(\epsilon_p\cdot\epsilon_q)^1$ is the sum of the $C^{(1)}(p,q)$ which reflects a substructure with a subcurrent $\phi_{\pmb\alpha_i|\,\pmb\rho_i}$ attached to the loop
\bea
\begin{minipage}{2cm}\includegraphics[width=2cm]{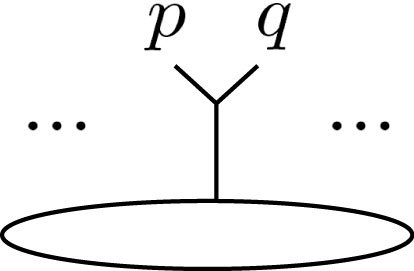}\end{minipage},
\eea
 and $C^{(1)}(p\text{-}q)$  which reflects the substructure with $p$, $q$, attached to the loop via a four-point vertex structure
 \bea
\begin{minipage}{2cm}\includegraphics[width=2cm]{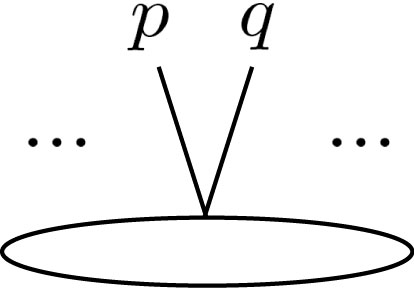}\end{minipage}.
\eea
\end{itemize}

\section{Understanding the tree-level expansion formulas (\ref{tree-yms-expansion})}

According to the tree-level recursive expansion formula \cite{Fu:2017uzt}, when single-trace YMS (or EYM) amplitude with only one gluon (or graviton) $p$ is expanded in terms of BS (or YM) amplitudes, the expansion coefficients have the form $\epsilon_p\cdot X_p$ which agree with (\ref{Eq:TreeCF}). 

For YMS (or EYM) amplitude with two gluons (or two gravitons) $p$, $q$, we apply the recursive expansion relation \cite{Fu:2017uzt} to express the full YMS (or EYM) amplitude in terms of BS or YM amplitudes with the reference order $p\prec q$. The resulted coefficients can be written as
\bea
C(1,\pmb\alpha)&=&C^{(0)}(1,\pmb\alpha)+C^{(1)}(1,\pmb\alpha), \label{Eq:appTree1}
\eea
where the two terms corresponding to the $(\epsilon_p\cdot\epsilon_q)^0$ and $(\epsilon_p\cdot\epsilon_q)^1$ are given by 
\bea
C^{(0)}(1,\pmb\alpha)&=&\epsilon_p\cdot X_p(1,\pmb\alpha)\,\epsilon_q\cdot X_q(1,\pmb\alpha)\nn
C^{(1)}(1,\pmb\alpha)&=&\Bigg\{\begin{matrix}(-k_p\cdot X_p(1,\pmb\alpha))\,(\epsilon_p\cdot\epsilon_q),&~~\pmb\alpha\in\{2,...,r-1\}\shuffle\{p,q\}\\
0,&~~\pmb\alpha\in\{2,...,r-1\}\shuffle\{q,p\}
\end{matrix}.\label{Eq:appTree2}
\eea
To redefine the coefficients, we introduce the following combination of amplitudes
\bea
0&=&\left(-{\epsilon_p\cdot\epsilon_q\over k_p\cdot k_q}\right)\Sl_{\pmb\alpha\in\{2,...,r-1\}\shuffle\{p\}\shuffle\{q\}}k_q\cdot Y_q(1,\pmb\alpha)\,k_p\cdot X_p(1,\pmb\alpha)\,A^{\text{BS}}(1,\pmb\alpha,r\,|\,\pmb\rho),\nn
&&~~~~~~~~~~~~~~~~~~~~~~~~~~~~~~~~~~~~~~\left(\,\text{or $A^{\text{BS}}(1,\pmb\alpha,r\,|\,\pmb\rho)\to A^{\text{YM}}(1,\pmb\alpha,r)$ for EYM}\,\right).
\eea
where $Y^{\mu}_q(1,\pmb\alpha)$ denotes the total momentum of {\it scalars} (or {\it gluons}) between $1$ and $q$ (including $1$) for YMS (or EYM). The the rhs. vanishes due to the fundamental BCJ relation for $p$. When this vanishing term is added to the expansion relation (\ref{tree-yms-expansion}) with coefficients (\ref{Eq:appTree1}), (\ref{Eq:appTree2}), the coefficient for a given permutation agrees with that in (\ref{Eq:TreeCF}).

For tree-level YMS amplitude $A_{\text{tree}}^{\text{YMS}}(1,...,r||\mathsf{G}\,|\,\pmb\rho)$ with three gluons $p$, $q$, $s$, the expansion coefficients are given as 
\bea
C(1,\pmb\alpha)&=&C^{(0)}(1,\pmb\alpha)+C_{\epsilon_p\cdot\epsilon_q}^{(1)}(1,\pmb\alpha)+C_{\epsilon_p\cdot\epsilon_s}^{(1)}(1,\pmb\alpha)+C_{\epsilon_q\cdot\epsilon_s}^{(1)}(1,\pmb\alpha), \label{Eq:appTree3}
\eea
where $C^{(0)}(1,\pmb\alpha)$ denotes the coefficient with no $\epsilon\cdot\epsilon$, while $C_{\epsilon_p\cdot\epsilon_q}^{(1)}$,  $C_{\epsilon_p\cdot\epsilon_s}^{(1)}$, $C_{\epsilon_q\cdot\epsilon_s}^{(1)}$ denote the coefficent with a factor $\epsilon_p\cdot\epsilon_q$, $\epsilon_p\cdot\epsilon_s$, $\epsilon_q\cdot\epsilon_s$, respectively. If we choose the reference order as $p\prec q\prec s$, these coefficients, according to \cite{Fu:2017uzt} are displayed as follows:
\bea
C^{(0)}(1,\pmb\alpha)&=&\epsilon_p\cdot X_p(1,\pmb\alpha)\,\epsilon_q\cdot X_q(1,\pmb\alpha)\,\epsilon_s\cdot X_s(1,\pmb\alpha),~~~~\pmb\alpha\in\{2,...,r-1\}\shuffle\{p\}\shuffle\{q\}\nn
C_{\epsilon_p\cdot\epsilon_q}^{(1)}(1,\pmb\alpha)&=&\left\{\begin{matrix}(-k_p\cdot X_p(1,\pmb\alpha))\,(\epsilon_p\cdot\epsilon_q)\,\Big[\epsilon_s\cdot Y_s(1,\pmb\alpha)+\epsilon_s\cdot k_q\Big],&~~\pmb\alpha\in\{2,...,r-1\}\shuffle\{p,q,s\}\\
(-k_p\cdot X_p(1,\pmb\alpha))\,(\epsilon_p\cdot\epsilon_q)\,(\epsilon_s\cdot k_p),&~~\pmb\alpha\in\{2,...,r-1\}\shuffle\{q,p,s\}\\
(-k_p\cdot X_p(1,\pmb\alpha))\,(\epsilon_p\cdot\epsilon_q)\,(\epsilon_s\cdot Y_s(1,\pmb\alpha)),&~~\pmb\alpha\in\{2,...,r-1\}\shuffle\{p,s,q\}\\
0,&~~\pmb\alpha\in\{2,...,r-1\}\shuffle\{q,s,p\}\\
(-k_p\cdot X_p(1,\pmb\alpha))\,(\epsilon_p\cdot\epsilon_q)\,(\epsilon_s\cdot X_s(1,\pmb\alpha)),&~~\pmb\alpha\in\{2,...,r-1\}\shuffle\{s,p,q\}\\
0,&~~\pmb\alpha\in\{2,...,r-1\}\shuffle\{s,q,p\}\\
\end{matrix}\right.\nn
C_{\epsilon_p\cdot\epsilon_s}^{(1)}(1,\pmb\alpha)&=&\left\{\begin{matrix}(-k_p\cdot X_p(1,\pmb\alpha))\,(\epsilon_p\cdot\epsilon_s)\,(\epsilon_q\cdot X_q(1,\pmb\alpha)),&~~\pmb\alpha\in\{2,...,r-1\}\shuffle\{p,q,s\}\\
\Big[(-k_p\cdot X_p(1,\pmb\alpha))\,(\epsilon_p\cdot\epsilon_s)\,(\epsilon_q\cdot X_q(1,\pmb\alpha))&\\
~~~~~~~~~~~~\,+(-k_q\cdot X_q(1,\pmb\alpha))(-\epsilon_q\cdot k_p)(\epsilon_p\cdot\epsilon_s)\Big],&~~\pmb\alpha\in\{2,...,r-1\}\shuffle\{q,p,s\}\\
(-k_p\cdot X_p(1,\pmb\alpha))\,(\epsilon_p\cdot\epsilon_s)\,(\epsilon_q\cdot X_q(1,\pmb\alpha)),&~~\pmb\alpha\in\{2,...,r-1\}\shuffle\{p,s,q\}\\
0,&~~\pmb\alpha\in\{2,...,r-1\}\shuffle\{q,s,p\}\\
0,&~~\pmb\alpha\in\{2,...,r-1\}\shuffle\{s,p,q\}\\
0,&~~\pmb\alpha\in\{2,...,r-1\}\shuffle\{s,q,p\}\\
\end{matrix}\right.\nn
C_{\epsilon_q\cdot\epsilon_s}^{(1)}(1,\pmb\alpha)&=&C_{\epsilon_p\cdot\epsilon_s}^{(1)}(1,\pmb\alpha)\big|_{p\leftrightarrow q}.
\label{Eq:appTree4}
\eea
The vanishing terms that are added to the full amplitude are presented by
\bea
I&=&I_{\epsilon_p\cdot\epsilon_q}+I_{\epsilon_p\cdot\epsilon_s}+I_{\epsilon_q\cdot\epsilon_s}\nn
0&=&I_{\epsilon_p\cdot\epsilon_q}\equiv\left(-{\epsilon_p\cdot\epsilon_q\over k_p\cdot k_q}\right)\,\Biggl[\,\Sl_{\pmb\alpha\in\{2,...,r-1\}\shuffle\{p\}\shuffle\{q,s\}}\,\,(k_q\cdot Y_q(1,\pmb\alpha))\,(\epsilon_s\cdot k_q)\,(k_p\cdot X_p(1,\pmb\alpha))\,A(1,\pmb\alpha,r|\,\pmb\rho)\nn
&&~~~~~~~~~~~~~~+\Sl_{\pmb\alpha\in\{2,...,r-1\}\shuffle\{p\}\shuffle\{q\}\shuffle\{s\}}(k_q\cdot Y_q(1,\pmb\alpha))\,(\epsilon_s\cdot Y_s(1,\pmb\alpha))\,(k_p\cdot X_p(1,\pmb\alpha))\,A(1,\pmb\alpha,r|\,\pmb\rho)\nn
&&~~~~~~~~~~~~~~\,+\Sl_{\pmb\alpha\in\{2,...,r-1\}\shuffle\{q\}\shuffle\{p,s\}}(k_p\cdot Y_p(1,\pmb\alpha))\,(\epsilon_r\cdot k_p)\,(k_q\cdot X_q(1,\pmb\alpha))\,A(1,\pmb\alpha,r|\,\pmb\rho)\nn
&&~~~~~~~~~~~~~~\,+\Sl_{\pmb\alpha\in\{2,...,r-1\}\shuffle\{s,q\}\shuffle\{p\}}(\epsilon_r\cdot Y_r(1,\pmb\alpha))\,(k_q\cdot k_r)\,(k_p\cdot X_p(1,\pmb\alpha))\,A(1,\pmb\alpha,r|\,\pmb\rho)\,\Biggr],\nn
0&=&I_{\epsilon_p\cdot\epsilon_s}\equiv\left(-{\epsilon_p\cdot\epsilon_s\over k_p\cdot k_s}\right)\,\Biggl[\,\Sl_{\pmb\alpha\in\{2,...,r-1\}\shuffle\{p\}\shuffle\{q\}\shuffle\{s\}}\,\,(\epsilon_q\cdot Y_q(1,\pmb\alpha))\,(k_s\cdot Y_s(1,\pmb\alpha))\,(k_p\cdot X_p(1,\pmb\alpha))\,A(1,\pmb\alpha,r|\,\pmb\rho)\nn
&&~~~~~~~~~~~~~~+\Sl_{\pmb\alpha\in\{2,...,r-1\}\shuffle\{p\}\shuffle\{s,q\}}\,\,(\epsilon_q\cdot k_s)\,(k_s\cdot Y_s(1,\pmb\alpha))\,(k_p\cdot X_p(1,\pmb\alpha))\,A(1,\pmb\alpha,r|\,\pmb\rho)\nn
&&~~~~~~~~~~~~~~\,+\Sl_{\pmb\alpha\in\{2,...,r-1\}\shuffle\{p\}\shuffle\{q,s\}}\,(\epsilon_q\cdot Y_q(1,\pmb\alpha))\,(k_s\cdot k_q)\,(k_p\cdot X_p(1,\pmb\alpha))\,A(1,\pmb\alpha,r|\,\pmb\rho)\nn
&&~~~~~~~~~~~~~~\,+\Sl_{\pmb\alpha\in\{2,...,r-1\}\shuffle\{p,q\}\shuffle\{s\}}\,(k_p\cdot Y_p(1,\pmb\alpha))\,(\epsilon_q\cdot k_p)\,(k_s\cdot X_s(1,\pmb\alpha))\,A(1,\pmb\alpha,r|\,\pmb\rho)
\,\Biggr].\nn
0&=&I_{\epsilon_q\cdot\epsilon_s}=\left(I_{\epsilon_p\cdot\epsilon_s}\right)|_{p\leftrightarrow q}.\label{Eq:appTree5}
\eea
All the above terms vanish due to BCJ relations. When we add (\ref{Eq:appTree5}) to the original expansion formula (\ref{tree-yms-expansion}) of YMS amplitude with coefficients(\ref{Eq:appTree3}), (\ref{Eq:appTree4}), recollect the coefficient corresponding to each permutation, and apply the graphic BCJ relation \cite{Hou:2018bwm} with three elements to the $\epsilon_p\cdot\epsilon_s$ and $\epsilon_q\cdot\epsilon_s$ parts, we get the expected symmetric coefficients (\ref{Eq:TreeCF}) for the YMS amplitude with three gluons. The coefficients for EYM amplitude with three gravitons follow from the same construction.




\end{document}